\documentclass[universe,article,accept,moreauthors,pdftex]{Definitions/mdpi} 

\graphicspath{{./Definitions/}}
\usepackage{relsize}
\usepackage{blindtext}
\usepackage{graphicx}
\usepackage{bm}
\usepackage{cancel}
\usepackage{epsfig}
\usepackage{amsmath}
\usepackage{amssymb}
\usepackage{hhline}
\usepackage{adjustbox}
\usepackage{caption}
\usepackage{longtable}
\usepackage{rotating}
\usepackage{tablefootnote}
\usepackage{float}

\def\aaps{{Astron.\@  \@Astroph.\ }}
\def\apss{{Astroph.\@ \& \@Space \@Science\ }}

\def \beq  {\begin{equation}}
\def \eeq  {\end{equation}}
\def \ber  {\begin{eqnarray}}
\def \eer  {\end{eqnarray}}

\firstpage{1} 
\pubvolume{1}
\issuenum{1}
\articlenumber{0}
\pubyear{2022}
\copyrightyear{2022}
\externaleditor{Academic Editor: {Firstname Lastname}}
\datereceived{} 
\dateaccepted{} 
\datepublished{} 
\hreflink{https://doi.org/} 

\Title{A reanalysis of the latest SH0ES data for $H_0$: \\ Effects of new degrees of freedom on the Hubble tension.} 

\TitleCitation{title}

\Author{Leandros Perivolaropoulos$^{1,\dagger}$ \orcidA{} and Foteini Skara$^{1,\ddagger}$ \orcidB{}}

\AuthorNames{Leandros Perivolaropoulos, Foteini Skara}

\address{$^{1}$ \quad Department of Physics, University of Ioannina, 45110 Ioannina, Greece}

\AuthorCitation{Perivolaropoulos, L.;Skara, F.}

\firstnote{leandros@uoi.gr} 
\secondnote{f.skara@uoi.gr}
\newcommand{\newc}{\newcommand}

\newcommand{\be}{\begin{equation}}
\newcommand{\ee}{\end{equation}}
\newcommand{\ba}{\begin{eqnarray}}
\newcommand{\ea}{\end{eqnarray}}
\newcommand{\bea}{\begin{eqnarray*}}
\newcommand{\eea}{\end{eqnarray*}}
\newc{\D}{\partial}
\newc{\ie}{{i.e.,} }
\newc{\eg}{{e.g.,} }
\newc{\etc}{{etc.} }
\newc{\etal}{{et al.}}
\newc{\lcdm}{$\Lambda$CDM }
\newc{\lcdmnospace}{$\Lambda$CDM}
\newc{\wcdm}{$w$CDM }
\newc{\plcdm}{Planck18/$\Lambda$CDM }
\newc{\plcdmnospace}{Planck18/$\Lambda$CDM}
\newc{\omom}{$\Omega_{0m}$ }
\newc{\omomnospace}{$\Omega_{0m}$}
\newcommand{\nn}{\nonumber}
\newc{\ra}{\Rightarrow}
\newc{\baodv}{$\frac{D_V}{r_s}$ }
\newc{\baodvnospace}{$\frac{D_V}{r_s}$}
\newc{\baoda}{$\frac{D_A}{r_s}$ } 
\newc{\baodanospace}{$\frac{D_A}{r_s}$}
\newc{\baodh}{$\frac{D_H}{r_s}$ }
\newc{\baodhnospace}{$\frac{D_H}{r_s}$}

\abstract{We reanalyze in a simple and comprehensive manner the recently released SH0ES data for the determination of $H_0$. We focus on testing the homogeneity of the Cepheid+SnIa sample and the  robustness of the results in the presence of new degrees of freedom in the modeling of Cepheids and SnIa. We thus focus on the four modeling parameters of the analysis: the fiducial luminosity of SnIa $M_B$ and Cepheids $M_W$ and the two parameters ($b_W$ and $Z_W$) standardizing Cepheid luminosities with period  and metallicity. After reproducing the SH0ES baseline model results, we allow for a transition of the value of any one of these parameters at a given distance $D_c$ or cosmic time $t_c$ thus adding a single degree of freedom in the analysis.  When the SnIa absolute magnitude $M_B$ is allowed to have a transition at $D_c\simeq 50Mpc$ (about $160Myrs$ ago), the best fit value of the Hubble parameter drops from $H_{0}=73.04\pm1.04\,km\,s^{-1}\,Mpc^{-1}$ to $H_0=67.32\pm 4.64\, km\,s^{-1}\,Mpc^{-1}$  in full consistency with the Planck value. Also,  the best fit SnIa absolute magnitude $M_B^>$ for $D>D_c$ drops to the Planck inverse distance ladder value $M_{B}^>=-19.43\pm 0.15$ while the low distance best fit $M_B^<$ parameter remains close to the original distance ladder calibrated value $M_{B}^<=-19.25\pm 0.03$. Similar hints for a transition behavior is found for the other three main parameters of the analysis ($b_W$, $M_W$ and $Z_W$) at the same critical distance $D_c\simeq 50\,Mpc$ even though in that case the best fit value of $H_0$ is not significantly affected. When the inverse distance ladder constraint on $M_B^>$ is included in the analysis, the uncertainties for $H_0$ reduce dramatically ($H_0= 68.2\pm 0.8\, km\,s^{-1}\,Mpc^{-1}$) and the $M_B$ transition model is strongly preferred over the baseline SH0ES model ($\Delta \chi^2 \simeq -15$, $\Delta AIC \simeq -13$) according to AIC and BIC model selection criteria.}

\keyword{Hubble tension; cosmology; galaxies; Cepheid calibrators; gravitational transition; SH0ES data}

\begin{document}
\section{Introduction}

\subsection{The current status of the Hubble tension and its four assumptions}

Measurements of the Hubble constant using observations of type Ia supernovae (SnIa) with Cepheid calibrators  by the SH0ES Team has lead to a best fit value $H_{0}^{R21}=73.04\pm1.04$~km~s$^{-1}$~Mpc$^{-1}$ \cite{Riess:2021jrx} (hereafter R21). This highly precise but not necessarily accurate measurement is consistent with a wide range of other less precise local measurements of $H_0$ using alternative SnIa calibrators \cite{Freedman:2021ahq,Gomez-Valent:2018hwc,Pesce:2020xfe,Freedman:2020dne}, gravitational lensing \cite{Wong:2019kwg,Chen:2019ejq,Birrer:2020tax,Birrer:2018vtm}, gravitational waves \cite{LIGOScientific:2018gmd,Hotokezaka:2018dfi,LIGOScientific:2017adf,DES:2020nay,DES:2019ccw}, gamma-ray bursts as standardizable candles \cite{Cao:2022wlg,Cao:2022yvi,Dainotti:2022rea,Dainotti:2022wli,Dainotti:2013cta}, quasars as distant standard candles \cite{Risaliti:2018reu}, type II supernovae \cite{deJaeger:2022lit,deJaeger:2020zpb}, $\gamma-$ray attenuation \cite{Dominguez:2019jqc} etc. (for recent reviews see Refs. \cite{DiValentino:2021izs, Perivolaropoulos:2021jda}). This measurement is based on two simple assumptions:
\begin{itemize}
    \item There are no significant systematic errors in the measurements of the properties (period, metallicity) and luminosities of Cepheid calibrators and SnIa.
    \item The physical laws involved and the calibration properties of Cepheids and SnIa in all the rungs of the distance ladder are well understood and modelled properly.
\end{itemize}
This measurement however is at $5\sigma$ tension (Hubble tension) with the corresponding measurement from {\em Planck} observations of the CMB angular power spectrum (early time inverse distance ladder measurement) $H_0^{P18}=67.36\pm0.54$~km~s$^{-1}$~Mpc$^{-1}$  \cite{Planck:2018vyg} (see also Refs. \cite{Perivolaropoulos:2021jda,Abdalla:2022yfr,DiValentino:2021izs,Shah:2021onj,Knox:2019rjx,Vagnozzi:2019ezj,Ishak:2018his,Mortsell:2018mfj,Huterer:2017buf,Bernal:2016gxb} for relevant recent reviews). This inverse distance ladder measurement is also based on two basic assumptions:
\begin{itemize}
    \item The scale of the sound horizon at the last scattering surface is the one calculated in the context of the standard cosmological model with the known degrees of freedom (cold dark matter, baryons and radiation) and thus it is a reliable distance calibrator.
    \item The evolution of the Hubble free expansion rate $E(z)\equiv H(z)/H_0$ from the time of recombination (redshift $z=z_{rec}$) until the present time ($z=0$) is the one predicted by the standard \lcdm model  as defined by the best fit Planck parameters  (Planck18$/\Lambda$CDM) \cite{Planck:2018vyg,eBOSS:2020yzd}.
\end{itemize}

A wide range of approaches have been implemented in efforts to explain this Hubble tension (for reviews see Refs. \cite{Bernal:2016gxb,Perivolaropoulos:2021jda,Verde:2019ivm,DiValentino:2021izs,Schoneberg:2021qvd,Abdalla:2022yfr,Krishnan:2020obg,Jedamzik:2020zmd}). These approaches introduce new degrees of freedom that violate at least one of the above four assumptions and may be classified in accordance with the particular assumption they are designed to violate.

Thus, early time sound horizon models introduce new degrees of freedom at the time just before recombination (e.g. early dark energy \cite{Poulin:2018cxd,Niedermann:2019olb,Verde:2019ivm,Smith:2022hwi,Smith:2020rxx,Chudaykin:2020acu,Fondi:2022tfp,Sabla:2022xzj,Herold:2021ksg,McDonough:2021pdg,Hill:2020osr,Sakstein:2019fmf,Niedermann:2020qbw,Rezazadeh:2022lsf}, 
radiation \cite{Green:2019glg,Schoneberg:2022grr,Seto:2021xua,CarrilloGonzalez:2020oac} or modified gravity \cite{Braglia:2020auw,Abadi:2020hbr,Renk:2017rzu,Nojiri:2022ski,Lin:2018nxe,CANTATA:2021ktz}) to change the expansion rate at that time and thus decrease the sound horizon scale $r_s$ (early time distance calibrator) to increase $H_0$, which is degenerate with $r_s$, to a value consistent with local measurements. 

The mechanism proposed by these models attempts to decrease the  scale of the sound horizon at recombination which can be calculated as
\be
 r_s =\int_0^{t_*} \frac{c_s(a)}{a(t)}dt=\int_{z_{*}}^\infty \frac{c_s(z)}{H(z;\rho_b,\rho_{\gamma},\rho_c)}dz
=\int_0^{a_*}\frac{c_s(a)}{a^2H(a;\rho_b,\rho_{\gamma},\rho_c)}da
\label{rsdef}
\ee
where the recombination redshift $z_*$ corresponds to time $t_*$, $\rho_b$, $\rho_c$ and $\rho_\gamma$ denote the densities for baryon, cold dark matter and radiation (photons) respectively and $c_s$ is the  sound  speed  in  the  photon-baryon  fluid.
The angular scale of the sound horizon is measured by the peak locations of the CMB perturbations angular power spectrum and may be expressed in terms of $r_s$ as
\be 
\theta_s=\frac{r_s}{d_A}=\frac{H_0 r_s}{c\int_0^{z_{*}} \frac{dz'}{E(z')}}
\label{thetas}
\ee 
where $d_A$ is the comoving angular diameter distance to last scattering  (at redshift $z\approx 1100$) and $E(z)$ is the dimensionless normalized Hubble parameter which for a flat $\Lambda$CDM model is given by
\be
E(z)\equiv \frac{H(z)}{H_0}=\left[\Omega_{0m}(1+z)^3 +(1-\Omega_{0m})\right]^{1/2}
\ee
Eq. (\ref{thetas}) indicates that there is a degeneracy between $r_s$, $H_0$ and $E(z)$ given the measured value of $\theta_s$. A decrease of $r_s$ would lead to an increase of the predicted value of $H_0$ (early time models) and a late time deformation of $E(z)$ could lead to an increase of the denominator of Eq. (\ref{thetas}) leading also to an increase of $H_0$ (late time models).

Early dark energy models have the problem of predicting stronger growth of perturbations than implied by dynamical probes like redshift space distortion (RSD) and weak lensing (WL) data and thus may  worsen the $\Omega_m$-$\sigma_8$ growth tension \cite{Benisty:2020kdt,Heymans:2020gsg,Kazantzidis:2018rnb,Joudaki:2017zdt,Kazantzidis:2019nuh,Skara:2019usd,Avila:2022xad, Kohlinger:2017sxk,Nunes:2021ipq,Clark:2021hlo}  and reduce consistency with growth data and with other cosmological probes and conjectures \cite{Ivanov:2020ril,Hill:2021yec,Hill:2020osr,Clark:2021hlo,Jedamzik:2020zmd,Herold:2021ksg,Vagnozzi:2021gjh,Krishnan:2020obg,Philcox:2022sgj,McDonough:2021pdg}. Thus, a compelling and full resolution of the Hubble tension may require multiple (or other) modifications beyond the scale of the sound horizon predicted by $\Lambda$CDM cosmology. Even though these models are severely constrained by various cosmological observables, they currently constitute the most widely studied class of models  \cite{Smith:2020rxx,Chudaykin:2020igl,Sakstein:2019fmf,Reeves:2022aoi,Chudaykin:2020acu,Smith:2022hwi}.

Late time $H(z)$ deformation models introduce new degrees of freedom (e.g. modified gravity \cite{SolaPeracaula:2020vpg,Braglia:2020iik,Pogosian:2021mcs,Bahamonde:2021gfp} dynamical late dark energy \cite{DiValentino:2020naf,Alestas:2020mvb,DiValentino:2019jae,Pan:2019hac,Li:2019yem,Zhao:2017cud,Keeley:2019esp,SolaPeracaula:2018wwm,Yang:2018qmz,Krishnan:2020vaf, Dainotti:2021pqg,Colgain:2022nlb,Colgain:2022rxy,Zhou:2021xov} or interacting dark energy with matter \cite{DiValentino:2019ffd,Yang:2018euj,Vattis:2019efj,DiValentino:2019jae,Yang:2018uae,Ghosh:2019tab}) to deform $E(z)$ at redshifts $z\sim O(1)$ so that the present time value of $H(z=0)=H_0$ increases and becomes consistent with the local measurements. This class of models is even more severely constrained \cite{Brieden:2022lsd,Alestas:2020mvb,Alestas:2021xes,Keeley:2022ojz,Clark:2020miy,DES:2020mpv,Anchordoqui:2022gmw,DES:2022doi,Cai:2022dkh,Heisenberg:2022gqk,Vagnozzi:2021tjv,Davari:2022uwd}, by other cosmological observables (SnIa, BAO and growth of perturbations probes) which tightly constrain any deformation \cite{Alam:2020sor} from the \plcdm shape of $E(z)\equiv H(z)/H_0$.

The third approach to the resolution of the Hubble tension is based on a search for possible unaccounted systematic effects including possible issues in modelling 
the Cepheid data such as non-standard dust induced color correction \cite{Mortsell:2021nzg}, the impact of outliers \cite{Efstathiou:2013via,Efstathiou:2020wxn,Efstathiou:2021ocp}, blending effects, SnIa color properties \cite{Wojtak:2022bct},   robustness of the constancy of the SnIa absolute magnitude in the Hubble flow \cite{Benisty:2022psx,Martinelli:2019krf,1968ApJ...151..547T,Kang:2019azh,Rose:2019ncv,Jones:2018vbn,Rigault:2018ffm,2018ApJ...854...24K,Colgain:2019pck,Kazantzidis:2019nuh,Kazantzidis:2020tko,Sapone:2020wwz,Koo:2020ssl,Kazantzidis:2020xta,Lukovic:2019ryg,Tutusaus:2018ulu,Tutusaus:2017ibk,Drell:1999dx} etc.  There is currently a debate about the importance of these potential systematic effects \cite{Kenworthy:2019qwq,Riess:2022mme,Yuan:2022kxa,Riess:2018byc}. Also the possibility for redshift evolution of Hubble constant was studied by Ref. \cite{Dainotti:2022bzg}. In  \cite{Dainotti:2022bzg} the Hubble tension has been analyzed with a binned analysis on the Pantheon sample of SNe Ia through a multidimensional MCMC analysis. Finally the need for new standardizable candles with redshift values far beyond the SnIa ($3\lesssim z\lesssim 9$) has been studied (see Refs. \cite{Cao:2022wlg,Cao:2022yvi,Dainotti:2022rea,Dainotti:2022wli,Dainotti:2013cta} for gamma-ray bursts and Refs. \cite{Bargiacchi:2021hdp,Dainotti:2022rfz} for quasars).


A fourth approach related to the previous one is based on a possible change of the physical laws (e.g. a gravitational transition\cite{Khosravi:2021csn,Perivolaropoulos:2022txg}) during the past $200\,Myrs$ ($z\lesssim 0.01$) when the light of the Cepheid calibrator hosts was emitted \cite{Alestas:2021luu,Desmond:2019ygn,Perivolaropoulos:2021bds,Odintsov:2022eqm,Alestas:2020zol,Marra:2021fvf,Alestas:2021nmi,Perivolaropoulos:2022vql,Perivolaropoulos:2022txg,Odintsov:2022umu}. In this context, new degrees of freedom should be allowed for the modeling of either Cepheid calibrators or/and SnIa to allow for the possibility of this physics change. If these degrees of freedom are shown not to be excited by the data then this approach would also be severely constrained. It is possible however that nontrivial values of these new parameters are favored by the data while at the same time the best fit value of $H_0$ shifts to a value consistent with the inverse distance ladder measurements of $H_0$ using the sound horizon at recombination as calibrator. In this case \cite{Perivolaropoulos:2021bds}, this class of models would be favored especially in view of the severe constraints that have been imposed on the other three approaches. 

The possible new degrees of freedom that could be allowed in the Cepheid+SnIa modeling analysis are clearly infinite but the actual choices to be implemented in a generalized analysis may be guided by three principles: {\it simplicity, physical motivation and improvement of the quality of fit to the data}. 

In a recent analysis \cite{Perivolaropoulos:2021bds} using a previous release of the SH0ES data \cite{Riess:2016jrr,Riess:2019cxk,Riess:2020fzl} we showed that a physically motivated new degree of freedom in the Cepheid calibrator analysis allowing for a transition in one of the Cepheid modelling parameters $R_W$ or $M_W$, is mildly favored by the data and can lead to a reduced best fit value of $H_0$. Here we extend that analysis in a detailed and comprehensive manner, to a wider range of transition degrees of freedom using the latest publicly available SH0ES data described in R21.   

\subsection{The new SH0ES Cepheid+SnIa data}

\begin{table}
\caption{A comparison of the latest SH0ES data release (R21) with previous data updates.}
\label{tab:sh0es} 
\vspace{1.3mm}
\setlength{\tabcolsep}{0.5em}
\begin{adjustwidth}{0.cm}{1cm}
{\footnotesize\begin{tabular}{ccrcc} 
\hhline{=====}
 &&&& \\
SH0ES &  Cepheid + SnIa  & Cepheids\qquad \qquad &  Calibrator & Hubble flow  \\ 
Year/Ref. & host galaxies  &   & SnIa &  SnIa\\  &&&& \\
\hhline{=====}
 &&&& \\
 &  &MW\qquad \quad \quad 15\,      & &  \\
& &LMC$^a$\qquad \quad 785\,   & &  \\
2016& 19  &N4258\quad \quad \quad 139\,   &19  & 217 \\
R16 \cite{Riess:2016jrr}& $ z < 0.01$  & M31\quad \quad \quad\, 372\,  & $ z < 0.01$&  $0.0233< z < 0.15$  \\
\cline{3-3}
 & &Total\qquad \quad1311\, &  &  \\
 & &In SnIa hosts\quad\,  975\, &  &    \\
\cline{3-3}
 & &Total All\quad \quad2286\,&    & \\
  &&&& \\
 \hline 
 &&&& \\ 
 &  &MW\qquad \quad \quad 15\,     & &  \\
&  &LMC$^b$\quad\,  785+70\,   &  &  \\
2019&19&N4258 \quad \quad \quad 139\,   &19&217 \\
R19 \cite{Riess:2019cxk}& $ z < 0.01$  & M31\quad \quad \quad\, 372\,  &$ z < 0.01$  & $0.0233< z < 0.15$    \\
\cline{3-3}
& &Total\qquad \quad 1381\, &  &  \\
 & &In SnIa hosts\quad\, 975\,&  &  \\
\cline{3-3}
 & &Total All\quad \quad 2356\,&    & \\
  &&&& \\
  \hline 
  &&&& \\ 
 &  &MW\qquad \quad \quad 75\,      & &  \\
&  &LMC$^b$ \quad\,785+70\,   & &  \\
2020& 19 &N4258\quad \quad \quad 139\,   &19  &217  \\
R20 \cite{Riess:2020fzl}& $ z < 0.01$ & M31\quad \quad \quad\, 372\,  &$ z < 0.01$ & $0.0233< z < 0.15$   \\
\cline{3-3}
 & &Total\qquad \quad 1441\, &  &    \\
 & &In SnIa hosts\quad\, 975\, &  &   \\
\cline{3-3}
 & &Total All \quad \quad2416\,&    & \\
  &&&& \\
  \hline
  &&&& \\ 
   &   &LMC$^b$\quad\, 270+69\,&&   \\
& &SMC$^a$\qquad \quad 143\, &  &    \\

2021 &37&N4258\quad \quad \quad 443\,& 42 & 277\\
 R21 \cite{Riess:2021jrx}&$0.0015\lesssim z < 0.011 $  &M31\quad \quad \quad\,\,\,\, 55\,&$0.0015\lesssim z < 0.011 $& $0.023< z < 0.15$\\
\cline{3-3}
  &&Total\qquad \quad\,\, 980\,&& \\ 
   &&In SnIa hosts\quad\,2150\,&(77 lightcurve  meas.)& \\  
\cline{3-3}   
   &&Total All\quad \quad 3130\,&& \\
   &&&& \\ 
\hhline{=====}
 &&&& \\
\end{tabular} }
\end{adjustwidth}
{\footnotesize NOTE - (a) From the ground. (b) From the ground+HST.}
\end{table}

The new Cepheid+SnIa data release and analysis  from the SH0ES collaboration in R21 includes a significant increase of the sample of SnIa calibrators from 19 in Ref. \cite{Riess:2016jrr} to 42. These SnIa reside in 37 hosts observed between 1980 and 2021 in a redshift range $0.0015\lesssim z<0.011$ (see Table \ref{tab:sh0es} for a more detailed comparison of the latest SH0ES data release with previous updates). These SnIa are calibrated using Cepheids in the SnIa host galaxies. In turn, Cepheid luminosities are calibrated using geometric methods in calibrator nearby galaxies (anchors). These anchor galaxies include the megamaser host NGC$\,$4258\footnote{At a distance $D=7.6Mpc$ \cite{Reid:2019tiq}  NGC$\,$4258 is the closest galaxy, beyond the
Local Group, with a geometric distance measurement.}, the Milky Way (MW) where distances are measured with several parallaxes, and the Large Magellanic Cloud (LMC) where distances are measured  via detached eclipsing binaries \cite{Riess:2019cxk} as well as two supporting anchor galaxies ($M31$ \cite{Li:2021qkc} and the Small Magellanic Cloud (SMC). These supporting galaxies are pure Cepheid hosts and do not host SnIa but host large and well measured Cepheid samples. However, geometric measurements of their distances are not so reliable and thus are not directly used in the analysis\footnote{A differential distance measurement of the SMC with respect to the LMC is used and thus LMC+SMC are considered in a unified manner in the released data.}. The calibrated SnIa in the Hubble flow ($z\gtrsim 0.01$) are used to measure $H_0$  due to to their high precision (5\% in distance per source) and high luminosity which allows deep reach and thus reduces the impact of local velocity flows. 

The new SH0ES data release includes a factor of 3 increase in the sample of Cepheids within NGC$\,$4258. In total it has 2150 Cepheids in SnIa hosts\footnote{45 Cepheids in N1365 are mentioned in R21 but there are 46 in the fits files of the released dataset at Github repository: \href{https://github.com/PantheonPlusSH0ES/DataRelease}{PantheonPlusSH0ES/DataRelease}.}, 980 Cepheids in anchors or supporting galaxies\footnote{A total of 3130 Cepheids have been released in the data fits files but 3129 are mentioned in R21 (see Table \ref{tab:props}). These data are also shown concisely in Table \ref{tab:hoscep} of Appendix \ref{AppendixE} and may be download in electronic form.}, 42 SnIa (with total 77 lightcurve dataset measurements) in 37 Cepheid+SnIa hosts with redshifts $0.0015\lesssim z<0.011$ and 277 SnIa in the Hubble flow in the redshift range $0.023<z<0.15$. In addition 8 anchor based constraints (with uncertainties included) constrain the following Cepheid modeling parameters: $M_W$ (the Cepheid absolute magnitude zeropoint), $b_W$ (the slope of the Cepheid Period-Luminosity P-L relation), $Z_W$ (the slope of the Cepheid Metallicity-Luminosity M-L relation), a zeropoint parameter $zp$ used to refine the Cepheid P-L relation by describing the difference between the ground and HST zeropoints in LMC Cepheids (zp is set to 0 for HST observations), the distance moduli of the anchors NGC$\,$4258 and LMC and a dummy parameter we call $X$ which has been included in the R21 data release and is set to 0 with uncertainty $10^{-9}$ \footnote{This parameter is not defined in R21 but is included in the data release fits files. We thank A. Riess for clarifying this point.}. 

The parameters fit with these data include the four modeling parameters $M_W$, $b_W$, $Z_W$, $M_B$ (the SnIa absolute magnitude),  the 37 distance moduli of SnIa/Cepheid hosts, the distance moduli to the 2 anchors (NGC$\,$4258, LMC) and to the supporting Cepheid host M31, the zeropoint $zp$ of the Cepheid P-L relation in the LMC ground observations, the Hubble parameter and the dummy parameter mentioned above (tightly constrained to 0). This is a total of 47 parameters (46 if the dummy parameter $X$ is ignored). 

In addition to these parameters, there are other modeling parameters like the color and shape correction slopes of SnIa (usually denoted as $\beta$ and $\alpha$) as well as the Wesenheit dust extinction parameter $R_W$ which have been incorporated in the released SnIa and Cepheid apparent magnitudes and thus can not be used as independent parameters in the analysis, in contrast to the previous data release. 

The provided in R21 Cepheid Wesenheit dust corrected dereddened apparent magnitudes  $m_H^W$ are connected with the Wesenheit dust extinction parameter $R_W$ as \cite{1982ApJ...253..575M} (see also Refs. \cite{Riess:2016jrr,Riess:2019cxk})
\be 
m_H^W\equiv m_H-R_W(V-I)
\label{wesmag}
\ee
where $m_H$ is the observed apparent magnitude in the near-infrared $H$ (F160W) band, $V$ (F555W) and $I$ (F814W) are optical mean apparent magnitudes in the corresponding bands. The empirical parameter $R_W$ is also called 'the reddening-free "Wesenheit" color ratio' and is different from $R_H$ which  can be derived from a dust law (e.g. the Fitzpatrick law \cite{Fitzpatrick:1998pb}). The parameter $R_W$ corrects for both dust and intrinsic variations applied to observed blackbody colors $V-I$.

The provided in R21 SnIa apparent magnitudes $m_B^0$,  standardized using light curve color c and shape $x_1$ corrections are defined as
\be
m_B^0 \equiv m_B-\alpha\; x_{1}-\beta\; c = \mu+M_{B}
\label{mhwdef}
\ee
where $m_B$ is the peak apparent magnitude, $\mu$ is the SnIa distance modulus while the B-band absolute magnitude, $M_{B}$, and correction coefficients $\alpha$ and $\beta$ are fit directly using the SnIa data. The latest SH0ES data release provides the measured values of $m_H^W$ and $m_B^0$ for Cepheid and SnIa respectively which are also used in the corresponding analysis while the parameters $R_W$, $\alpha$ and $\beta$ are fit previously and independently by the SH0ES team to provide the values of $m_H^W$ and $m_B^0$. 

\subsection{The prospect of new degrees of freedom in the SH0ES data analysis}

The homogeneity of the SH0ES data with respect to the parameters $R_W$, $\alpha$ and $\beta$ has been analysed in previous studies with some interesting results. In particular, using the data of the previous SH0ES data release \cite{Riess:2016jrr,Riess:2019cxk,Riess:2020fzl}, it was shown \cite{Perivolaropoulos:2021bds} (see also \cite{Mortsell:2021nzg} for a relevant study) that if the parameter $R_W$ is allowed to vary among Cepheid and SnIa hosts then the fit quality is significantly improved and the best fit value of $H_0$ is lowered to a level consistent with the inverse distance ladder best fit. In addition, a more recent analysis has allowed the parameter $\beta$ to have a different value in Hubble flow SnIa ($\beta=\beta_{HF}$ for $z>0.02$)  compared to calibrating SnIa ($\beta=\beta_{cal}$ for $z<0.01$).  A reanalysis allowing for this new degree of freedom has indicated a tension between the the two best fit values ($\beta_{HF}$ and $\beta_{cal}$) at a level of up to $3.8\sigma$ \cite{Wojtak:2022bct}.

Motivated by these hints for inhomogeneities in the SH0ES data, in what follows we introduce new degrees of freedom in the analysis that are designed to probe the origin of these inhomogeneities. We thus accept three of the four above mentioned assumptions that have lead to the Hubble tension and test the validity of the fourth assumption. In particular we keep the following assumptions:
\begin{enumerate}
   \item 
    There are no significant systematics in the SH0ES data and thus they are reliable.
    \item 
    The CMB sound horizon scale used as a calibrator in the inverse distance ladder approach is correctly obtained in the standard model using the known particles.
    \item 
    The Hubble expansion history from the time of recombination up to $z=0.01$ (or even $z=0$) used in the inverse distance ladder measurement of $H_0$ is provided correctly by the standard Planck18$/\Lambda$CDM cosmological model.
\end{enumerate}

As discussed above there are several studies in the literature that support the validity of these assumptions (e.g.  \cite{Fondi:2022tfp,Keeley:2022ojz}). If these assumptions are valid then the most probable source of the Hubble tension is the violation of the fourth assumption stated above namely  {\it 'the physical laws involved and the calibration properties of Cepheids and SnIa in all the rungs of the distance ladder are well understood and modelled properly'}. 

If this assumption is violated then the modeling of Cepheids+SnIa should be modified to take into account possible changes of physics by introducing new degrees of freedom that were suppressed in the original (baseline) SH0ES analysis.  In the context of this approach, if these degrees of freedom are properly introduced in the analysis then the best fit value of $H_0$ will become consistent with the corresponding inverse distance ladder value of $H_{0}=67.36\pm0.54$~km~s$^{-1}$~Mpc$^{-1}$.

In an effort to pursue this approach for the resolution of the Hubble tension we address the following questions:
\begin{itemize}
    \item How can new degrees of freedom (new parameters) be included in the SH0ES data analysis for the determination of $H_0$?
    \item What are the new degrees of freedom that can expose internal tensions and inhomogeneities in the Cepheid/SnIa data?
    \item What new degrees of freedom can lead to a best fit value of $H_0$ that is consistent with Planck?
\end{itemize}
The main goal of the present analysis is to address these questions. The new degree of freedom we allow and investigate is a transition of any one of the four Cepheid/SnIa modeling parameters at a specific distance $D_c$ or equivalently (in the context of the cosmological principle) at a given cosmic time $t_c$ such that $t_0-t_c=D_c/c$ where $t_0$ is the present cosmic time (age of the Universe). In the context of this new degree of freedom we reanalyse the SH0ES data to find how does the quality of fit to the data and the best fit value of $H_0$ change when the new degree of freedom is excited. The possible introduction of new constraints included in the analysis is also considered.

The structure of this paper is the following: In the next section \ref{sec:standard analysis} we describe the standard analysis of the SH0ES Cepheid+SnIa data in a detailed and comprehensive manner stressing some details of the released dataset that are not described in R21. We also describe some tension between the values of the best fit Cepheid modeling parameters $b_W$ and $Z_W$ obtained in anchor or pure Cepheid host galaxies and the corresponding mean values obtained in SnIa host galaxies. In section \ref{sec:Generalized analysis} we present our generalized analysis with new degrees of freedom that allow a transition of the main modeling parameters at specific distances (cosmic times of radiation emission). We also investigate the effect of the inverse distance ladder constraint on $M_B$ \cite{Camarena:2021jlr, Marra:2021fvf,Gomez-Valent:2021hda}  for both the baseline SH0ES analysis and for our analysis involving the $M_B$ transition degree of freedom. Finally in section \ref{sec:Conclusion} we conclude, discuss the implications of our results and the possible extensions of our analysis.

\section{The new SH0ES data and their standard analysis: Hints for intrinsic tensions}
\label{sec:standard analysis}
\subsection{The original baseline SH0ES analysis: a comprehensive presentation}
\label{sub:baseline}

The main equations used to model the Cepheid SnIa measured apparent magnitudes with parameters that include $H_0$ are described as follows: 
\begin{itemize}
    \item The equation that connects the measured Wesenheit magnitude of the $j$th Cepheid in the $i$th galaxy,  with the host distance moduli $\mu_i$ and the modeling parameters $M_W$, $b_W$ and $Z_W$ is of the form\footnote{For Cepheids in the LMC/SMC anchor observed from the ground the zeropoint parameter $zp$ is added on the RHS and thus Eq. (\ref{wesmagcep}) becomes $m_{H,i,j}^W 
=\mu_i+M_{H,i}^W+b_W[P]_{i,j}+Z_W[O/H]_{i,j}+zp$ to allow for a different P-L zeropoint between ground and HST observations.}
    \be
m_{H,i,j}^W 
=\mu_i+M_{H}^W+b_W[P]_{i,j}+Z_W[O/H]_{i,j}
\label{wesmagcep}
\ee
where $\mu_i$ is the inferred distance modulus to the galaxy, $M_H^W$ is the zeropoint Cepheid absolute magnitude of a period $P = 10\,d$  Cepheid ($d$ for days), and $b_W$-$Z_W$ are the slope parameters that represent  the dependence of magnitude on both period and metallicity. The $[O/H]$ is a measure of the metallicity of the Cepheid. The usual bracket shorthand notation for the metallicity $[O/H]$ represents the Cepheid metal abundance compared to that of the Sun
\be
[O/H]\equiv \log(O/H)-\log(O/H)_{\odot}=\Delta \log(O/H)
\ee

Here O and H is the number of oxygen and hydrogen atoms per unit of volume respectively. The unit often used for metallicity is the dex (decimal exponent) defined as $n\, dex \equiv 10^n$. Also, the bracket shorthand notation for the period $[P]$ is used as  ($P$ in units of days)  
\be
[P]\equiv \log P-1    
\ee
\item
The color and shape corrected SnIa B-band peak magnitude in the $i$th host is connected with the distance modulus $\mu_i$ of the $i$th host and with the SnIa absolute magnitude $M_B$ as shown in Eq. (\ref{mhwdef}) i.e.
\be
m_{B,i}^0=\mu_i+M_B
\label{magsnia}
\ee
The distance modulus is connected with the luminosity distance $d_L$ in $Mpc$ as
\be
\mu= 5 \log (d_L/Mpc) + 25
\label{mudef}
\ee
where in a flat universe 
\be 
d_L(z)=c (1+z) \int_0^z \frac{dz'}{H(z')}=c H_0^{-1} (1+z)\int_0^z \frac{H_0 \; dz'}{H(z')} \equiv  H_0^{-1} \;D_L(z)
\label{dlhz}
\ee
where $D_L(z)$ is the Hubble free luminosity distance which is independent of $H_0$.
\item
Using Eqs. (\ref{magsnia})-(\ref{dlhz}) it is easy to show that that $H_0$ is connected with the SnIa absolute magnitude and the Hubble free luminosity distance as
\be 
5 \log H_0=M_B + 5 \log D_L(z) - m_B^0(z) +25
\label{logh0}
\ee
In the context of a cosmographic expansion of $H(z)$ valid for $z<<1$ we have
\be
\log D_L(z)_c \simeq \log \left[cz\left(1+\frac{1}{2}(1-q_0)z 
-\frac{1}{6}(1-q_0-3q_0^2+j_0)z^2+\mathcal{O}(z^3)\right) \right]
\label{dlcosmogr}
\ee
where $q_0\equiv -\frac{1}{H_0^2}\frac{d^2a(t)}{dt^2}\Big|_{t=t_0}$ and $j_0\equiv \frac{1}{H_0^3}\frac{d^3a(t)}{dt^3}\Big|_{t=t_0}$ are the deceleration and jerk parameters respectively.
Thus Eqs. (\ref{logh0}) and (\ref{dlcosmogr}) lead to the equation that connects $H_0$ with the SnIa absolute magnitude $M_B$ which may be expressed as
\be 
5 \log H_0=M_B + 5 \log D_L(z) - m_B^0(z) +25 \equiv M_B +5\; a_B +25 
\label{abdef}
\ee
where we have introduced the parameter $a_B\equiv \log D_L(z) - 0.2 m_B^0(z)$ as defined in the SH0ES analysis \cite{Riess:2016jrr}.
\end{itemize}

Thus the basic modeling equations used in the SH0ES analysis for the measurement of $H_0$ are Eqs. (\ref{wesmagcep}), (\ref{magsnia}) and  (\ref{abdef}). In these equations the input data are the measured apparent magnitudes (luminosities) of Cepheids $ m_{H,i,j}^W$  and the SnIa apparent magnitudes $m_{B,i}^0$ (in Cepheid+SnIa hosts and in the Hubble flow). The parameters to be fit using a maximum likelihood method are the distance moduli $\mu_i$ (of the anchors and supporting hosts, the Cepheid+SnIa hosts and Hubble flow SnIa), the four modeling parameters ($M_H^W$, $b_W$, $Z_W$ and $M_B$), the Hubble constant $H_0$, the zeropoint $zp$ of the Cepheid P-L relation in the LMC ground measurements and the dummy parameter $X$. This is a total of 47 parameters. The actual data have been released by the SH0ES team as a .fits file in the form of a column vector $Y$ with 3492 entries which includes 8 constraints on the parameters obtained from measurements in anchor galaxies where the distance moduli are measured directly with geometric methods.

The entries of the provided $Y$ data column vector do not include the pure measured apparent magnitudes. Instead its entries are residuals defined by subtracting specific quantities. In particular:
\begin{itemize}
    \item The Cepheid Wesenheit magnitudes are presented as residuals with respect to a fiducial P-L term as
    \be 
    {\bar m}_{H,i,j}^W \equiv m_{H,i,j}^W - b_W^0 [P]
    \label{residmw}
    \ee
    where $b_W^0=-3.286$ is a fiducial Cepheid P-L slope. As a result of this definition the derived best fit slope is actually a residual slope $\Delta b_W \equiv b_W - b_W^0$.
    \item The residual Cepheid Wesenheit magnitudes of the Cepheids in the anchors $N4258$, $LMC$ and the supporting pure Cepheid host $SMC$ (non SnIa hosts), are presented after subtracting a corresponding fiducial distance modulus obtained with geometric methods.\footnote{In the case of SMC a differential distance with respect to LMC is used.}
    \item The SnIa standardized apparent magnitudes in the Hubble flow are presented as residuals after subtracting the Hubble free luminosity distance with cosmographic expansion $5\log D_L(z)_c+25$ (see Eq. (\ref{dlcosmogr})).
\end{itemize}

Thus the released data vector $Y$ has the following form
\be
\nonumber
\begin{tabular}{ccc}
\(\bf {Y}=
\begin{pmatrix}
{\bar m}_{H,1}^W\\
\ldots\\
{\bar m}_{H,2150}^W\\
\hline
{\bar m}_{H,N4258,1}^W-\mu_{0,N4258}\\
\ldots\\
{\bar m}_{H,N4258,443}^W-\mu_{0,N4258}\\
{\bar m}_{H,M31,1}^W\\
\ldots\\
{\bar m}_{H,M31,55}^W\\
{\bar m}_{H,LMC,ground,1}^W-\mu_{0,LMC}\\
\ldots\\
{\bar m}_{H,LMC,ground,270}^W-\mu_{0,LMC}\\
{\bar m}_{H,SMC,ground,1}^W-\mu_{0,SMC}\\
\ldots\\
{\bar m}_{H,SMC,ground,143}^W-\mu_{0,SMC}\\
{\bar m}_{H,LMC,HST,1}^W-\mu_{0,LMC}\\
\ldots\\
{\bar m}_{H,LMC,HST,69}^W-\mu_{0,LMC}\\
\hline
{\bar m}_{B,1}^0\\
\ldots\\
m_{B,77}^0\\
\hline
-5.803 \;  (M_{H,HST}^W)\\
-5.903 \; (M_{H,Gaia}^W)\\
-0.21 \; (Z_{W,Gaia})\\
0  \; (X) \\
0  \; (\Delta zp)\\
0  \; (\Delta b_W)\\
0  \; (\Delta \mu_{N4258}) \\
0  \; (\Delta \mu_{LMC}) \\
\hline
m_{B,1}^0-5\log [cz_1(...)]-25\\
\ldots\\
m_{B,277}^0-5\log [cz_{277}(...)]-25\\
\end{pmatrix}\)  
 &   &
$\begin{matrix} 
\left.\begin{matrix}
\\
\\
\\
\end{matrix}\right\}\,2150\,Cepheids\, in \,37 \,SnIa\, hosts\\
\left.\begin{matrix}
\\
\\
\\
\\
\\
\\
\\
\\
\\
\\
\\
\\
\\
\\
\\
\\
\end{matrix}\right\}\,980\,Cepheids\, in\,non\, SnIa\, hosts \qquad\\
\left.\begin{matrix}
\\
\\
\\
\end{matrix}\right\} \,77\,SnIa\, in \,Cepheid\, hosts\; \quad \; \; \;\;\;\;\\
\left.\begin{matrix}
\\
\\
\\
\\
\\
\\
\\
\\
\end{matrix}\right\}\,8\, External\, constraints \quad\;\; \; \; \;\; \;\;\;\;\\
\left.\begin{matrix}
\\
\\
\\
\end{matrix}\right\}\,277\, SnIa\, in\, Hubble\, flow \;\quad\;\; \; \; \; \\
\end{matrix}$
\\
\end{tabular}
\ee
The 8 external anchor constraints on the parameters that appear in this vector are the following:
\ba
M_H^W&=&-5.803\pm0.082 \nn \\
M_H^W&=&-5.903\pm 0.025 \nn \\
Z_W&=&-0.21\pm 0.12 \nn \\
X&=&0\pm0.00003 \label{constr} \\
\Delta zp &=&0\pm 0.1 \nn \\
\Delta b_W&=&0\pm 10 \nn  \\
\Delta \mu_{N4258}&=&0\pm 0.03 \nn \\
\Delta \mu_{LMC}&=&0\pm 0.026 \nn
\ea
The parameters to be fit using the $Y$ vector data may also be expressed as a vector $q$ with 47 entries of the following form
\begin{center}
\begin{tabular}{ccc}
\bf{q}=
$\begin{pmatrix}
\mu_1\\
\ldots\\
\mu_{37}\\
\Delta\mu_{N4258}\\
M_H^W\\
\Delta\mu_{LMC}\\
\mu_{M31}\\
\Delta b_W\\
M_B\\
Z_W\\
X\\
\Delta zp\\
5\log H_0 
\end{pmatrix}$& &$\left.\begin{matrix}
\\
\\
\\
\\
\\
\\
\\
\\
\\
\\
\\
\\
\\
\end{matrix}\right\}$\,47\, parameters \;\quad \;\; \;\;\;\; \;\;\; \\
\end{tabular}
\end{center}

Using the column vectors {\bf Y} and {\bf q}, Eqs. (\ref{wesmagcep}), (\ref{magsnia}) and  (\ref{abdef}) and the constraints stated above, can be expressed in matrix multiplication form as
\be
\bf{Y} = \bf{Lq }
\label{syst1}
\ee
with $\bf{Y}$ the matrix of measurements (data vector), $\bf{q}$ the matrix of parameters and  $\bf{L}$ a model (or design) matrix which has 3492 rows corresponding to the entries of the $\bf{Y}$ data vector and 47 columns corresponding to the entries of the parameter vector $\bf{q}$.  The model matrix $\bf{L}$ also includes some data (Cepheid period and metallicity) and in the context of this baseline modeling of data has the form 
\begin{adjustwidth}{-4.3cm}{1cm}


\setlength{\tabcolsep}{1pt}
\begin{tabular}{ccc}
\(\bf {L}={\footnotesize
\left( \begin{array}[c]{ccccccccccccc}
1&\ldots&0&0&1&0&0& [P]_1&0&[O/H]_1&0&0&0\\
\ldots&\ldots&\ldots&\ldots&\ldots&\ldots&\ldots&\ldots&\ldots&\ldots&\ldots&\ldots&\ldots\\
0&\ldots&1&0&1&0&0&[P]_{2150}&0&[O/H]_{2150}&0&0&0\\
\hline
0&\ldots&0&1&1&0&0&[P]_{N4258,1}&0&[O/H]_{N4258,1}&0&0&0\\
\ldots&\ldots&\ldots&\ldots&\ldots&\ldots&\ldots&\ldots&\ldots&\ldots&\ldots&\ldots&\ldots\\
0&\ldots&0&1&1&0&0&[P]_{N4258,443}&0&[O/H]_{N4258,443}&0&0&0\\
0&\ldots&0&0&1&0&1&[P]_{M31,1}&0&[O/H]_{M31,1}&0&0&0\\
\ldots&\ldots&\ldots&\ldots&\ldots&\ldots&\ldots&\ldots&\ldots&\ldots&\ldots&\ldots&\ldots\\
0&\ldots&0&0&1&0&1&[P]_{M31,55}&0&[O/H]_{M31,55}&0&0&0\\
0&\ldots&0&0&1&1&0&[P]_{LMC,ground,1}&0&[O/H]_{LMC,ground,1}&0&1&0\\
\ldots&\ldots&\ldots&\ldots&\ldots&\ldots&\ldots&\ldots&\ldots&\ldots&\ldots&\ldots&\ldots\\
0&\ldots&0&0&1&1&0&[P]_{LMC,ground,270}&0&[O/H]_{LMC,ground,270}&0&1&0\\
0&\ldots&0&0&1&1&0&[P]_{SMC,ground,1}&0&[O/H]_{SMC,ground,1}&0&1&0\\
\ldots&\ldots&\ldots&\ldots&\ldots&\ldots&\ldots&\ldots&\ldots&\ldots&\ldots&\ldots&\ldots\\
0&\ldots&0&0&1&1&0&[P]_{SMC,ground,143}&0&[O/H]_{SMC,ground,143}&0&1&0\\
0&\ldots&0&0&1&1&0&[P]_{LMC,HST,1}&0&[O/H]_{LMC,HST,1}&0&0&0\\
\ldots&\ldots&\ldots&\ldots&\ldots&\ldots&\ldots&\ldots&\ldots&\ldots&\ldots&\ldots&\ldots\\
0&\ldots&0&0&1&1&0&[P]_{LMC,HST,69}&0&[O/H]_{LMC,HST,69}&0&0&0\\
\hline
1&\ldots&0&0&0&0&0&0&1&0&0&0&0\\
\ldots&\ldots&\ldots&\ldots&\ldots&\ldots&\ldots&\ldots&\ldots&\ldots&\ldots&\ldots&\ldots\\
0&\ldots&1&0&0&0&0&0&1&0&0&0&0\\
\hline
0&\ldots&0&0&1&0&0&0&0&0&0&0&0\\
0&\ldots&0&0&1&0&0&0&0&0&0&0&0\\
0&\ldots&0&0&0&0&0&0&0&1&0&0&0\\
0&\ldots&0&0&0&0&0&0&0&0&1&0&0\\
0&\ldots&0&0&0&0&0&0&0&0&0&1&0\\
0&\ldots&0&0&0&0&0&1&0&0&0&0&0\\
0&\ldots&0&1&0&0&0&0&0&0&0&0&0\\
0&\ldots&0&0&0&1&0&0&0&0&0&0&0\\
\hline
0&\ldots&0&0&0&0&0&0&1&0&0&0&-1\\
\ldots&\ldots&\ldots&\ldots&\ldots&\ldots&\ldots&\ldots&\ldots&\ldots&\ldots&\ldots&\ldots\\
0&\ldots&0&0&0&0&0&0&1&0&0&0&-1\\
\end{array} \right) } 
$ & {\footnotesize}  &
{\footnotesize $\begin{matrix} 
\left.\begin{matrix}
\\
\\
\\
\end{matrix}\right\} \,2150\,Cepheids\, in \,37 \,SnIa\, hosts\\
\left.\begin{matrix}

\\
\\
\\
\\
\\
\\
\\
\\
\\
\\
\\
\\
\\
\\
\\
\end{matrix}\right\}\,980\,Cepheids\, in\,non\,SnIa\,hosts \qquad\\
\left.\begin{matrix}
\\
\\
\\
\end{matrix}\right\} \,77\,SnIa\, in \,Cepheid\, hosts\,\, \quad \; \; \;\;\;\; \\
\left.\begin{matrix}
\\
\\
\\
\\
\\
\\
\\
\\
\end{matrix}\right\}\,8\, External\, constraints \;\quad \;\; \;\;\;\; \;\;\; \\
\left.\begin{matrix}
\\
\\
\\
\end{matrix}\right\}\,277\, SnIa\, in\, Hubble\, flow \;\quad\;\; \; \; \; \\
\end{matrix}$}
\\
\end{tabular}
\end{adjustwidth}

The system (\ref{syst1}) has 3492 equations and 47 unknown parameter values. Thus it is overdetermined and at best it can be used in the context of the maximum likelihood analysis to find the best fit parameter values that have maximum likelihood and thus minimum $\chi^2$. For the definition of $\chi^2$ the measurement error matrix (covariance matrix) $\bf{C}$ is needed and provided in the data release as square matrix with dimension $3492\times 3492$\footnote{The  $\bf{Y}$, $\bf{L}$ and  $\bf{C}$ matrices are publicly available as fits files by SH0ES team at Github repository: \href{https://github.com/PantheonPlusSH0ES/DataRelease}{PantheonPlusSH0ES/DataRelease}.}. In Appendix \ref{AppendixA} we present the schematic form of the matrix $\bf{C}$ which also includes the standard uncertainties of the constraints as diagonal elements. Using the covariance matrix that quantifies the data uncertainties and their correlation, the  $\chi^2$ statistic may be constructed as  
\be
\chi^2=(\bf{Y}-\bf{Lq})^T\bf{C}^{-1}(\bf{Y}-\bf{Lq})
\label{chi21}
\ee
The numerical minimization of $\chi^2$ in the presence of 47 parameters that need to be fit would be very demanding computationally even with the use of Markov chain Monte Carlo (MCMC) methods. Fortunately, the linear form of the system (\ref{syst1}) allows the analytical minimization of $\chi^2$ and the simultaneous analytic evaluation of the uncertainty of each parameter. In Appendix \ref{AppendixB}  we show that the analytic minimization of $\chi^2$ of Eq. (\ref{chi21}) leads to the best fit parameter maximum likelihood vector\footnote{The results for  the parameters in $\bf{q_{best}}$ are the same as the results obtained using numerical minimization of $\chi^2$ (see the absolute matching of the results in the numerical analysis file "Baseline1 structure of system" in the \href{https://github.com/FOTEINISKARA/A-reanalysis-of-the-SH0ES-data-for-H_0}{A reanalysis of the SH0ES data for $H_0$} GitHub repository).}
\be
\bf{q_{best}}=(\bf{L}^T\bf{C}^{-1}\bf{L})^{-1}\bf{L}^T\bf{C}^{-1}\bf{Y}
\label{bfpar1}
\ee
The $1\sigma$ standard errors for the parameters in $\bf{q_{best}}$ are obtained as the square roots of the 47 diagonal elements of the transformed error matrix
\be
\bf{\varSigma}=(\bf{L}^T\bf{C}^{-1}\bf{L})^{-1}
\label{errmat}
\ee
For example the best fit of the parameter\footnote{We use the standard notation $\log\equiv \log_{10}$ which is important in the error propagation to $H_0$.} $5\log H_0$  is obtained as the 47th entry of the best fit parameter vector $\bf{q_{best}}$ and the corresponding $1\sigma$ standard error is the $\sqrt{\bf{\varSigma_{47,47}}}$ element of the error matrix. Using  equations (\ref{bfpar1}) and (\ref{errmat}) and the latest released data of the SH0ES team presented in R21 we find full agreement with all values of the best fit parameters. For example for $H_0$ we find (after implementing error propagation) $H_0=73.04\pm 1.04km\; s^{-1}\; Mpc^{-1}$ fully in agreement with the published result of R21.

\begin{table}
\caption{ Best fit parameter values in the absence and in the presence of the inverse distance ladder constraint for baseline model.}
\label{tab:resall} 
\vspace{2.5mm}
\setlength{\tabcolsep}{1.8em}
\begin{adjustwidth}{0.3cm}{1.2cm}
{\footnotesize\begin{tabular}{cc ccc} 
\hhline{=====}
   & \\
Parameter & Best fit value&$\sigma$ & Best fit value$^{a}$& $\sigma\,^{a}$\\
     & \\
   \hhline{=====}
     & \\
 $\mu_{M101}$& 29.16& 0.04& 29.20 & 0.04\\
 $\mu_{M1337}$& 32.92& 0.08 & 32.97 & 0.08 \\
 $\mu_{N0691}$& 32.82& 0.09& 32.87 & 0.09\\
 $\mu_{N1015}$& 32.62& 0.069& 32.67 & 0.06\\
 $\mu_{N0105}$& 34.49& 0.12& 34.56 & 0.12\\    
  $\mu_{N1309}$& 32.51& 0.05& 32.56& 0.05\\
 $\mu_{N1365}$& 31.33& 0.05& 31.37& 0.05\\
 $\mu_{N1448}$& 31.3& 0.04& 31.33& 0.03\\
 $\mu_{N1559}$& 31.46& 0.05& 31.51& 0.05\\
 $\mu_{N2442}$& 31.47& 0.05& 31.51& 0.05\\
 $\mu_{N2525}$& 32.01& 0.06& 32.08& 0.06\\
 $\mu_{N2608}$& 32.63& 0.11& 32.69& 0.11\\
 $\mu_{N3021}$& 32.39& 0.1& 32.45& 0.1\\
 $\mu_{N3147}$& 33.09& 0.09& 33.16& 0.08\\
 $\mu_{N3254}$& 32.4& 0.06& 32.46& 0.05\\
 $\mu_{N3370}$& 32.14& 0.05& 32.19& 0.04\\
 $\mu_{N3447}$& 31.94& 0.03& 31.98& 0.03\\
 $\mu_{N3583}$& 32.79& 0.06& 32.84& 0.06\\
 $\mu_{N3972}$& 31.71& 0.07& 31.76& 0.07\\
 $\mu_{N3982}$& 31.64& 0.06& 31.69& 0.05\\
 $\mu_{N4038}$& 31.63& 0.08& 31.69& 0.08\\
 $\mu_{N4424}$& 30.82& 0.11& 30.87& 0.11\\
 $\mu_{N4536}$& 30.84& 0.05& 30.87& 0.05\\
 $\mu_{N4639}$& 31.79& 0.07& 31.83& 0.07\\
 $\mu_{N4680}$& 32.55& 0.15& 32.61& 0.15\\
 $\mu_{N5468}$& 33.19& 0.05& 33.25& 0.05\\
 $\mu_{N5584}$& 31.87& 0.05& 31.91& 0.04\\
 $\mu_{N5643}$& 30.51& 0.04& 30.56& 0.04\\
 $\mu_{N5728}$& 32.92& 0.1& 32.98& 0.1\\
 $\mu_{N5861}$& 32.21& 0.08& 32.26& 0.07\\
 $\mu_{N5917}$& 32.34& 0.08& 32.4& 0.08\\
 $\mu_{N7250}$& 31.61& 0.1& 31.65& 0.1\\
 $\mu_{N7329}$& 33.27& 0.07& 33.33& 0.07\\
 $\mu_{N7541}$& 32.58& 0.08& 32.64& 0.08\\
 $\mu_{N7678}$& 33.27& 0.08& 33.33& 0.08\\
 $\mu_{N0976}$& 33.54& 0.08& 33.61& 0.08\\
 $\mu_{U9391}$& 32.82& 0.05& 32.86& 0.05\\
 $\Delta \mu_{N4258}$& -0.01& 0.02& 0.01& 0.02\\
 $M_H^W$& -5.89& 0.02& -5.92& 0.02\\
 $\Delta \mu_{LMC}$& 0.01& 0.02& 0.03& 0.02\\
 $\mu_{M31}$& 24.37& 0.07& 24.4& 0.07\\
 $\Delta b_W$& -0.013& 0.015& -0.026& 0.015\\
 $M_B$& -19.25& 0.03& -19.33& 0.02\\
 $Z_W$& -0.22& 0.05& -0.22& 0.05\\
$ X$& 0.& 0.& 0.& 0.\\
$ \Delta zp$& -0.07& 0.01& -0.07& 0.01\\
$ 5\log H_0$& 9.32& 0.03& 9.24& 0.02\\
  $H_0$& 73.04& 1.04& 70.5& 0.7\\
 & \\
 \hhline{=====} 
 \\
\end{tabular} }
{\footnotesize NOTE - (a) With constraint $M_B=-19.401 \pm 0.027 $ included in the data vector $\bf{Y}$ and in the model matrix $\bf{L}$ with included uncertainty in the extended covariance matrix $\bf{C}$.}
\end{adjustwidth}
\end{table}

In Table \ref{tab:resall} we show the best fit values and uncertainties for all 47 parameters of the vector $\bf{q}$ including the four Cepheid modeling parameters ($b_W=b_W^0+\Delta b_W\equiv -3.286+\Delta b_W$, $M_H^W$, $Z_W$ and $M_B$) along with best fit value of $H_0$ for the baseline SH0ES model. The corresponding best fit values and uncertainties with an additional constraint on $M_B$ from the inverse distance ladder is also included and discussed in detail in the next section. The agreement with the corresponding results of R21 (left three columns) is excellent.

Before closing this subsection we review a few points that are not mentioned in R21 but are useful for the reproduction of the results and the analysis
\begin{itemize}
    \item The number of entries of the parameter vector ${\bf q}$ is 47 and not 46 as implied in Ref. R21 due to the extra dummy parameter $X$ which is included in the released data $\bf{Y}$, $\bf{L}$ and  $\bf{C}$ but not mentioned in R21.
    \item The entry referred  as $b_W$ in R21 in the parameter vector ${\bf q}$ should be $\Delta b_W$ because it is actually the residual $b_W$ ($\Delta b_W \equiv b_W - b_W^0$) as stated above and not the original slope of the P-L relation (its best fit value is $\Delta b_W=-0.014\pm 0.015$).
    \item The number of constraints shown in the definition of the ${\bf Y}$ vector in R21 is 5 while in the actual released data $\bf{Y}$, $\bf{L}$ and  $\bf{C}$ we have found the 8 constraints defined above.
\end{itemize}

\subsection{Homogeneity of the Cepheid modeling parameters} \label{seccephomog}

Before generalizing the model matrix $\bf{L}$ with new degrees of freedom and/or constraints, it is interesting to test the self-consistency of the assumed universal modeling parameters $b_W$ and $Z_W$ of the Cepheids. These parameters can be obtained separately for each one of the $i=1,...,40$ Cepheid hosts\footnote{37 SnIa/Cepheid hosts and 3 pure Cepheid hosts of which 2 are anchors (N4258 and LMC).} by fitting linear relations. 

In particular, in order to obtain the best fit P-L slope $b_{W,i}$ of the $ith$ Cepheid host we fit the  $m_{H,i,j}^W-\log P_{i,j}$ data (where $P_{i,j}$ is the period in units of days of the $jth$ Cepheid in the $ith$ host) with a linear relation of the form 
\be
m_{H,i,j}^W=s_i+b_{W,i} \log P_{i,j}
\label{bwi}
\ee
with parameters to be fit in each host $s_i-b_{W,i}$ (intercept and slope).
These equations may be expressed as a matrix equation of the form
\be
\bf{Y_i} = \bf{A_iX_i }
\label{syst}
\ee
with $\bf{Y_i}$ the vector of measurements, $\bf{X_i}$ the vector of parameters and  $\bf{A_i}$ the model (or design) matrix. These are defined as
\be
\begin{tabular}{ccccc}
\(\bf {Y_i}=
\begin{pmatrix}
m_{H,i,1}^W\\
m_{H,i,2}^W\\
\vdots\\
m_{H,i,N}^W
\end{pmatrix}\)  
 &,   &
\(
\bf{X_i}=
\begin{pmatrix}
s_i\\
b_{W,i}
\end{pmatrix}\)  
 &,   &
 \(
\bf{A_i}=
\begin{pmatrix}
1& log P_{i,1}\\
1&log P_{i,2}\\
\vdots&\vdots\\
1&log P_{i,N}
\end{pmatrix}\)  \\
&&&&
\end{tabular}
\ee

The analytically (along the lines of the previous subsection and of Appendix \ref{AppendixB}) obtained minimum of 
\be
\chi_i^2=(\bf{Y_i}-\bf{A_iX_i})^T\bf{C_i}^{-1}(\bf{Y_i}-\bf{A_iX_i})
\label{chi2i}
\ee
with respect to the slope $b_{W,i}$ and intercept $s_i$ leads to the best fit values and standard errors for these parameters. For all Cepheids we adopt the $ \mathrm{N\times N}$ covariance matrix $\bf{C_i}$ of standard errors of the magnitude measurements from R21.

Thus, the analytic minimization of $\chi^2$ of Eq. (\ref{chi2i}) leads to the best fit parameter maximum likelihood vector
\be
\bf{X_{i,best}}=(\bf{A_I}^T\bf{C_i}^{-1}\bf{X_i})^{-1}\bf{A_i}^T\bf{C_i}^{-1}\bf{Y_i}
\label{bfpari}
\ee
The $1\sigma$ standard errors for $b_{W,i}$ slope and intercept in $\bf{X_{i,best}}$ are obtained as the square roots of the 2 diagonal elements of the error matrix
\be
\bf{\varSigma}_i=(\bf{A_i}^T\bf{C_i}^{-1}\bf{A_i})^{-1}
\label{errmati}
\ee

A similar analysis was implemented for the other Cepheid modeling parameter, the metallicity slopes $Z_{W,i}$. In this case the linear fit considered was of the form
\be
m_{H,i,j}^W=s_i+Z_{W,i}[O/H]_{i,j} 
\label{zwi}
\ee

\begin{figure*}
\begin{centering}
\setlength{\headheight}{23.60004pt}
\addtolength{\topmargin}{-8.40002pt}
\includegraphics[width=1\textwidth]{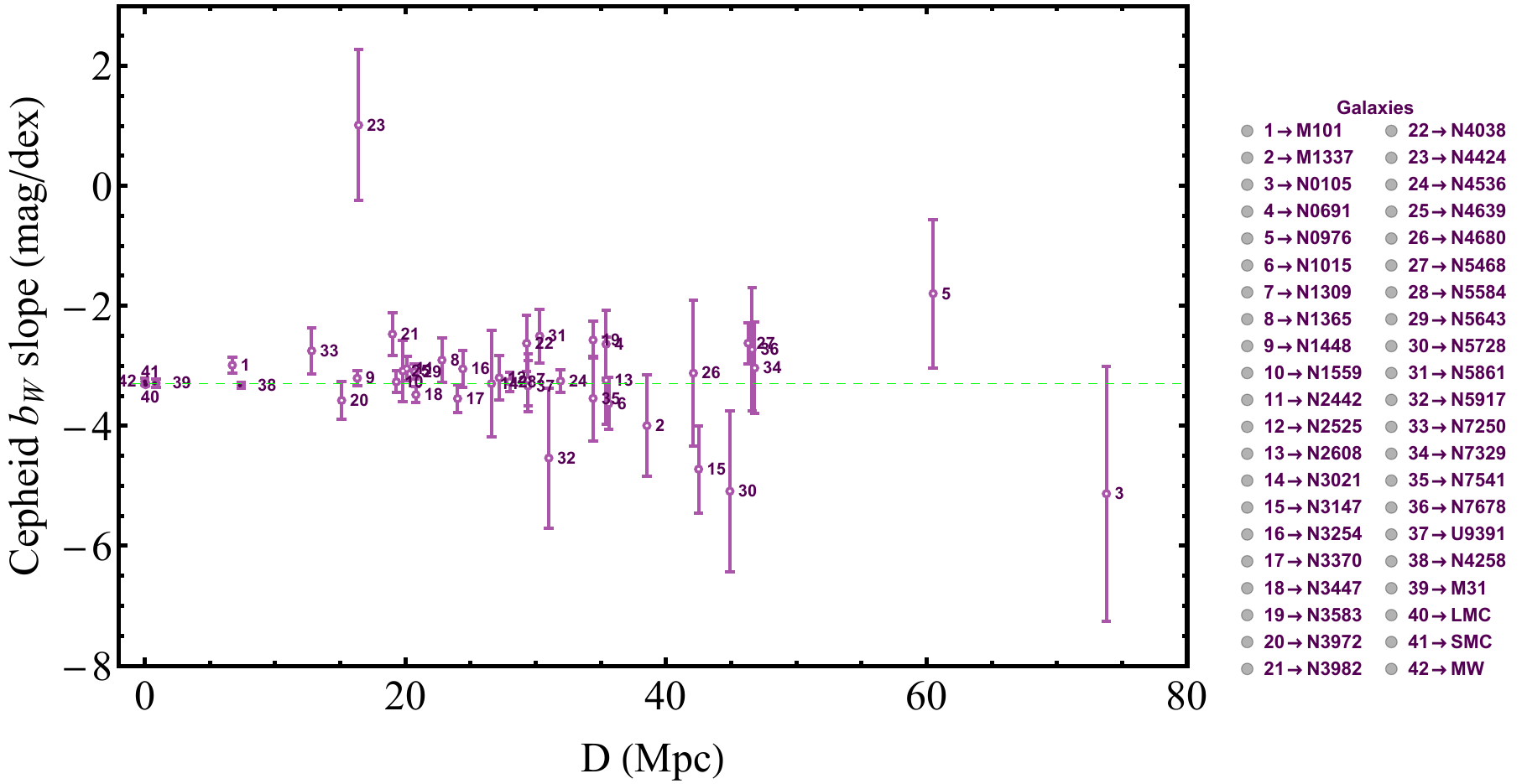}
\par\end{centering}
\caption{Independently fitted slopes $b_{W,i}$ of the Cepheid P-L relations. Here we have ignored non-diagonal terms of the covariance as was done in R21 in the corresponding Figure 10. There is a consistent trend of most non-anchor hosts to have a best fit absolute slope $\vert b_W \vert$ that is smaller than corresponding best fit $b_W$ of the anchor hosts (above the dotted line). The dotted line corresponds to the best fit value of $b_W=b_W^0+\Delta b_W=-0.286-0.013=-0.299$ in the context of the baseline SH0ES analysis as shown in Table \ref{tab:resall}.} 
\label{figb10dnocov} 
\end{figure*}

\begin{figure*}
\begin{centering}
\includegraphics[width=1\textwidth]{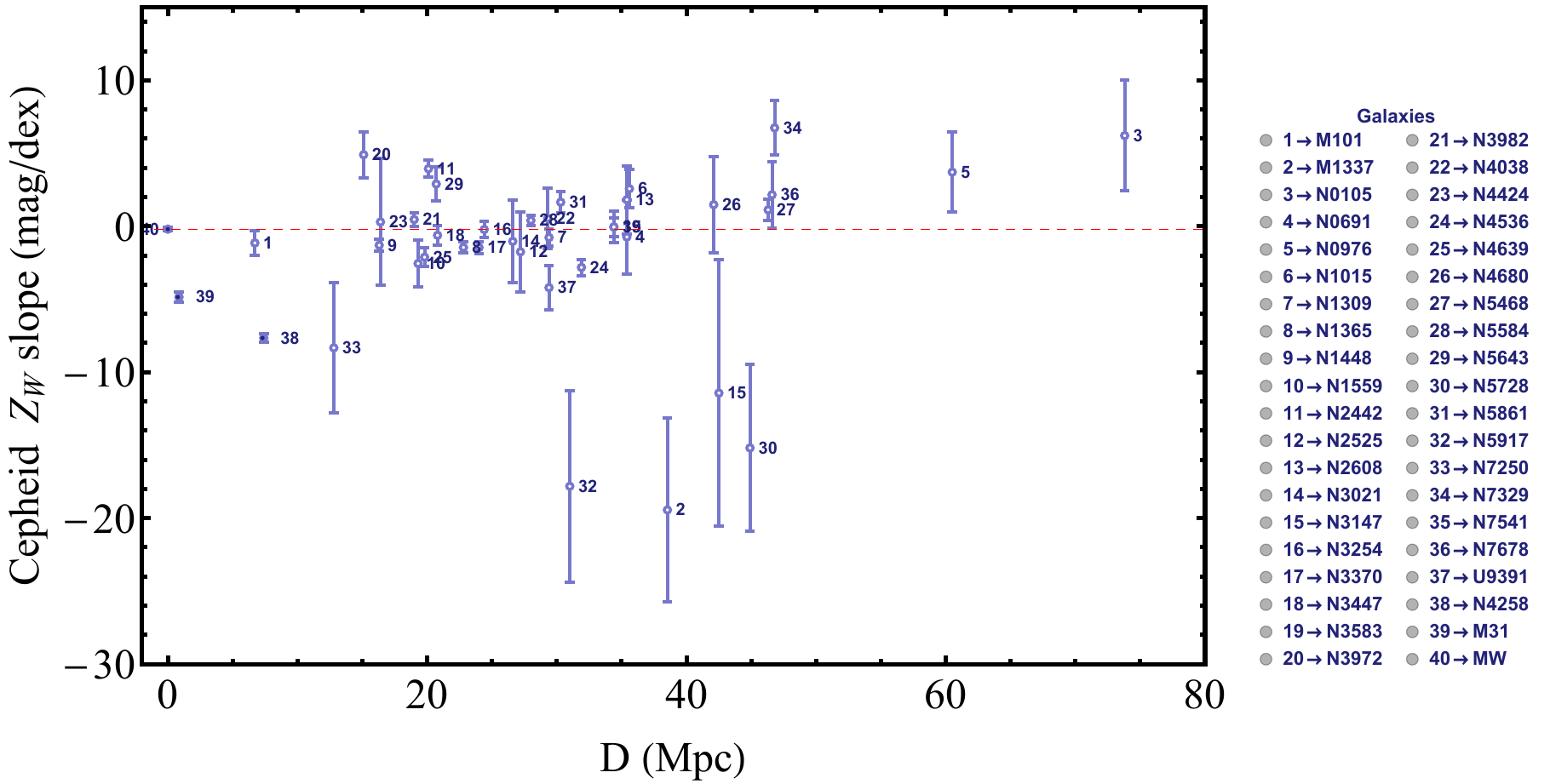}
\par\end{centering}
\caption{Independently fitted metallicity slopes $Z_{W,i}$ in terms of the distance of the Cepheid host. Clearly the spread of many points is larger than the corresponding uncertainties indicating either inhomogeneities of the sample or underestimation of the uncertainties. In fact, there seems to be a trend of most $\vert Z_{W,i} \vert$ absolute slopes to be larger than the corresponding slope of the the Milky Way (MW) (below the dotted line). The dotted line corresponds to the best fit value of $Z_W=-0.21$ in the context of the baseline SH0ES analysis as shown in Table \ref{tab:resall}. } 
\label{figz10dnocov} 
\end{figure*}

The best fit values of the slopes $b_{W,i}$ and $Z_{W,i}$ for each one of the 40 Cepheid hosts are shown in Figs. \ref{figb10dnocov} and \ref{figz10dnocov}  in terms of the host distance respectively. The actual Cepheid data in each host with the best fit $\log P_{i,j}-m_{H,i,j}^W$ and $[O/H]_{i,j}-m_{H,i,j}^W$ straight lines are shown in Figs. 
\ref{figballnocov} and \ref{figzallnocov} in Appendix \ref{AppendixC} for each one of the 40 Cepheid hosts $i$.  The $b_{W,i}$ slopes shown in Fig. \ref{figb102bnocov} for each host in sequence of increasing uncertainties, is in excellent agreement with a corresponding Figure 10 shown in R21 \footnote{As discussed in Appendix \ref{AppendixC}, 2 slopes corresponding to the hosts N4038 and N1365 are slightly shifted in our analysis compared to R21 due a small disagreement in the best fit slope and a typo of R21 in transferring the correct slope to Fig. 10 while the slope corresponding to the host N4424 is missing in Fig. 10 of R21.}. The corresponding numerical values of the best fit $b_{W,i}$ and $Z_{W,i}$ are shown in Table \ref{tab:slopes}. 

Some inhomogeneities are visible in the distance distribution of both the individual Cepheid slopes in Figs. \ref{figb10dnocov} and \ref{figz10dnocov} as well as in Fig. \ref{figb102bnocov}. For example in Figs. \ref{figb10dnocov} and \ref{figb102bnocov} there is a consistent trend of most non-anchor hosts to have an absolute best fit slope $b_W$ that is smaller than the corresponding best fit $b_W$ of the anchor hosts (above the dotted line). Similarly, in Fig. \ref{figz10dnocov} there seems to be a trend of most $Z_{W,i}$ absolute slopes to be larger than the corresponding slope of the the Milky Way (MW) (below the dotted line). 

A careful inspection of Figs. \ref{figb10dnocov} and \ref{figz10dnocov} indicates that the scatter is significantly larger than the standard uncertainties of the slopes. Indeed a $\chi^2$ fit to a universal slope  for $Z_W$ of the form
\be
\chi^2(Z_{W})=\sum_{i=1}^{N}\frac{(Z_{W,i}-Z_{W})^2}{\sigma_{Z_{W,i}}^2}
\label{chizw1}
\ee
leads to a minimum $\chi_{min}^2$ per degree of freedom ($dof$)\footnote{The number of degrees of freedom is $dof=N-M=42-1=41$ for $b_W$ and $dof=N-M=40-1=39$ for $Z_W$. The smaller number of $N$ for $Z_W$ is due to the fact that metallicities were not provided for individual Cepheids in the LMC and SMS in R21 and thus we evaluated a smaller number of slopes for $Z_W$. See also Table \ref{tab:slopes}. The $N=42$ points shown in the plot \ref{figb10dnocov} include the two additional points corresponding to MW and SMC (which is degenerate with LMC).} $\frac{\chi^2_{ZW,min}}{dof}=22$ (with a best fit $Z_{W,bf}\simeq -1$) while $\frac{\chi^2_{min}}{dof}=O(1)$ would be expected for an acceptable fit. Also for $b_W$ we find $\frac{\chi^2_{bW,min}}{dof}=1.55$ ($b_{W,bf}=-3.3$) which is more reasonable but still larger than 1 indicating a relatively poor quality of fit to a universal slope. 

There are two possible causes for this poor quality of fit to universal slopes: either many of uncertainties of the individual host slopes have been underestimated or the universal slope model is not appropriate. In view of the fact that the uncertainties of the Cepheid periods and metallicities have not been included in the $\chi^2$ fit because they were not provided by R21 released data we make the working hypothesis that the uncertainties have been underestimated and thus we add a scatter error adjusted so that $\frac{\chi^2_{bW,min}}{dof}\simeq \frac{\chi^2_{ZW,min}}{dof} \simeq 1$. Thus for $\chi^2(Z_{W})$ we have
\be
\chi^2(Z_{W})=\sum_{i=1}^{N}\frac{(Z_{W,i}-Z_{W})^2}{\sigma_{Z_{W,i}}^2+\sigma_{Z,scat}^2}
\label{chizw2}
\ee
For $\frac{\chi^2_{ZW,min}}{dof} \simeq 1$ we must set $\sigma_{Z,scat}\simeq 3.2$ which is significantly larger than most uncertainties of individual host $Z_W$ slopes. Similarly, for $\frac{\chi^2_{bW,min}}{dof} \simeq 1$ we must set $\sigma_{b,scat}\simeq 0.18$ which is comparable or smaller than most uncertainties of individual host $b_W$ slopes as shown in Table \ref{tab:slopes}. 

\begin{figure*}
\begin{centering}
\includegraphics[width=0.9\textwidth]{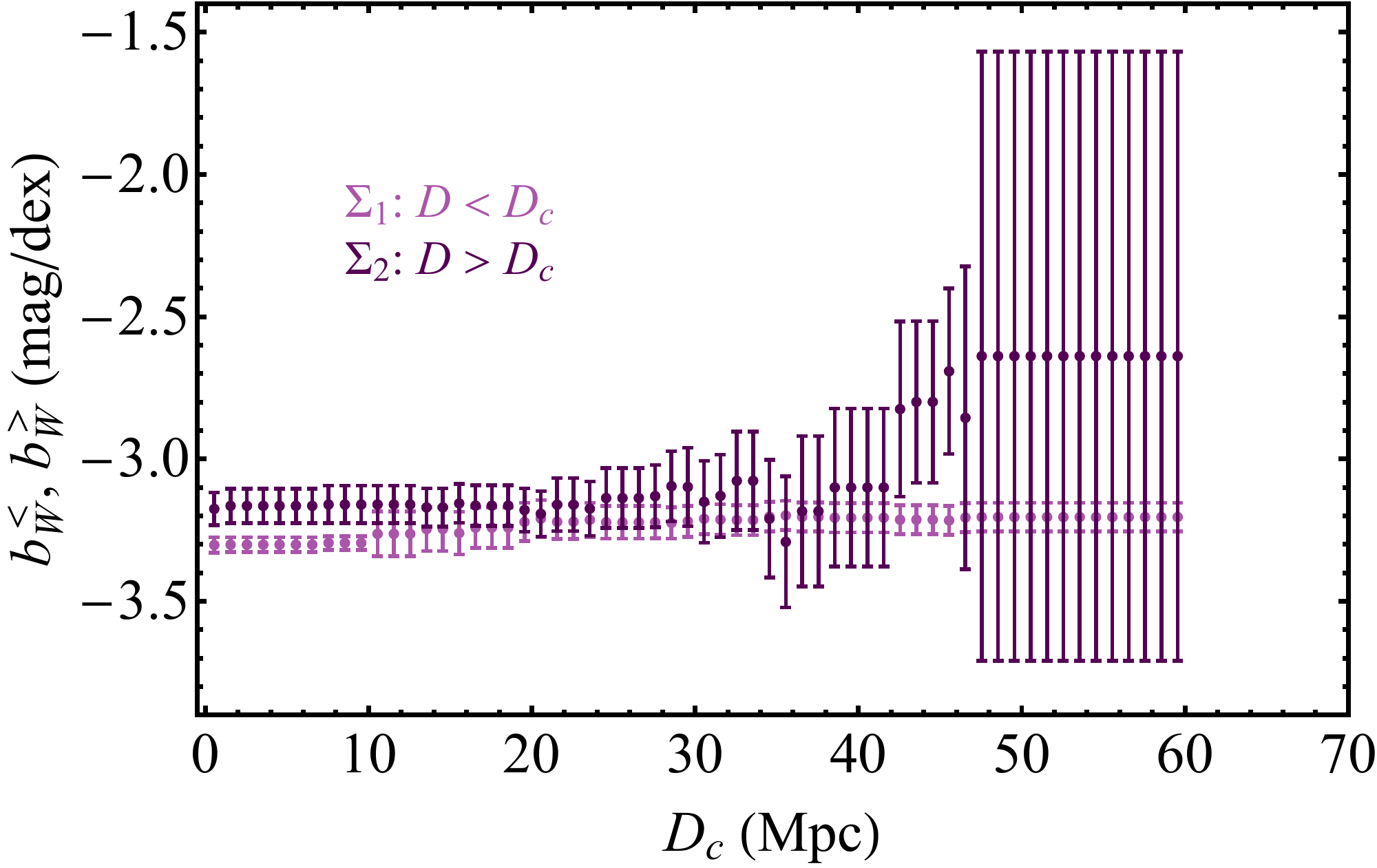}
\par\end{centering}
\caption{Binned Cepheid P-L slopes $b_{W}^>$ and $b_{W}^<$ for the high distance bin $\Sigma_2$ ($D>D_c$) and the low distance bin $\Sigma_1$  ($D<D_c$) respectively. In each bin $\sigma_{scat}$ was used and adjusted so that $\chi_{bW,min}^2 \simeq 1$. Notice that for $D_c\simeq 50Mpc$, the high-low distance bins are statistically consistent with each other due to small number of Cepheid in the high distance bin,  but the difference of their slope best fit values is maximized. } 
\label{figbfsets} 
\end{figure*}

\begin{figure*}
\begin{centering}
\includegraphics[width=0.9\textwidth]{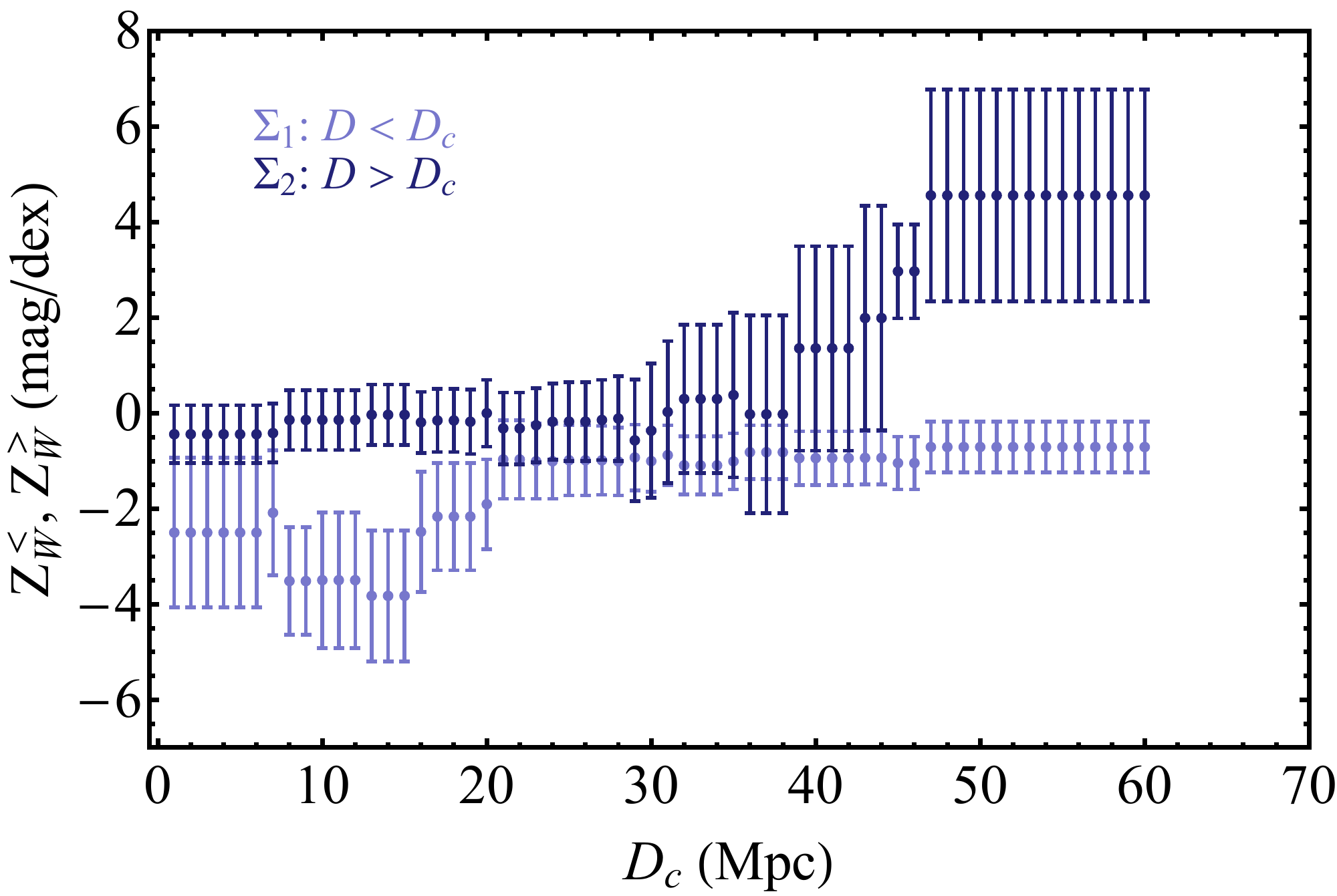}
\par\end{centering}
\caption{Binned Cepheid metallicity slopes $Z_{W}^>$ and $Z_{W}^<$ for the high distance bin $\Sigma_2$ ($D>D_c$) and the low distance bin $\Sigma_1$ ($D<D_c$) respectively.  In each bin the additional scattering uncertainty $\sigma_{scat}$ was used and adjusted so that $\chi_{ZW,min}^2 \simeq 1$. For $D_c\simeq 50Mpc$ the difference between the best fit bin slopes is maximized and the tension between them is larger than $2\sigma$ (see also Fig. \ref{figsdistnocovzb}).} 
\label{figwslnocovz} 
\end{figure*}

In order to quantify possible inhomogeneities of Cepheid hosts $b_{W,i}$ and $Z_{W,i}$ slopes, we have split each sample of slopes in two  bins: a low distance bin with Cepheid host distances $D$ smaller than a critical distance $D_c$ and a high distance bin with $D>D_c$. Given $D_c$, for each bin we find the best fit slope and its standard $1\sigma$ error using the maximum likelihood method.   For example for a low distance $b_W$ bin we minimize 
\be
\chi^2(b_{W}^<)=\sum_{i=1}^{N^<}\frac{(b_{W,i}-b_{W}^<)^2}{\sigma_{b_{W,i}}^2+\sigma_{scat}^{2\,<}}
\label{chibw}
\ee
where $N^<$ is the number of hosts in the low distance bin and $\sigma_{scat}$ is the additional uncertainty (scatter error) chosen in each bin so that $\chi^2\simeq 1$ so that the fit becomes consistent with a constant $b_W$ in each bin. Thus we find the best fit low distance bin slope $b_{W}^<$ and its standard error and similarly for the high distance bin $D>D_c$, $b_W^>$  and for the slopes $Z_W^<$, $Z_W^>$. Thus for each $D_c$ we obtain two best fit slopes (one for each bin) and their standard errors. The level of consistency between the two binned slopes at each $D_c$ determines the level of homogeneity and self consistency of the full sample. The results for the best fit binned slopes for $b_W$ and $Z_W$ are shown in Figs. \ref{figbfsets}  and \ref{figwslnocovz} respectively as functions of the dividing critical distance $D_c$.

\begin{figure*}
\begin{centering}
\includegraphics[width=1\textwidth]{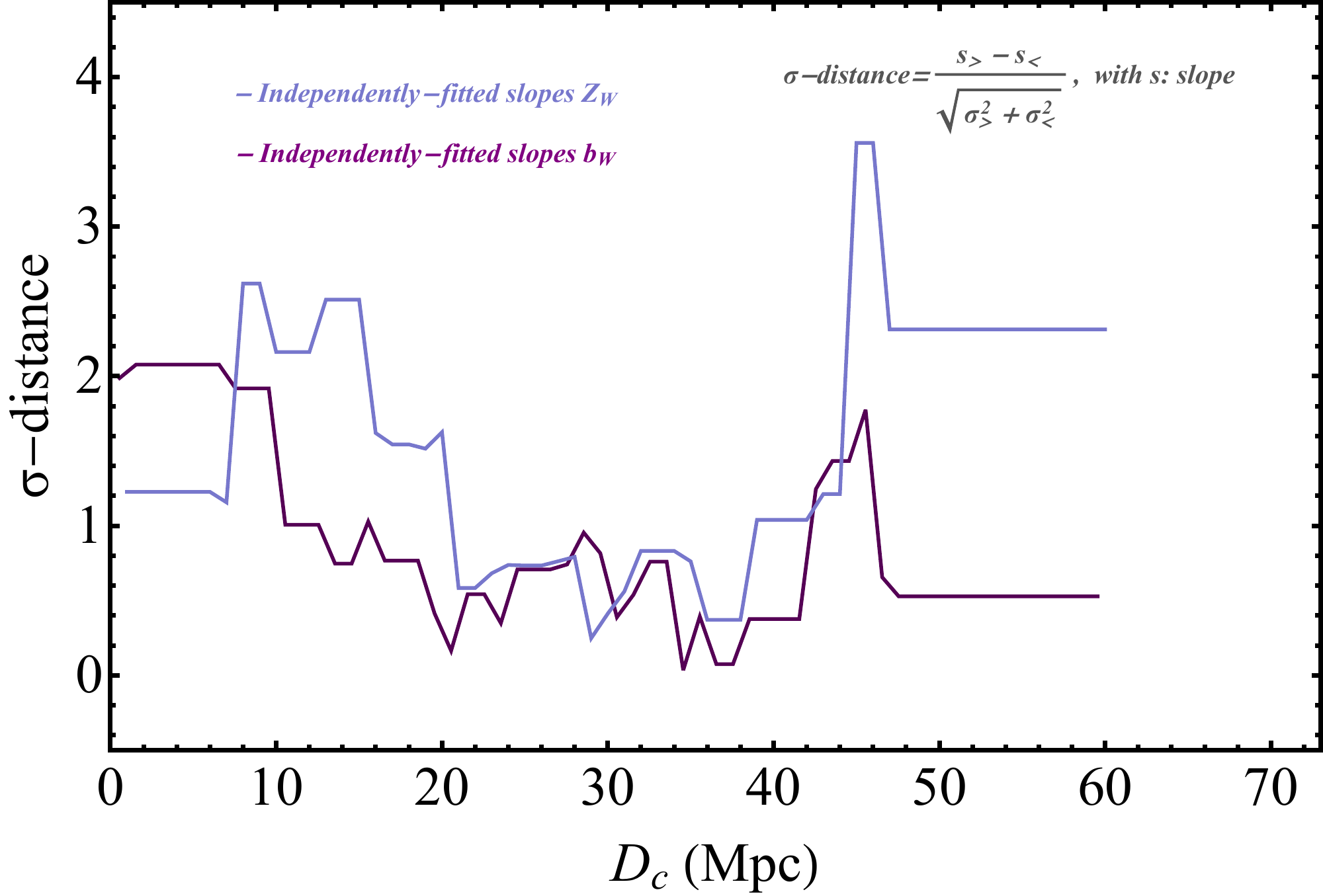}
\par\end{centering}
\caption{The $\sigma$ distance between the best fit high distance bin ($D>D_c$) and low distance bin slopes ($s^> (b_{W}^>, Z_{W}^>)$ and $s^< (b_{W}^<, Z_{W}^<)$) defined in Eq. (\ref{zsigdist}) for $Z_W$ and similarly for $b_W$.}
\label{figsdistnocovzb} 
\end{figure*}

Interestingly, for a range of $D_c$ there is a mild tension between the best fit values of the high and low distance bins which reaches levels of $2-3\sigma$ especially for the metallicity slopes $Z_W$. For both $b_W$ and $Z_W$, the absolute value of the difference between the high and low distance bin slopes is maximized for $D_c>47Mpc$. For the case of $Z_W$ this difference is significant statistically as it exceeds the level of $2\sigma$. The level of statistical consistency between high and low distance bins for both slopes $b_W$ and $Z_W$ is shown in Fig. \ref{figsdistnocovzb}. In the range of $D_c$ between $40Mpc$ and $50Mpc$ and also between $10Mpc$ and $20Mpc$, 
it is shown that in the case of $Z_W$ the $\sigma$-distance 
\be
\sigma_d \equiv \frac{\vert Z_W^>-Z_W^< \vert}{\sqrt{\sigma_{Z_W^>}^2+\sigma_{Z_W^<}^2}}
\label{zsigdist}
\ee
between the best fit binned $Z_W$ metallicity slopes can reach a level beyond $2\sigma$ (see Figs \ref{figwslnocovz} and \ref{figsdistnocovzb}).

\begin{figure*}
\begin{centering}
\includegraphics[width=1\textwidth]{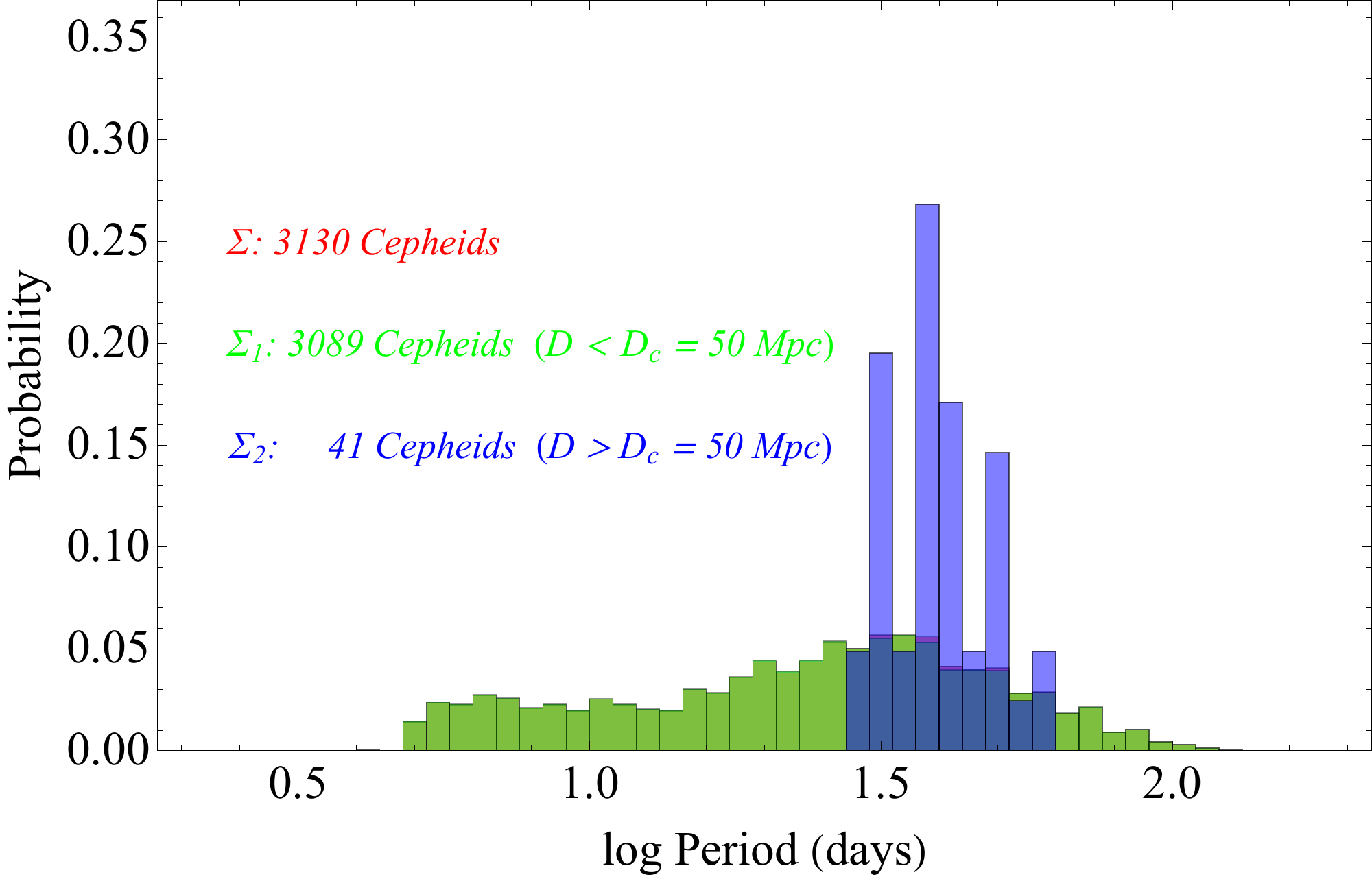}
\par\end{centering}
\caption{Probability distributions of the Cepheid period for the whole Cepheid sample and for the high ($D>D_c$) and low distance ($D<D_c$) subsamples for $D_c=50Mpc$. Due to the small number of Cepheids in the $D>50Mpc$ bin (41), the probability distribution of the full sample (3130 Cepheids) is almost identical with the the probability distribution of the $D<50Mpc$ bin and thus the light green bars overlap with the red bars.}
\label{fighisp} 
\end{figure*}

\begin{figure*}
\begin{centering}
\includegraphics[width=1\textwidth]{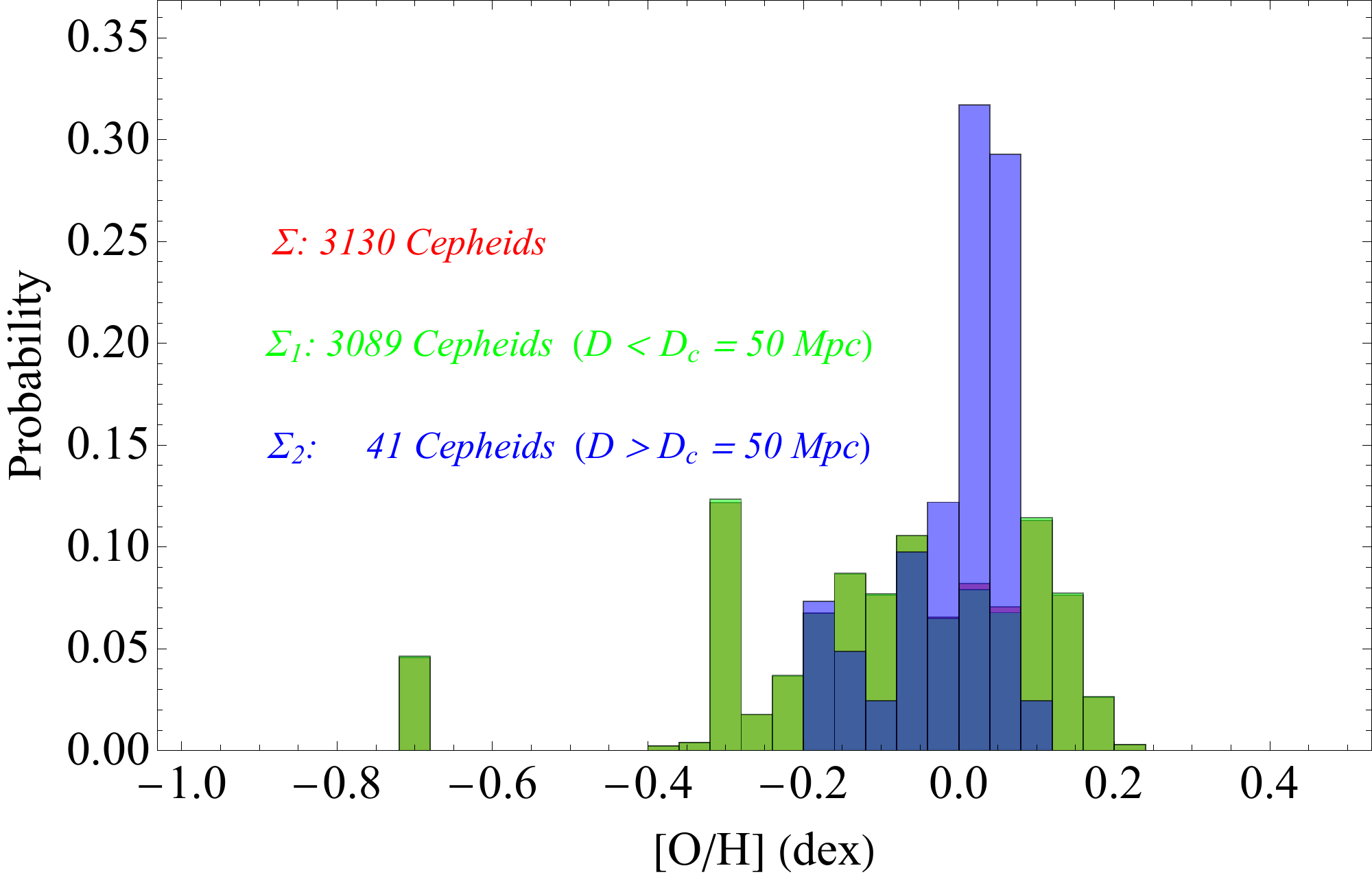}
\par\end{centering}
\caption{Probability distributions of the Cepheid metallicity for the whole Cepheid sample and for the high ($D>D_c$) and low distance ($D<D_c$) subsamples for $D_c=50Mpc$. Due to the small number of Cepheids in the $D>50Mpc$ bin (41), the probability distribution of the full sample (3130 Cepheids) is almost identical with the the probability distribution of the $D<50Mpc$ bin and thus the light green bars overlap with the red bars.}
\label{fighism} 
\end{figure*}

An additional test of possible intrinsic tensions of the Cepheid properties in the SH0ES Cepheid sample is obtained by comparing the probability distributions of the Cepheid period and metallicity for the full sample of the 3130 Cepheid with high and low distance subsamples. In Figs. \ref{fighisp} and \ref{fighism} we show histograms of the probability distributions of the Cepheid period and metallicity for the whole Cepheid sample and for the high ($D>D_c$) and low distance ($D<D_c$) subsamples for $D_c=50Mpc$. The two subsamples for each observable are clearly inconsistent with each other and with the full sample. This is demonstrated visually and also through the Kolmogorov-Smirnov consistency test which quantifies the inconsistency and gives a p-value very close to 0 for the three sample pairs. However, as communicated to us by members of the SH0ES team (private communication) this inconsistency can be justified by observational selection effects and does not necessarily indicate a physics change at $D_c=50 Mpc$. For example bright Cepheids have longer periods and they are more easily observed at high distances. Thus it is expected that there will be higher period (brighter) Cepheids observed at higher distances. Other variables such as the timespan of the observations also play a role.  For more distant galaxies there is a trend to allow a longer window of observations so that longer period Cepheids can be found. Also the star formation  history of the galaxies dictates whether one will have very long period Cepheids which come from massive, short lived stars. However, even though the observed inconsistency in the Cepheid properties probability distributions may be explained using observational biases and anticipated galactic properties, it may also be a hint of interesting new physics and/or systematic effects. 

In view of the above identified level of mild inhomogeneities identified is the SH0ES data, it would be interesting to extend the SH0ES modeling of the Cepheids and SnIa with new degrees of freedom that can model the data taking into account these inhomogeneities. This is the goal of the next section.

\section{Generalized analysis: New degrees of freedom allowing for a physics transition.}
\label{sec:Generalized analysis}

An obvious generalization of the SH0ES analysis described in subsection \ref{sub:baseline} which models the  Cepheid-SnIa luminosities using four main parameters is to allow for a transition of any one of these parameters at a particular distance or equivalently cosmic time when radiation was emitted. Such a transition could be either a result of a sudden change in a physics constant (e.g. the gravitational constant) during the last $200Myrs$ in the context e.g. of a first order phase transition or a result of the presence of an unknown systematic. 

Another modification of the standard SH0ES analysis in R21 would be to extend the number of constraints imposed on the data vector $\bf{Y}$ taking into account other cosmological data like the inverse distance ladder estimate of $M_B$ \cite{Camarena:2021jlr,Marra:2021fvf,Gomez-Valent:2021hda}. Both of these generalizations will be implemented in the present section. 

\subsection{Allowing for a transition of a Cepheid calibration parameter}

A transition of one of the Cepheid calibration parameters can be implemented by replacing one of the four modeling parameters in the $\bf{q}$ vector by two corresponding parameters: one for high and one for low distances (recent cosmic times). Thus, in this approach the number of parameters and entries in the parameter vector $\bf{q}$ increases by one from 47 to 48. Since one of the entries of $\bf{q}$ is replaced by two entries, the corresponding column of the modeling matrix $\bf{L}$ should be replaced by two columns. The high distance parameter column should be similar to the original column it replaced but with zeros in entries corresponding to low distance data (or constraints) and the reverse should happen for the low distance parameter column. This process is demonstrated in the following schematic diagram where the $q_j$ parameter is replaced by the two parameters $q_j$ (or $q_j^<$) and $q_{j+1}$ (or $q_j^>$)   and the $j$ column of $\bf{L}$ is replaced by the $j$ and $j+1$ corresponding columns which have zeros in the low or high distance data (or constraint) entries.

\begin{adjustwidth}{-3.9cm}{0.5cm}
\be
\setlength{\tabcolsep}{0.2em}
\begin{tabular}{ccccc}
\(\bf {L_{(3492\times47)}}=
\left( \arraycolsep=1.6 pt\def\arraystretch{1.8}\begin{array}[c]{ccc}
\ldots&L_{1,j}&\ldots\\
\ldots&L_{2,j}&\ldots\\
\ldots&L_{3,j}&\ldots\\
\ldots&\ldots&\ldots\\
\ldots&\ldots&\ldots\\
\ldots&L_{3491,j}&\ldots\\
\ldots&L_{3492,j}&\ldots\\
 \end{array}\right)$
 &\arraycolsep=1.6 pt\def\arraystretch{1.8}$\left.\begin{matrix}
&\\
&\\
&\\
&\\
&\\
&\\
&\\
\end{matrix}\right\}$ &\arraycolsep=1.6 pt\def\arraystretch{1.8} $\begin{array}[c]{c}
 D_{Y_1}<D_c\\
 D_{Y_2}>D_c\\
 D_{Y_3}>D_c\\
 \ldots\\
 \ldots\\
 D_{Y_{3491}}<D_c\\ 
 D_{Y_{3492}}>D_c\\
\end{array}$ & $\implies$  &$
{\bf L_{(3492\times48)}}=\arraycolsep=1.6 pt\def\arraystretch{1.8}\left( \begin{array}[c]{cccc}
\ldots&L_{1,j}&0&\ldots\\
\ldots&0&L_{2,j+1}&\ldots\\
\ldots&0&L_{3,j+1}&\ldots\\
\ldots&\ldots&\ldots&\ldots\\
\ldots&\ldots&\ldots&\ldots\\
\ldots&L_{3491,j} &0&\ldots\\
\ldots&0&L_{3492,j+1}  &\ldots\\
 \end{array}\right)\)
\end{tabular}
\ee
\end{adjustwidth}

\begin{adjustwidth}{0cm}{0.6cm}
\be
\begin{tabular}{ccc}
\({\bf q_{(47\times 1)}}=\arraycolsep=1.6 pt\def\arraystretch{1.8}
\begin{pmatrix}
\ldots\\
\ldots\\
q_j\\
\ldots\\
\ldots\\
\end{pmatrix}  
\rightarrow
\bf q_{(48\times  1)}= $\arraycolsep=1.6 pt\def\arraystretch{1.8}$\begin{pmatrix}
\ldots\\
\ldots\\
q_j\\
q_{j+1}\\
\ldots\\
\ldots\\
\end{pmatrix}\)
\end{tabular} 
\ee
\end{adjustwidth}

In this manner, if the parameter $b_W$ was to be split to $b_{W}^>$ and $b_{W}^<$ for example,  Eq. (\ref{wesmagcep}) would be replaced by 
\be
m_{H,i,j}^W (D)
=\mu_i+M_{H}^W+b_{W}^>\Theta(D-D_c) [P]_{i,j}+b_{W}^<\Theta(D_c-D) [P]_{i,j} +Z_W[O/H]_{i,j}
\label{wesmagcep1}
\ee
and similarly for splittings of each one the other three parameters $M_H^W$, $Z_W$ and $M_B$. In (\ref{wesmagcep1}) $D$ is a distance that may be assigned to every entry of the data-vector $\bf{Y}$ (Cepheids, SnIa and constraints). Notice that the splitting of any parameter to a high and low distance version does not affect the form of the $\bf{Y}$ data vector and the covariance matrix $\bf{C}$ of the data. In order to properly place the 0 entries of each one of the new columns, a distance must be assigned to every entry of $\bf{Y}$. We have specified this distance for each host using the literature resources or the best fit distance moduli of each host. These distances along with other useful properties of the Cepheids used in our analysis are shown in Table \ref{tab:props}.

\begin{table}
\caption{Properties of the Cepheid hosts considered in the analysis.}
\label{tab:props} 
\vspace{2mm}
\setlength{\tabcolsep}{0.2em}
\begin{adjustwidth}{-2. cm}{1.5cm}
{\footnotesize\begin{tabular}{cc ccccc  cc} 
\hhline{=========}
   & \\
Galaxy &SnIa &Ranking in&Ranking in& Ranking in& $D$ $^{a}$ & Number of fitted  &Initial point&Final point \\
 & &Fig. \ref{figb102bnocov} & Table 3 in R21& Data Vector Y &$[Mpc]$ &Cepheids&in Vector Y&in Vector Y\\
     & \\
   \hhline{=========}
     & \\
M101& 2011fe &8 &1 &1 &6,71 &259 &1 &259\\
M1337& 2006D &34 &2 &2 &38,53 &15 &260 &274\\
N0691&2005W &28 &4 &3 &35,4 &28 &275 &302\\
N1015&2009ig &24 &6 &4 &35,6 &18 &303 &320\\
N0105&2007A &42 &3 &5 &73,8 &8 &321 &328\\
N1309&2002fk &25 &7 &6 &29,4 &53 &329 &381\\
N1365&2012fr &20 &8 &7 &22,8 &46 &382 &427\\
N1448&2001el,2021pit &7 &9 &8 &16,3 &73 &428 &500\\
N1559&2005df &11 &10 &9 &19,3 &110 &501 &610\\
N2442&2015F &13 &11 &10 &20,1 &177 &611 &787\\
N2525&2018gv &21 &12 &11 &27,2 &73 &788 &860\\
N2608&2001bg &31 &13 &12 &35,4 &22 &861 &882\\
N3021&1995al &35 &14 &13 &26,6 &16 &883 &898\\
N3147&1997bq,2008fv,2021hpr &32 &15 &14 &42,5 &27 &899 &925\\
N3254&2019np &16 &16 &15 &24,4 &48 &926 &973\\
N3370&1994ae &14 &17 &16 &24 &73 &974 &1046\\
N3447&2012ht &9 &18 &17 &20,8 &101 &1047 &1147\\
N3583&2015so &15 &19 &18 &34,4 &54 &1148 &1201\\
N3972&2011by &17 &20 &19 &15,1 &52 &1202 &1253\\
N3982&1998aq &19 &21 &20 &19 &27 &1254 &1280\\
N4038&2007sr &29 &22 &21 &29,3 &29 &1281 &1309\\
N4424&2012cg &40 &23 &22 &16,4 &9 &1310 &1318\\
N4536&1981B &12 &24 &23 &31,9 &40 &1319 &1358\\
N4639&1990N &27 &25 &24 &19,8 &30 &1359 &1388\\
N4680&1997bp &38 &26 &25 &42,1 &11 &1389 &1399\\
N5468&1999cp,2002cr&18 &27 &26 &46,3 &93 &1400 &1492\\
N5584&2007af &10 &28 &27 &28 &165 &1493 &1657\\
N5643&2013aa,2017cbv &6 &29 &28 &20,7 &251 &1658 &1908\\
N5728&2009Y &41 &30 &29 &44,9 &20 &1909 &1928\\
N5861&2017erp &26 &31 &30 &30,3 &41 &1929 &1969\\
N5917&2005cf &37 &32 &31 &31 &14 &1970 &1983\\
N7250&2013dy &22 &33 &32 &12,8 &21 &1984 &2004\\
N7329&2006bh &33 &34 &33 &46,8 &31 &2005 &2035\\
N7541&1998dh &30 &35 &34 &34,4 &33 &2036 &2068\\
N7678&2002dp &36 &36 &35 &46,6 &16 &2069 &2084\\
N0976&1999dq &39 &5 &36 &60,5 &33 &2085 &2117\\
U9391&2003du &23 &37 &37 &29,4 &33 &2118 &2150\\
\hline
Total&  &  &  &  &  &2150&\;\;\;\;\;1 &2150\\
\hline
N4258&Anchor &4 &38 &38 &7,4 &443 &2151 &2593\\
M31&Supporting &5 &39 &39 &0,86 &55 &2594 &2648\\
LMC$^b$&Anchor &2 &40 &40 &0,05 &270&2649 &2918\\
SMC$^b$& Supporting &3 &41 &41 &0,06 &143 &2919 &3061\\
LMC$^c$& Anchor &2 &40 &42 &0,05 &69 &3062&3130\\
\hline
Total&  &  &  &  &  &980&2151 &3130\\
\hline
\hline
Total All& &  &  &  &  &3130&\;\;\;\;\;1 &3130\\
\hline
 \hhline{=========} 
 \\
 
\end{tabular} }
\\
{\footnotesize NOTE - (a) Distances from \href{https://ned.ipac.caltech.edu/}{NASA/IPAC Extragalactic Database}. (b) From the ground. (c) From HST.}
\end{adjustwidth}
\end{table}

\begin{figure*}
\begin{centering}
\setlength{\headheight}{23.60004pt}
\addtolength{\topmargin}{-8.40002pt}
\includegraphics[width=0.9\textwidth]{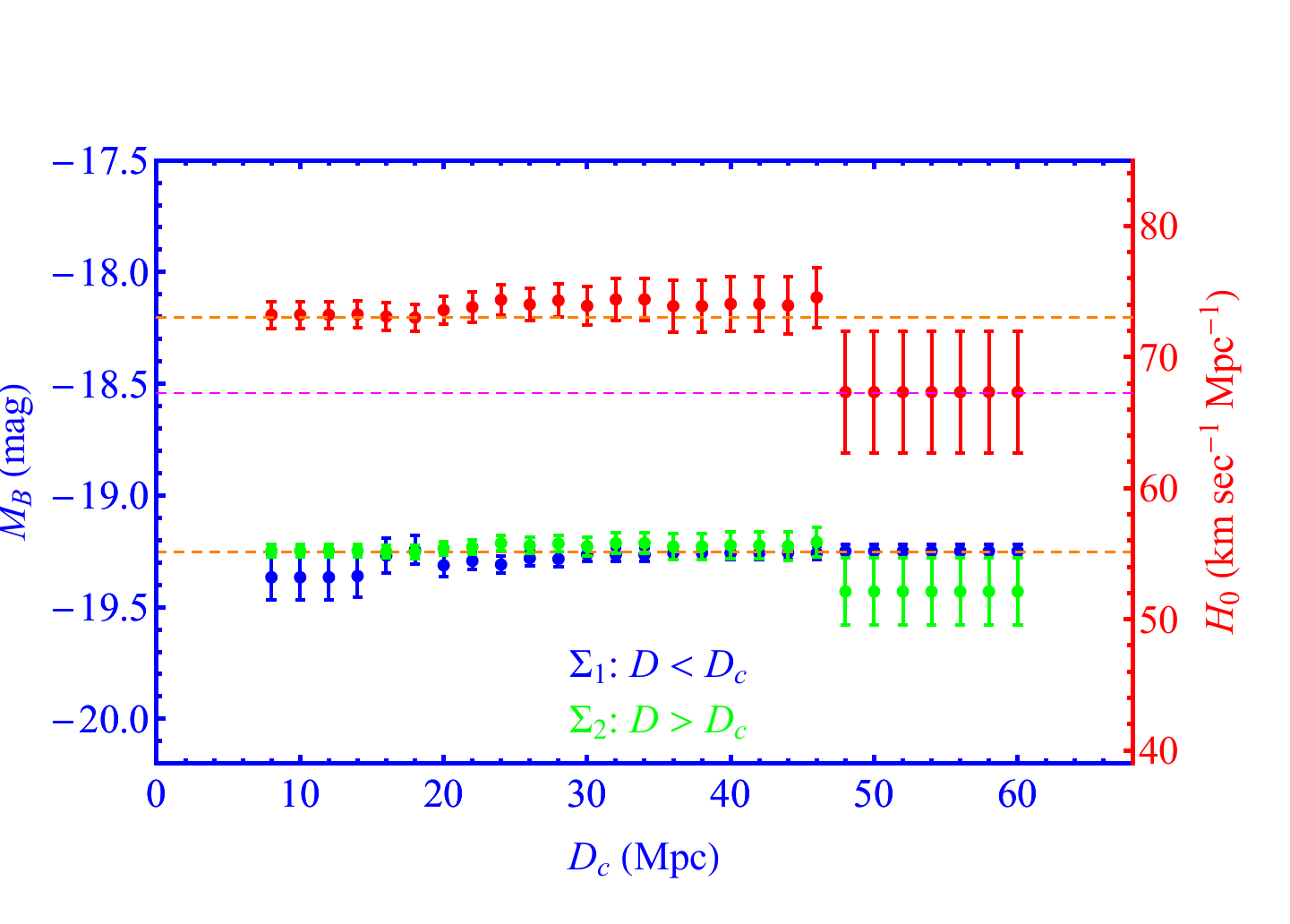}
\par\end{centering}
\caption{The best fit values with uncertainties for the high-low $D$ bin of SnIa absolute magnitude $M_B$ (green and blue points)  and for $H_0$ (red points). This generalization of the SH0ES baseline analysis allows for a transition of $M_B$. The best fit parameter values as shown as functions of the critical transition $D_c$. Dotted lines correspond to the SH0ES R21 best fit and to the \plcdm best fit for $H_0$.} 
\label{figmb} 
\end{figure*}

\begin{figure*}
\begin{centering}
\setlength{\headheight}{23.60004pt}
\addtolength{\topmargin}{-8.40002pt}
\includegraphics[width=0.9\textwidth]{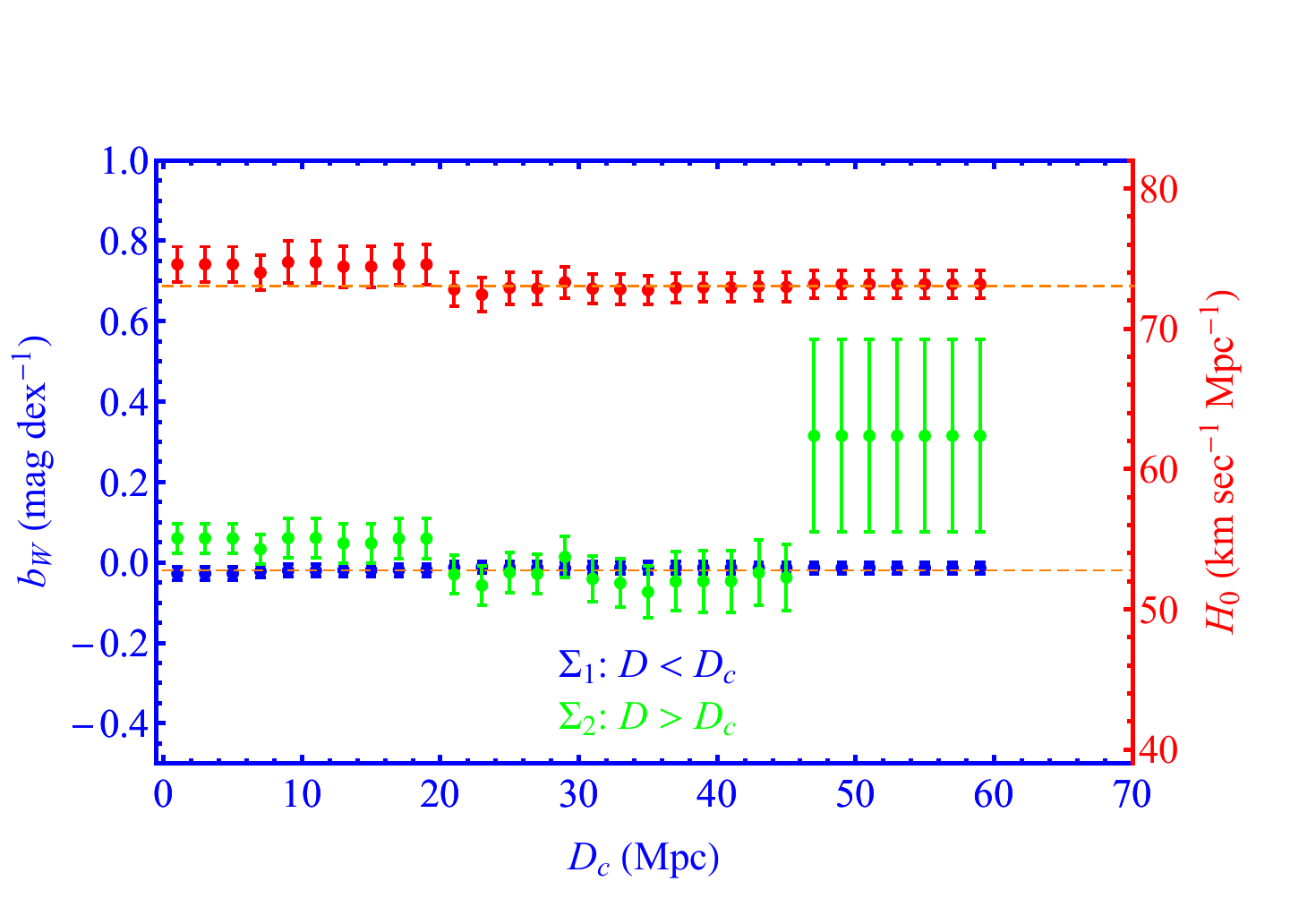}
\par\end{centering}
\caption{Same as Fig. \ref{figmb} in an analysis generalization where the parameter $b_W$ is allowed to have a transition at a distance $D_c$. Notice the sharp increase in the difference between best fit parameters high of $D$ and low $D$ bin $\vert b_{W}^>-b_{W}^<\vert$ for $D_c>47Mpc$.} 
\label{figbw} 
\end{figure*}

\begin{figure*}
\begin{centering}
\setlength{\headheight}{23.60004pt}
\addtolength{\topmargin}{-8.40002pt}
\includegraphics[width=0.9\textwidth]{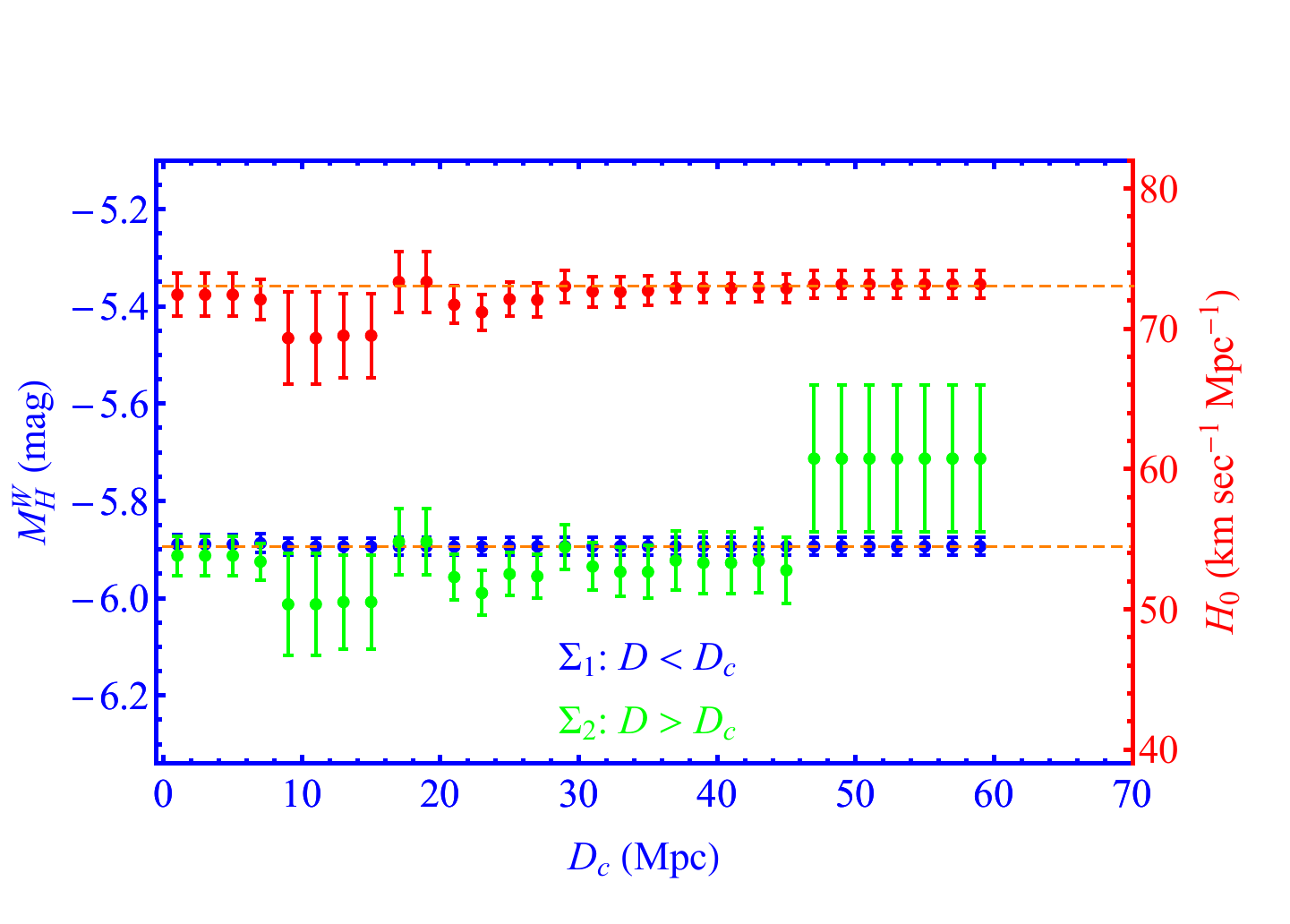}
\par\end{centering}
\caption{Same as Fig. \ref{figmb} in an analysis generalization where the parameter $M_W$ is allowed to have a transition at a distance $D_c$. Notice the sharp increase in the difference between best fit parameters of high $D$ and low $D$ bin $\vert M_{W}^>-M_{W}^<\vert$ for $D_c>47Mpc$.} 
\label{figmw} 
\end{figure*}

\begin{figure*}
\begin{centering}
\setlength{\headheight}{23.60004pt}
\addtolength{\topmargin}{-8.40002pt}
\includegraphics[width=0.9\textwidth]{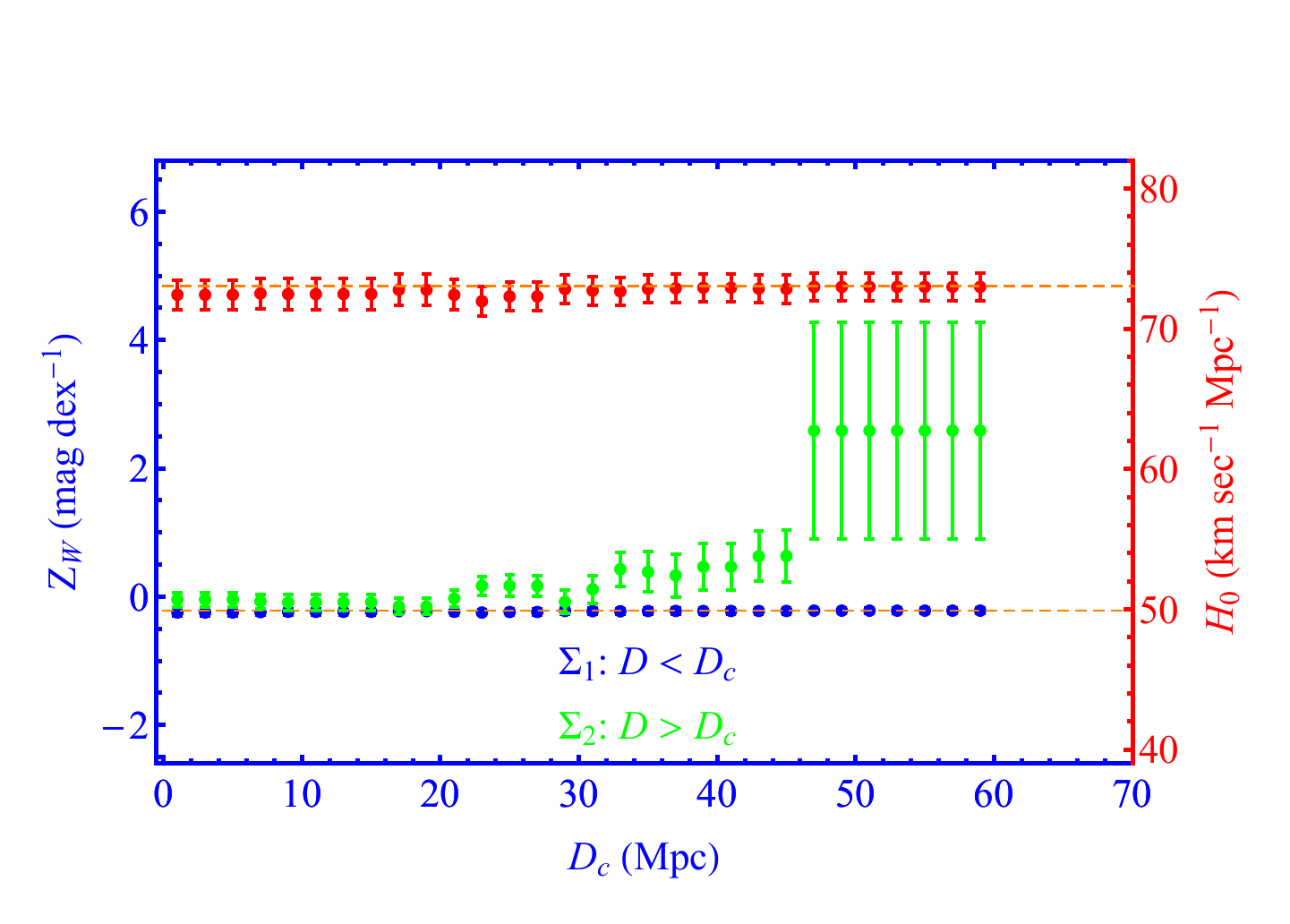}
\par\end{centering}
\caption{Same as Fig. \ref{figmb} in an analysis generalization where the parameter $Z_W$ is allowed to have a transition at a distance $D_c$. Notice the sharp increase in the difference between best fit parameters of high $D$  and low $D$ bin $\vert Z_{W}^>-Z_{W}^<\vert$ for $D_c>47Mpc$. The best fit value of the Hubble parameter (red points) is not significantly affected by this type of degree of freedom perhaps due to the indirect connection of $Z_W$ with $H_0$. } 
\label{figzw} 
\end{figure*}

\begin{figure*}
\begin{centering}
\setlength{\headheight}{23.60004pt}
\addtolength{\topmargin}{-8.40002pt}
\includegraphics[width=0.9\textwidth]{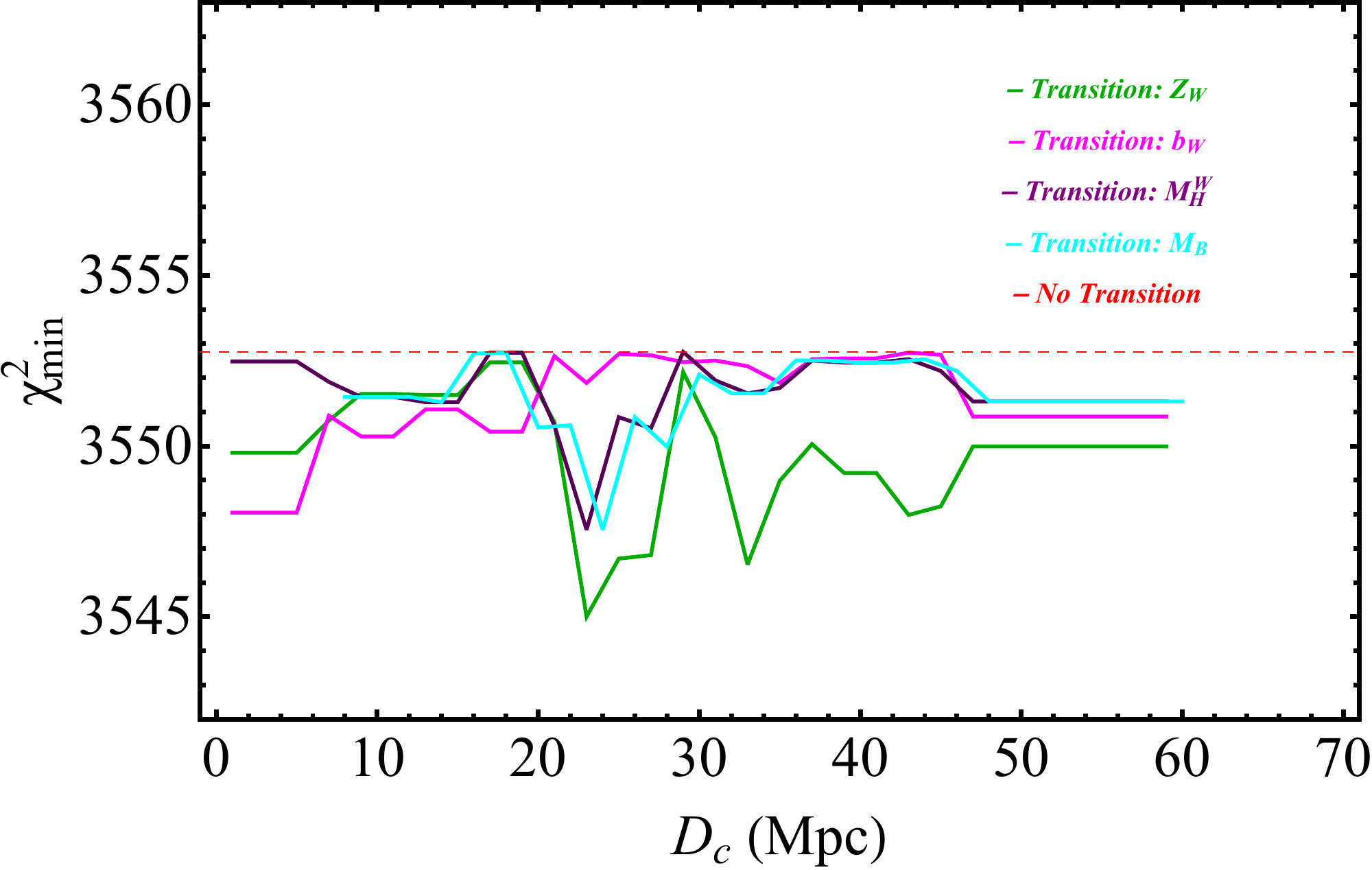}
\par\end{centering}
\caption{The values of $\Delta \chi^2_{min}$ in terms of $D_c$ for the four transition models considered with respect to the SH0ES baseline model. No inverse distance ladder constraint on $M_B$ has been used for either the baseline or the transition models. The small improvement of the fit to the data is not enough to justify any model preference.} 
\label{figchiminall} 
\end{figure*}

\begin{figure*}
\begin{centering}
\includegraphics[width=0.9\textwidth]{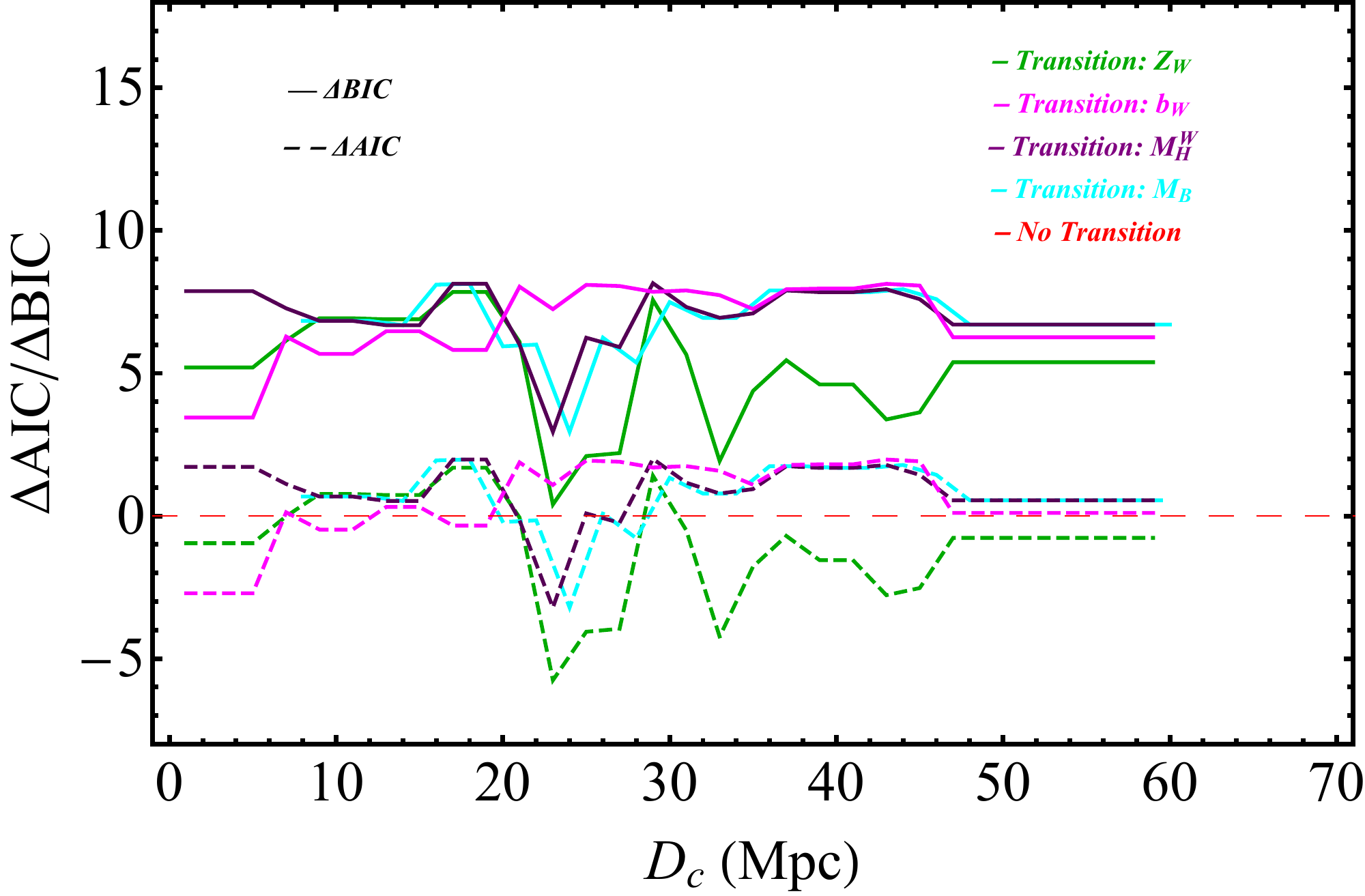}
\par\end{centering}
\caption{The values of $\Delta AIC$ and $\Delta BIC$  in terms of $D_c$ for the four transition models considered with respect to the SH0ES baseline model. No inverse distance ladder constraint on $M_B$ has been used for either the baseline or the transition models. The small improvement of the $\chi^2$ fit to the data is not enough to justify any model preference. In fact the BIC which strongly penalizes any additional parameter mildly disfavors the transition models.} 
\label{figdaicdbicall} 
\end{figure*}

We have thus considered four generalizations of the baseline SH0ES analysis each one corresponding to a high-low distance split of each one of the four modeling parameters. For each generalization we obtained the best fit values and uncertainties of all 48 parameters of the generalized parameter vector $\bf{q}$ for several values of the critical distance $D_c\in[1,60Mpc]$ which defined in each case the high-low distance data bins. The best fit values with uncertainties for the high-low $D$ split parameter (green and blue points)  and for $H_0$ (red points), for each generalization considered, is shown in Figs \ref{figmb}, \ref{figbw}, \ref{figmw} and \ref{figzw} in terms of $D_c$. Dotted lines correspond to the SH0ES R21 $H_0$ best fit and to the \plcdm best fit for $H_0$. The following comments can be made on the results of Figs  \ref{figmb}, \ref{figbw}, \ref{figmw} and \ref{figzw}. 
\begin{itemize}
    \item When the SnIa absolute magnitude $M_B$ is allowed to change at $D_c$ ($M_B$ transition) and for $D_c > 47Mpc$, the best fit value of $H_0$ drops spontaneously to the best fit \plcdm value albeit with larger uncertainty $67.33\pm 4.65$ (see Fig. \ref{figmb} and second row of Table \ref{tab:res}). This remarkable result appears with no prior or other information from the inverse distance ladder results. Clearly, there are increased uncertainties of the best fit parameter values for this range of $D_c$ because the most distant usable Cepheid hosts are at distances $46.8Mpc$ (N7329), $60.5Mpc$ (N0976) and $73.8Mpc$ (N0105) and only two of them are at distance beyond $47Mpc$. These hosts (N0975 and N0105) have a total of 41 Cepheids and 4 SnIa. Due to the large uncertainties involved there is a neutral model preference for this transition degree of freedom at $D_c>47Mpc$ (the small drop of $\chi^2$ by $\Delta \chi^2 \simeq -1.5$ is balanced by the additional parameter of the model). This however changes dramatically if the inverse distance ladder constraint on $M_B$ is included in the vector $\bf{Y}$ as discussed below.
    \item For all four modeling parameters considered, there is a sharp increase in the absolute difference between the high-low distance best fit parameter values for $D_c\gtrsim 47Mpc$. The statistical significance of this split however is low due to the relatively small number of available Cepheids at $D>47Mpc$.
    \item The best fit value of $H_0$ changes significantly when the SnIa absolute magnitude $M_B$ is allowed to make a transition at $D_c>47Mpc$ but is not significantly affected if the three Cepheid modeling parameters $M_W$, $b_W$ and $Z_W$ are allowed to make a transition at any distance. This is probably due to the large uncertainties involved and due to the fact that $H_0$ is only indirectly connected with the three Cepheid modeling parameters.
    \item Each one of the  transition degrees of freedom mildly reduces $\chi^2$ thus improving the fit to the data but these transition models are not strongly preferred by model selection criteria which penalize the extra parameter implied by these models. This is demonstrated in Figs. \ref{figchiminall}, \ref{figdaicdbicall}   which show the values of $\Delta \chi^2_{min}$, $\Delta AIC$ and $\Delta BIC$ of the four transition models with respect to the baseline SH0ES model for various $D_c$ transition distances. As discussed below this changes dramatically if the inverse distance ladder constraint on $M_B$ is introduced in the analysis.
\end{itemize}

\subsection{The effect of the inverse distance ladder constraint}

In view of the spontaneous transition of the SnIa absolute magnitude $M_B$ that appears to lead to a Hubble tension resolution, while being mildly favored by the SH0ES data ($\Delta \chi^2 \simeq -1.5$) without any inverse distance ladder information included in the fit, the following question arises: 
\begin{center}
{\it  How does the level of preference for the $M_B$ transition model (compared to the SH0ES baseline model) change if an additional constraint is included in the analysis obtained from the inverse distance ladder best fit for $M_B$?} 
\end{center}
The inverse distance ladder constraint on $M_B$ is \cite{Camarena:2021jlr, Marra:2021fvf,Gomez-Valent:2021hda} 
\be
M_B^{P18} = -19.401 \pm 0.027
\label{mbconstr}
\ee
In order to address this question we modify the analysis by adding one more constraint to the $\bf{Y}$ data-vector: after entry 3215 which corresponds to the 8th constraint we add the value -19.401 for the $M_B$ constraint. We also add a line to the model matrix $\bf{L}$ after line 3215 with all entries set to zero except the entry at column 43 corresponding to the parameter $M_{B}^<$. A column after column 43 is added in $\bf{L}$ to accommodate the high distance parameter $M_B^>$ as described above. The new constraint is assigned a large distance (larger than the distance of the most distant SnIa of the sample) so that it only affects the high distance parameter $M_{B}^>$ (the entry at line 3216 column 43 of $\bf{L}$ is set to 0 for all $D_c$ while the entry at column 44 of the same line is set to 1 for all $D_c$). To accommodate the corresponding uncertainty of the new constraint we also add a line at the covariance matrix after line 3215 with a single nonzero entry at the diagonal equal to $\sigma_{MB}^2=0.027^2=0.000729$. Thus after the implementation of the constraint in the $M_B$ transition model the $\bf{Y}$ vector has 3493 entries, the $\bf{L}$ model matrix has dimensions $3493\times 48$, the $\bf{q}$ parameter vector has $48$ entries and the covariance matrix $\bf{C}$ matrix has dimensions $3493\times 3493$. In a similar way we may implement the constraint (\ref{mbconstr}) in the SH0ES model without allowing for the additional transition degree of freedom and implement model selection criteria to compare the baseline with the transition model in the presence of the inverse distance ladder constraint. 

\begin{figure*}
\begin{centering}
\includegraphics[width=0.9\textwidth]{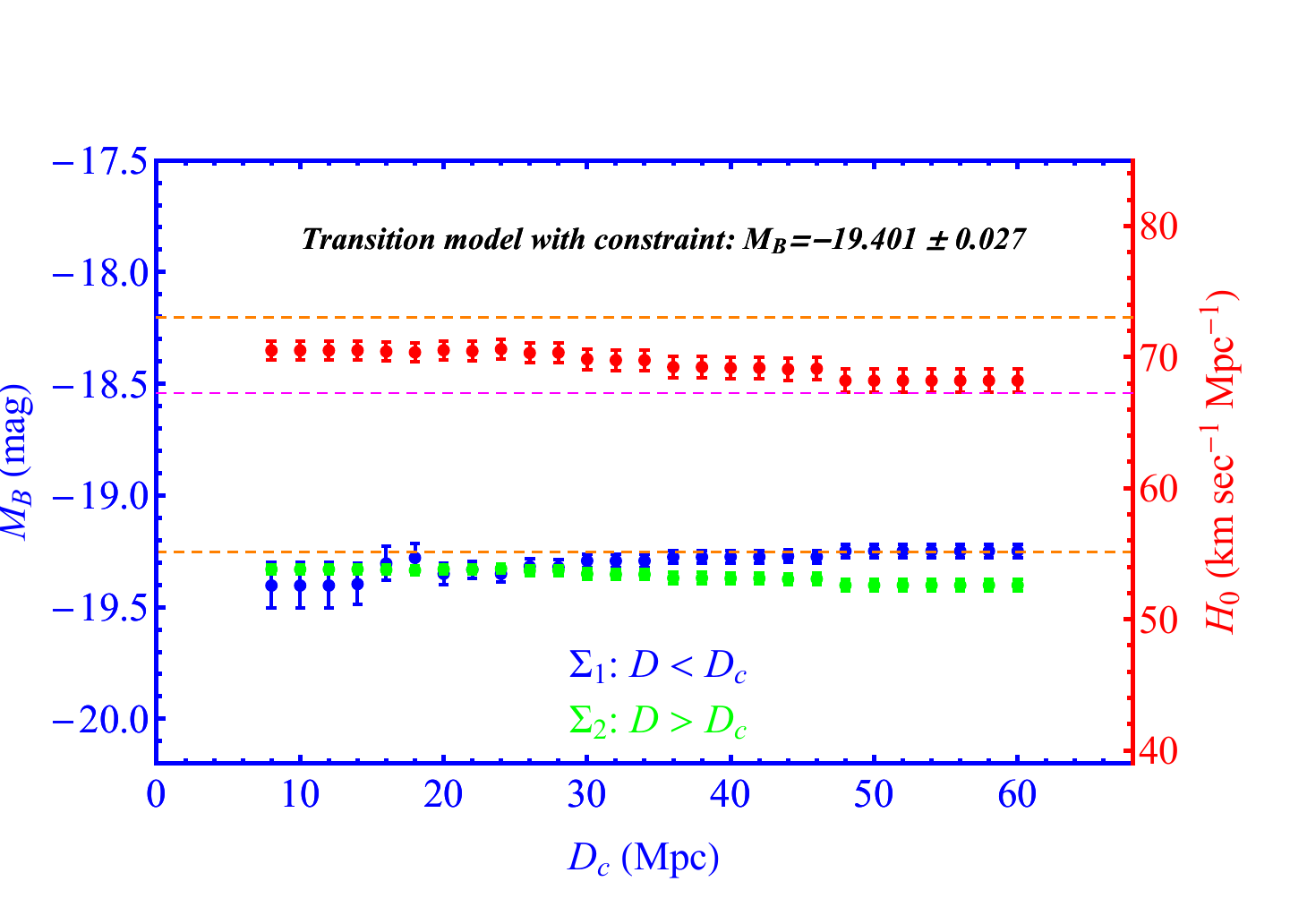}
\par\end{centering}
\caption{The new constraints on $H_0$ and on the parameters $M_{B}^<$, $M_{B}^>$ emerging after implementing the inverse distance ladder constraint (\ref{mbconstr}) on the high distance bin of the $M_B$ transition model.} 
\label{figmb2} 
\end{figure*}

The new constraints on $H_0$ and on the parameters $M_{B}^<$, $M_{B}^>$ emerging from this modified modeling analysis are shown in Fig. \ref{figmb2}  in terms of $D_c$. The corresponding quality of fit expressed via the value of $\chi_{min}^2$ and model selection \cite{Kerscher:2019pzk,Liddle:2007fy} expressed via the AIC \cite{akaike1974new} and BIC criteria is shown in Fig. \ref{figmbcon}. The definitions and properties of the AIC and BIC criteria are described in Appendix \ref{AppendixD}

\begin{figure*}
\begin{centering}
\includegraphics[width=0.9\textwidth]{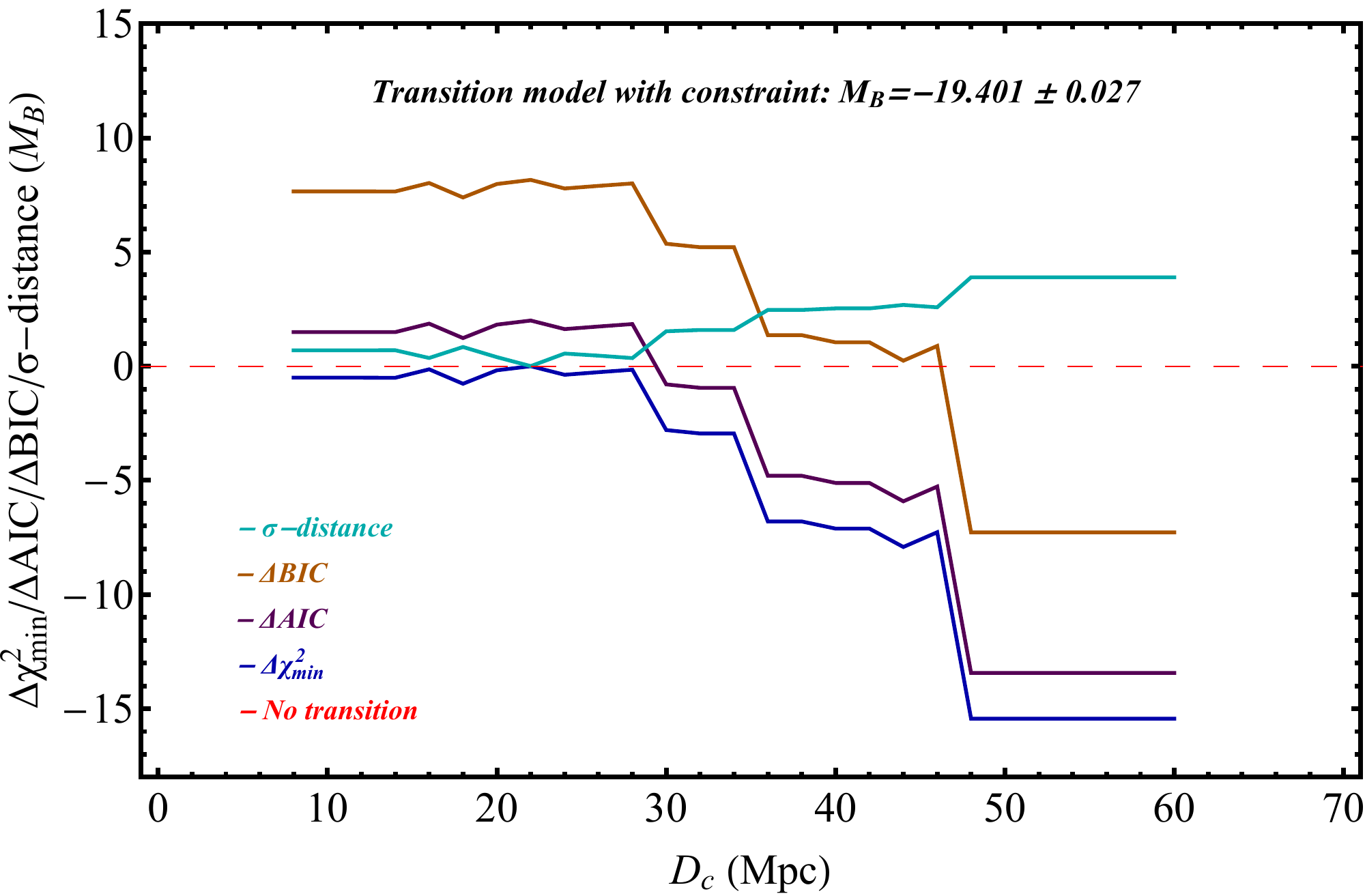}
\par\end{centering}
\caption{Model selection criteria of the $M_B$ transition model with respect to the baseline SH0ES model in the presence of the inverse distance ladder constraint (\ref{mbconstr}). The $\sigma$-distance between the low distance best fit $M_{B}^<$ and the high distance $M_{B}^>$ is also plotted in terms of $D_c$. Clearly the $M_B$ transition model is strongly preferred with respect to the SH0ES baseline (no transition) model especially for transition distances $D_c\simeq 50Mpc$.} 
\label{figmbcon} 
\end{figure*}

Clearly, the transition degree of freedom at $D_c\simeq 50Mpc$ which is mildly preferred by the data even in the absence of the inverse distance ladder constraint, is strongly preferred by the data in the presence of the constraint while the $\sigma$-distance between the low distance best fit parameter $M_{B}^<$ and the high distance $M_B^>$ reaches a level close to $4\sigma$. These results are described in more detail in Table \ref{tab:res}. The full list of the best fit values of all the parameters of the vector $\bf{q}$ for the SH0ES baseline model with and without the inverse distance ladder constraint in the data vector $\bf{Y}$ is shown in Table \ref{tab:resall} along with the corresponding uncertainties.

\begin{table}
\caption{A model comparison (baseline vs transitions) and best fit parameter values in the absence (top five rows) and in the presence (last two rows) of the inverse distance ladder constraint. Notice that in the presence of the inverse distance ladder constraint and the transition degree of freedom, the best fit value of $H_0$ is identical with the inverse distance ladder result of the completed SDSS-IV extended BAO survey \cite{eBOSS:2020yzd}.}
\label{tab:res} 
\vspace{2mm}
\begin{adjustwidth}{-3.8cm}{0.3cm}
\setlength{\tabcolsep}{0.25em}
{\footnotesize\begin{tabular}{cccccccccc} 
\hhline{==========}
   & \\
Model & $\chi^2_{min}$ & $\chi_{red}^2$ $^a$ & $\Delta AIC $& $\Delta BIC$ & $H_0$ &$M_B$&$M_H^W$&$\Delta b_W$&$Z_W$\\
 & &&& &$[Km\,s^{-1}\,Mpc^{-1}]$&$[mag]$&$[mag]$&$[mag/dex]$&$ [mag/dex]$\\
     & \\
   \hhline{==========}
     & \\
Baseline&3552.76&1.031&0 &0 &$73.043\pm 1.007$  &$-19.253\pm0.029 $&$-5.894\pm 0.018$ & $-0.013\pm0.015$ &$ -0.217\pm 0.045$\\
\\
\hline
\\
Transition$^b$ $M_B$&3551.31  &1.031&0.55 & 6.71&$67.326\pm 4.647$  & $-19.250\pm 0.029$& $-5.894\pm 0.018$&$-0.013\pm 0.015$&$-0.217 \pm 0.045$ \\
 & & & && &$ -19.430\pm  0.150$& & &  \\
& & & && &$1.2\sigma$& & &  \\  
\hline
\\
Transition$^b$  $M_H^W$&3551.31 & 1.031 &0.55&6.71&$73.162\pm 1.014$ & $-19.250\pm 0.029$&$-5.894\pm 0.018$  &$-0.013\pm 0.015$  &$-0.217\pm 0.045$  \\
 & & &&&  &&$-5.713 \pm 0.151$ & &  \\
& & & && & &$1.2\sigma$& &   \\  
\hline
\\
Transition$^b$  $Z_W$&3549.99 &1.030 &-0.77&5.39&$72.981\pm 1.007 $ & $-19.255 \pm 0.029 $ & $-5.894\pm 0.018 $ & $-0.014 \pm 0.015 $ & $-0.217 \pm 0.045 $\\
 & & &&&  && & & $\;\;2.588\pm 1.686 $ \\
 & & & && && & & $1.7\sigma$ \\ 
\hline
\\
Transition$^b$  $b_W$&3550.86 &1.030 &0.10& 6.26 & $73.173 \pm 1.013$  & $-19.249 \pm 0.029 $ & $-5.894 \pm 0.018$ & $-0.013\pm 0.015$ & $-0.217\pm 0.045$ \\
 & & & && && & $\;\;\:0.315\pm 0.239$ &  \\
& & & && && &$1.4\sigma$ &  \\
\hline
\\
{\bf Baseline+Constraint$^c$}&3566.78&1.035&0 &0 &$70.457\pm 0.696$  &$-19.332\pm0.020 $&$-5.920\pm 0.017$ & $-0.026\pm0.015$ &$ -0.220\pm 0.045$\\
\\
\hline
\\
{\bf Transition$^{b,c}$ $M_B$+Constraint}&3551.34  &1.031&-13.44 &-7.27&$68.202\pm 0.879$  & $-19.249\pm 0.029$& $-5.893\pm 0.018$&$-0.013\pm 0.015$&$-0.217 \pm 0.045$ \\
 & & & && &$ -19.402\pm 0.027$& & &  \\
 & & & && &$3.9\sigma$& & &  \\ 
 \hhline{==========} 
 \\
\end{tabular} }
{\footnotesize NOTE - (a) $\chi_{red}^2=\chi_{min}^2/dof$, where $dof=N-M$ is typically the number of degrees of freedom (with $N$ the number of datapoints used in the fit and $M$ the number of free  parameters) for each model. (b) At critical distance $D_c\simeq 50Mpc$. (c) With constraint $M_B=-19.401 \pm 0.027 $.}
\end{adjustwidth}
\end{table}

The following comments can be made on the results shown in Table \ref{tab:res}.
\begin{itemize}
    \item The $M_B$ transition degree of freedom resolves the $H_0$ tension both in the absence of the inverse distance ladder constraint (second raw of Table \ref{tab:res}) and in the presence of it (last row of Table \ref{tab:res}).
    \item In the presence of the inverse distance ladder constraint, the model with the $M_B$ transition degree of freedom at $D_c=50Mpc$ is strongly preferred over the baseline $SH0ES$ model as indicated by the model selection criteria despite the additional parameter it involves (comparison of the last two rows of Table \ref{tab:res}).
    \item The transition degree of freedom when allowed in each one of the other three modeling parameters does not lead to a spontaneous resolution of the Hubble tension since the best fit value of $H_0$ is not significantly affected. However, it does induce an increased absolute difference between the best fit values of the  high distance and low distance parameters which however is not statistically significant due to the large uncertainties of the bin with $D>D_c\simeq 50Mpc$ (see also Figs.  \ref{figbw}, \ref{figmw} and \ref{figzw}).
\end{itemize}

The above comments indicate that interesting physical and or systematic effects may be taking place at distances at or beyond $50Mpc$ in the SH0ES data and therefore more and better quality Cepheid/SnIa data are needed at these distances to clarify the issue. This point is further enhanced by the recent study of Ref. \cite{Wojtak:2022bct} indicating that SnIa in the Hubble flow (at distances $D>90Mpc$) appear to have different color calibration properties than SnIa in Cepheid hosts (at distances $D<75Mpc$).

\begin{figure*}
\begin{centering}
\setlength{\headheight}{23.60004pt}
\addtolength{\topmargin}{-8.40002pt}
\includegraphics[width=0.9\textwidth]{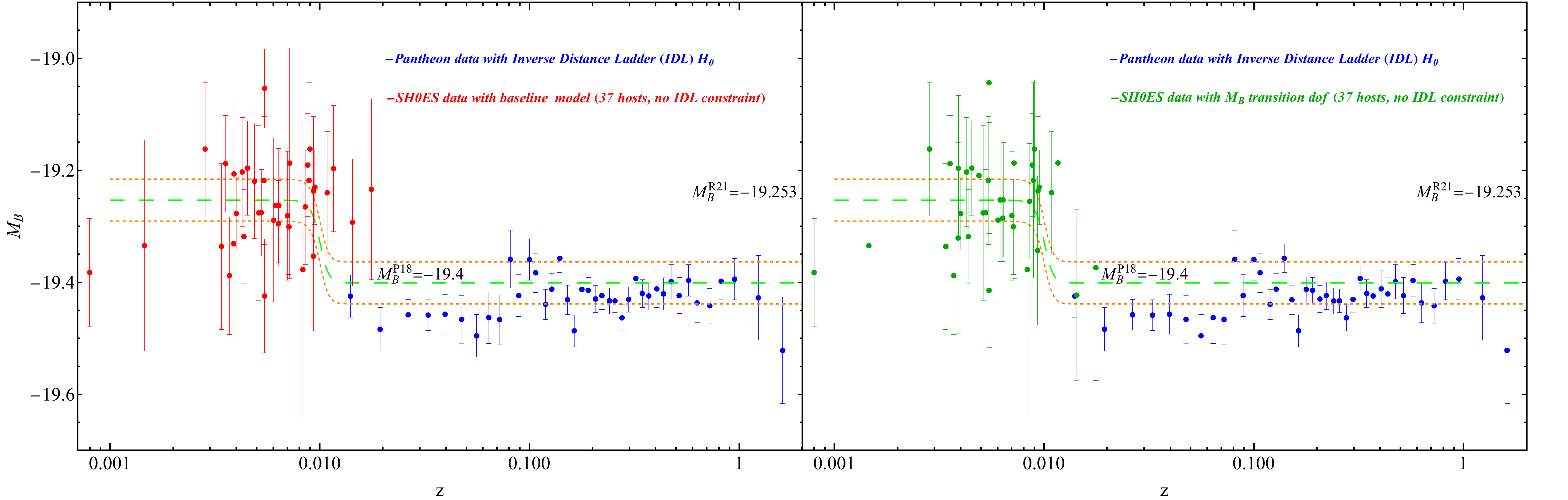}
\par\end{centering}
\caption{The mean SnIa absolute magnitude $M_{Bi}$ for each Cepheid+SnIa host $i$ obtained from Eq. (\ref{mbi}), obtained using the SH0ES baseline model (left panel) and the $M_B$ transition model (right panel) which allows (but does not enforce) an $M_B$ transition at $D_c=50 Mpc$. Notice that when the $M_B$ transition degree of freedom is allowed in the analysis the best fit values of $M_{Bi}$ for the more distant hosts N0976 ($D=60.5Mpc$) and N0105 ($D=73.8 Mpc$) spontaneously drop to the inverse distance ladder calibrated value range (green points with $z>0.01$). } 
\label{figMB2} 
\end{figure*}

The hints for a transition in the SnIa absolute luminosity and magnitude $M_B$ are also demonstrated in Fig. \ref{figMB2} where we show the mean\footnote{Some hosts have more than one SnIa and more than one light curves and thus averaging with proper uncertainties was implemented in these cases.} SnIa absolute magnitude $M_{Bi}$ for each Cepheid+SnIa host $i$ obtained from the equation 
\be
M_{Bi}=m_{Bi}^0-\mu_i 
\label{mbi}
\ee
where $m_{Bi}^0$ is the measured apparent magnitude of the SnIa and $\mu_i$ are the best fit distance host distance moduli obtained using the SH0ES baseline model (left panel of Fig. \ref{figMB2}) and the $M_B$ transition model (right panel) which allows (but does not enforce) an $M_B$ transition at $D_c=50 Mpc$. Notice that when the $M_B$ transition degree of freedom is allowed in the analysis the best fit values of $M_{Bi}$ for the more distant hosts N0976 ($D=60.5Mpc$) and N0105 ($D=73.8 Mpc$) spontaneously drop to the inverse distance ladder calibrated value range. 

\begin{figure*}
\begin{centering}
\setlength{\headheight}{23.60004pt}
\addtolength{\topmargin}{-8.40002pt}
\includegraphics[width=0.9\textwidth]{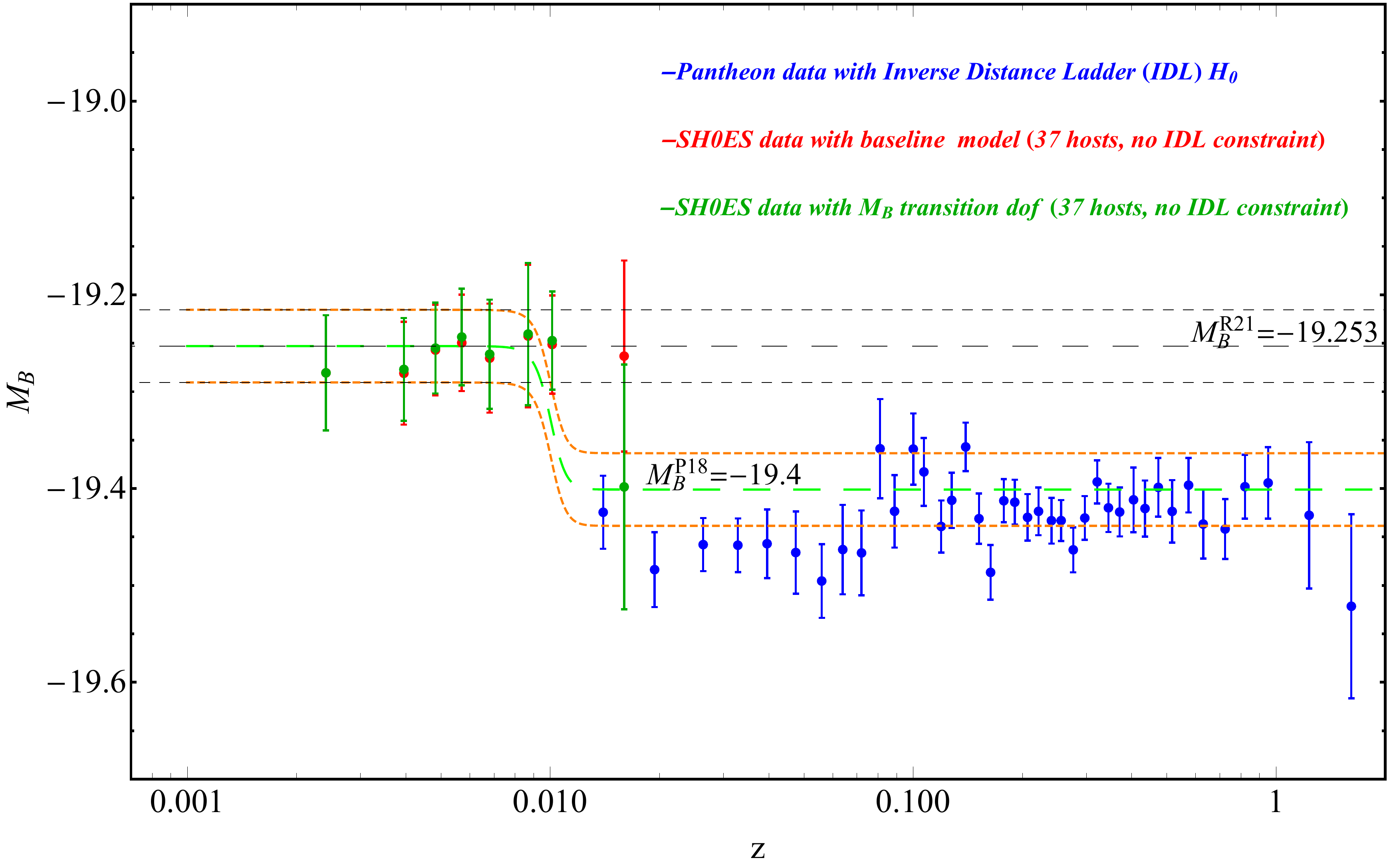}
\par\end{centering}
\caption{The binned (5 host bins with 2 hosts in last bin where $z>0.01$) Cepheid+SnIa host values of $M_{B}$ obtained assuming the baseline SH0ES model (red points) and the $M_B$ transition model ($D_c=50Mpc$, green points) are shown along with the inverse distance ladder calibrated binned $M_{B}$ of the Hubble flow SnIa of the Pantheon dataset (blue points). When the transition dof is allowed, the data excite it and a hint for a transition appears (the green data point of the transition model is clearly below the red point corresponding to the constant $M_B$ SH0ES baseline model).} 
\label{figMbbin} 
\end{figure*}

The inverse distance ladder calibrated values of the absolute magnitudes $M_{Bi}$ of SnIa in the Hubble flow are obtained by assuming $H_0=H_0^{P18}=67.36\pm0.54$~km~s$^{-1}$~Mpc$^{-1}$ and using the following equation
\be
M_{Bi}=m(z_{Bi}^0)+5\log_{10}\left[H_0^{P18}\cdot Mpc/c \right]-5\log_{10}\left[D_L(z_i) \right]-25
\ee
where $D_L(z_i)$ is the Hubble free luminosity distance in the context of \plcdm and $m(z_{Bi}^0)$ are the binned corrected SnIa apparent magnitudes of the Pantheon sample. The corresponding binned Cepheid+SnIa host values of $M_{B}$ obtained assuming the baseline Sh0ES model (red points) and the $M_B$ transition model ($D_c=50Mpc$, green points) are shown in Fig. \ref{figMbbin} along with the inverse distance ladder calibrated binned $M_{B}$ of the Hubble flow SnIa of the Pantheon dataset (blue points). When the transition dof is allowed, the data excite it and a hint for a transition appears (the green data point is of the transition model is clearly below the red point corresponding to the constant $M_B$ SH0ES baseline model) even though the statistical significance of the indicated transition is low due to the small number of Cepheids (41) included in the last bin with $D\in [50Mpc,75Mpc]$.

\section{Conclusion}
\label{sec:Conclusion}

We have described a general framework for the generalization of the Cepheid+SnIa modeling in the new SH0ES data. Such modeling generalization approach is motivated by the increased statistical significance of the Hubble tension and may hint towards new degrees of freedom favored by the data. Such degrees of freedom may be attributed to either new physics or to unknown systematics hidden in the data. 

In the present analysis we have focused on a particular type of new modeling degree of freedom allowing for a transition of any one of the four Cepheid/SnIa modeling parameters at a distance $D_c$. However, our analysis can be easily extended to different degrees of freedom with physical motivation. Examples include possible modeling parameter dependence on properties other than distance (e.g. dust extinction, color and stretch calibration properties etc.). In addition other degrees of freedom that could be excited by the data could involve the modeling parameters that have been absorbed in the R21 SH0ES data like the Cepheid dust extinction parameter $R_W$ and the color and stretch SnIa calibration parameters $\beta$ and $\alpha$. These degrees of freedom may also be probed and excited by the fit to the data if allowed by the modeling \cite{Wojtak:2022bct}.

We have demonstrated that our proposed transition degrees of freedom are mildly excited by the SH0ES data and in the case of the SnIa absolute magnitude $M_B$ transition degree of freedom at $D_c\simeq 50Mpc$, the best fit value of $H_0$ shifts spontaneously to a value almost identical with the \plcdm best fit value thus annihilating the Hubble  tension. However, the high distance bin in this case involves only 41 Cepheids and 4 SnIa and therefore the best fit value of $M_B^>$ which effectively also fixes $H_0$ involves significant uncertainties. 

In the presence of the inverse distance ladder constraint on the high distance bin parameter $M_B^>$ of the $M_B$ transition model and on the single $M_B$ parameter of the SH0ES model, the uncertainties reduce dramatically. The Hubble tension is fully resolved only in the presence of the $M_B$ transition degree of freedom at $D_c\simeq 50Mpc$. This transition model is also strongly  selected over the baseline SH0ES model with the constraint, by both model selection criteria AIC and BIC despite the penalty imposed by AIC and especially BIC on the additional parameter involved in the new modeling. This behavior along with other recent studies \cite{Wojtak:2022bct} hints towards the need of more detailed Cepheid+SnIa calibrating data at distances $D\gtrsim 50Mpc$ i.e. at the high end of  rung 2 on the distance ladder. 

Our comprehensively described generalized modeling approach opens a new direction in the understanding of the Hubble tension and may be extended by the introduction of a wide range of new degrees of freedom in the SH0ES data analysis and also by the introduction of new constraints motivated by other cosmological data. Such extensions could for example involve more distance bins in the distance dependence of the four main modeling parameters. In that case the relevant column of the modeling matrix $\bf{L}$ would have to be replaced by more than two columns (one for each distance bin). Such bins could also be defined not in terms of distance but in terms of other Cepheid/SnIa properties (e.g. Cepheid metallicity or period and/or SnIa light curve color or stretch). In addition, other parameters beyond the four basic modeling parameters may be considered including the dust extinction parameter $R_W$ and/or the SnIa light curve color and stretch parameters.  Also, similar modeling generalizations may be implemented on different distance calibrator data used for the measurement of $H_0$ such as the TRGB data \cite{Freedman:2019jwv}.

Physical motivation is an important factor for the evaluation of any new modeling degree of freedom especially if it is favored by the data. A transition of the SnIa luminosity at a particular recent cosmic time could be induced by a sudden change of the value of a fundamental physics constant e.g. the gravitational constant in the context of a recent first order phase transition \cite{Coleman:1977py,Callan:1977pt,Patwardhan:2014iha} to a new vacuum of a scalar-tensor theory or in the context of a generalization of the symmetron screening mechanism \cite{Perivolaropoulos:2022txg}. A similar first order transition is implemented in early dark energy models \cite{Niedermann:2020dwg} attempting to change the last scattering sound horizon scale without affecting other well constrained cosmological observables. Thus, even though no relevant detailed analysis has been performed so far, there are physical mechanisms that could potentially induce the SnIa luminosity transition degree of freedom.

The emerging new puzzles challenging our standard models may soon pave the way to exciting discoveries of new physical laws. The path to these discoveries goes through the deep and objective understanding of the true information that is hidden in the cosmological data. The present analysis may be a step in that direction.

-----------------------------------------------------

\funding{This work was supported by the Hellenic Foundation for Research and
Innovation (HFRI - Progect No: 789). }



\dataavailability{The numerical analysis files for the reproduction of the figures can be found in the \href{https://github.com/FOTEINISKARA/A-reanalysis-of-the-SH0ES-data-for-H_0}{A reanalysis of the SH0ES data for $H_0$} GitHub repository under the MIT license.}

\acknowledgments{We thank Adam Riess, Dan Scolnic,  Eoin Colgain, and Radoslaw Wojtak for useful comments.}

\conflictsofinterest{The authors declare no conflict of interest.}


\appendixtitles{yes} 

\newpage
\appendixstart

\appendix
\section[\appendixname~\thesection]{Covariance Matrix}
\label{AppendixA}

A schematic form of the non-diagonal covariance matrix used in our analysis is shown below. The symbols $\sigma_{tot,i}^2$ indicate the (non-diagonal in general) covariance submatrix  within the $i$ host while the symbols $Z_{cov}$ indicate submatrices that correlate the uncertainties between different hosts.

\begin{adjustwidth}{0.cm}{1cm}
\be
\nonumber
\begin{turn}{90}
{\bf {C}=\setcounter{MaxMatrixCols}{50}{\footnotesize
$\left(\arraycolsep=1.2pt\def\arraystretch{1.8}\begin{array}[c]{ccccccccccccccccccccc}
\sigma_{tot,1}^2&\ldots&Z_{cov}&Z_{cov}&0&0&0&\ldots&0&0&0&0&0&0&0&0&0&0&\ldots&0\\
\ldots&\ldots&\ldots&\ldots&\ldots&\ldots&\ldots&\ldots&\ldots&\ldots&\ldots&\ldots&\ldots&\ldots&\ldots&\ldots&\ldots&\ldots&\ldots&\ldots&\ldots\\
Z_{cov}&\ldots&\sigma_{tot,37}^2&Z_{cov}&0&0&0&\ldots&0&0&0&0&0&0&0&0&0&0&\ldots&0\\
\hline
Z_{cov}&\ldots&Z_{cov}&\sigma_{tot,N4258}^2&0&0&0&\ldots&0&0&0&0&0&0&0&0&0&0&\ldots&0\\
0&\ldots&0&0&\sigma_{tot,M31}^2&0&0&\ldots&0&0&0&0&0&0&0&0&0&0&\ldots&0\\
0&\ldots&0&0&0&\sigma_{tot,LMC}^2&0&\ldots&0&0&0&0&0&0&0&0&0&0&\ldots&0\\
\hline
0&\ldots&0&0&0&0&\sigma_{M_B,1}^2&\ldots&Sn_{cov}&0&0&0&0&0&0&0&0&Sn_{cov}&\ldots&Sn_{cov}\\
\ldots&\ldots&\ldots&\ldots&\ldots&\ldots&\ldots&\ldots&\ldots&\ldots&\ldots&\ldots&\ldots&\ldots&\ldots&\ldots&\ldots&\ldots&\ldots&\ldots&\ldots\\
0&\ldots&0&0&0&0&Sn_{cov}&\ldots&\sigma_{M_B,77}^2&0&0&0&0&0&0&0&0&Sn_{cov}&\ldots&Sn_{cov}\\
\hline
0&\ldots&0&0&0&0&0&\ldots&0&\sigma_{M_H^W,HST}^2&0&0&0&0&0&0&0&0&\ldots&0\\
0&\ldots&0&0&0&0&0&\ldots&0&0&\sigma_{M_H^W,Gaia}^2&0&0&0&0&0&0&0&\ldots&0\\
0&\ldots&0&0&0&0&0&\ldots&0&0&0&\sigma_{Z_W,Gaia}^2&0&0&0&0&0&0&\ldots&0\\
0&\ldots&0&0&0&0&0&\ldots&0&0&0&0&\sigma_x^2&0&0&0&0&0&\ldots&0\\
0&\ldots&0&0&0&0&0&\ldots&0&0&0&0&0&\sigma_{ground,zp}^2&0&0&0&0&\ldots&0\\
0&\ldots&0&0&0&0&0&\ldots&0&0&0&0&0&0&\sigma_{b_W}^2&0&0&0&\ldots&0\\
0&\ldots&0&0&0&0&0&\ldots&0&0&0&0&0&0&0&\sigma_{\mu,N4258}^2&0&0&\ldots&0\\
0&\ldots&0&0&0&0&0&\ldots&0&0&0&0&0&0&0&0&\sigma_{\mu,LMC}^2&0&\ldots&0\\
\hline
0&\ldots&0&0&0&0&Sn_{cov}&\ldots&Sn_{cov}&0&0&0&0&0&0&0&0&\sigma_{M_B,z,1}^2&\ldots&Sn_{cov}\\
\ldots&\ldots&\ldots&\ldots&\ldots&\ldots&\ldots&\ldots&\ldots&\ldots&\ldots&\ldots&\ldots&\ldots&\ldots&\ldots&\ldots&\ldots&\ldots&\ldots&\ldots\\
0&\ldots&0&0&0&0&Sn_{cov}&\ldots&Sn_{cov}&0&0&0&0&0&0&0&0&Sn_{cov}&\ldots&\sigma_{M_B,z,277}^2\\
\end{array}
 \right)$ }}
\end{turn}
\ee
\end{adjustwidth}

\appendix
\renewcommand\thefigure{\thesection.\arabic{figure}}
\section[\appendixname~\thesection]{: Analytic minimization of $\chi^2$.}
\label{AppendixB}
\setcounter{figure}{0} 

\begin{figure*}
\begin{centering}
\includegraphics[width=0.8\textwidth]{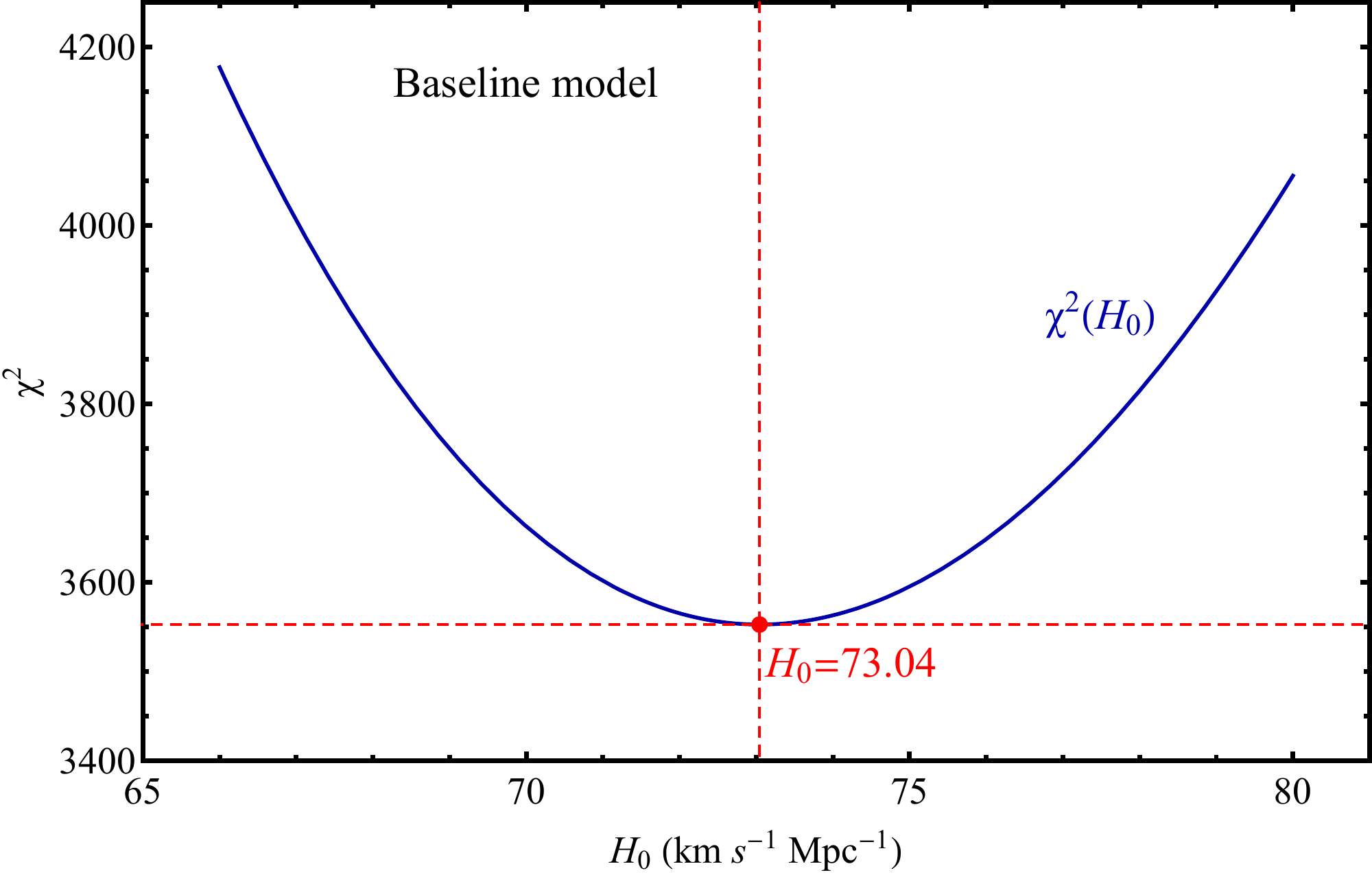}
\par\end{centering}
\caption{The form of $\chi^2(H_0$) for the SH0ES baseline model has a minimum at the analytically predicted value of $H_0$ provided by (\ref{qbest}) ( the rest of the parameters were fixed at their analytically predicted best fit values). } 
\label{plchi2h02} 
\end{figure*}

The proof of Eqs. (\ref{bfpar1}) and (\ref{errmat}) that lead to the best fit parameter values and their uncertainties through the analytic minimization of $\chi^2$ may be sketched as follows\footnote{For more details see \url{https://people.duke.edu/~hpgavin/SystemID/CourseNotes/linear-least-squres.pdf}}:\\
Using  the matrix of measurements (data vector) $\bf{Y}$,  the matrix of parameters $\bf{q}$ and   the equation (or design) matrix $\bf{L}$ with the measurement error matrix (covariance matrix) $\bf{C}$ the $\chi^2$ statistic is expressed as 
\be
\chi^2=(\bf{Y}-\bf{Lq})^T\bf{C}^{-1}(\bf{Y}-\bf{Lq})=\bf{q}^T\bf{L}^T\bf{C}^{-1}\bf{L}\bf{q}-2\bf{q}^T\bf{L}^T\bf{C}^{-1}\bf{Y}+\bf{Y}^T\bf{C}^{-1}\bf{Y}
\label{chi2p}
\ee
The $\chi^2$ is minimized with respect to the parameters $\bf{q}$, by solving the equation
\be 
\frac{\partial\chi^2}{\partial \bf{q}}\Big|_{{\bf q}_{best}}=0=>2\bf{L}^T\bf{C}^{-1}\bf{L}{\bf q}_{best}-2\bf{L}^T\bf{C}^{-1}\bf{Y}=0
\label{chi2m}
\ee
Thus the maximum-likelihood parameters are given as
\be
{\bf q}_{best}=(\bf{L}^T\bf{C}^{-1}\bf{L})^{-1}\bf{L}^T\bf{C}^{-1}\bf{Y}
\label{qbest}
\ee 
We have tested the validity of this equation numerically by calculating $\chi^2$ for the SH0ES baseline model and showing that indeed it has a minimum at the analytically predicted parameter values provided by (\ref{qbest}). This is demonstrated in Fig. \ref{plchi2h02} where we show $\chi^2(H_0)$ with the rest of the parameters fixed at their analytically predicted best fit values. As expected the minimum is obtained at the analytically predicted value of $H_0$.

The standard errors squared of the parameters in $\bf{q_{best}}$ are given as the diagonal elements of the transformed covariance matrix 
\be
\bf{\varSigma_{kl}}=\sum_{i}\sum_{j}\left[\frac{\partial  {\bf q}_{best,k}}{\partial Y_i}\right] \bf{C_{ij}}\left[\frac{\partial {\bf q}_{best,l}}{\partial Y_j}\right]
\ee
or
\be
\bf{\varSigma}=\left[\frac{\partial  {\bf q}_{best}}{\partial Y}\right] \bf{C}\left[\frac{\partial {\bf q}_{best}}{\partial Y}\right]^T
\ee

Thus
\begin{eqnarray}{}
\bf{\varSigma}&=(\bf{L}^T\bf{C}^{-1}\bf{L})^{-1}\bf{L}^T\bf{C}^{-1}C\left[(\bf{L}^T\bf{C}^{-1}\bf{L})^{-1}\bf{L}^T\bf{C}^{-1}\right]^T\\
&=(\bf{L}^T\bf{C}^{-1}\bf{L})^{-1}\bf{L}^T\bf{C}^{-1}C\bf{C}^{-1}\bf{L}(\bf{L}^T\bf{C}^{-1}\bf{L})^{-1}\\
&=(\bf{L}^T\bf{C}^{-1}\bf{L})^{-1}
\end{eqnarray}
The standard errors provided by the this equation for the SH0ES baseline model are consistent and almost identical with the published results of R21.

\renewcommand\thefigure{\thesection.\arabic{figure}}
\renewcommand\thetable{\thesection.\arabic{table}}
\section[\appendixname~\thesection]{: Reanalysis of Individual Cepheid $m_H^W$ slope data.}
\label{AppendixC}
\setcounter{figure}{0} 
\setcounter{table}{0} 

\begin{figure*}
\begin{centering}
\includegraphics[width=1\textwidth]{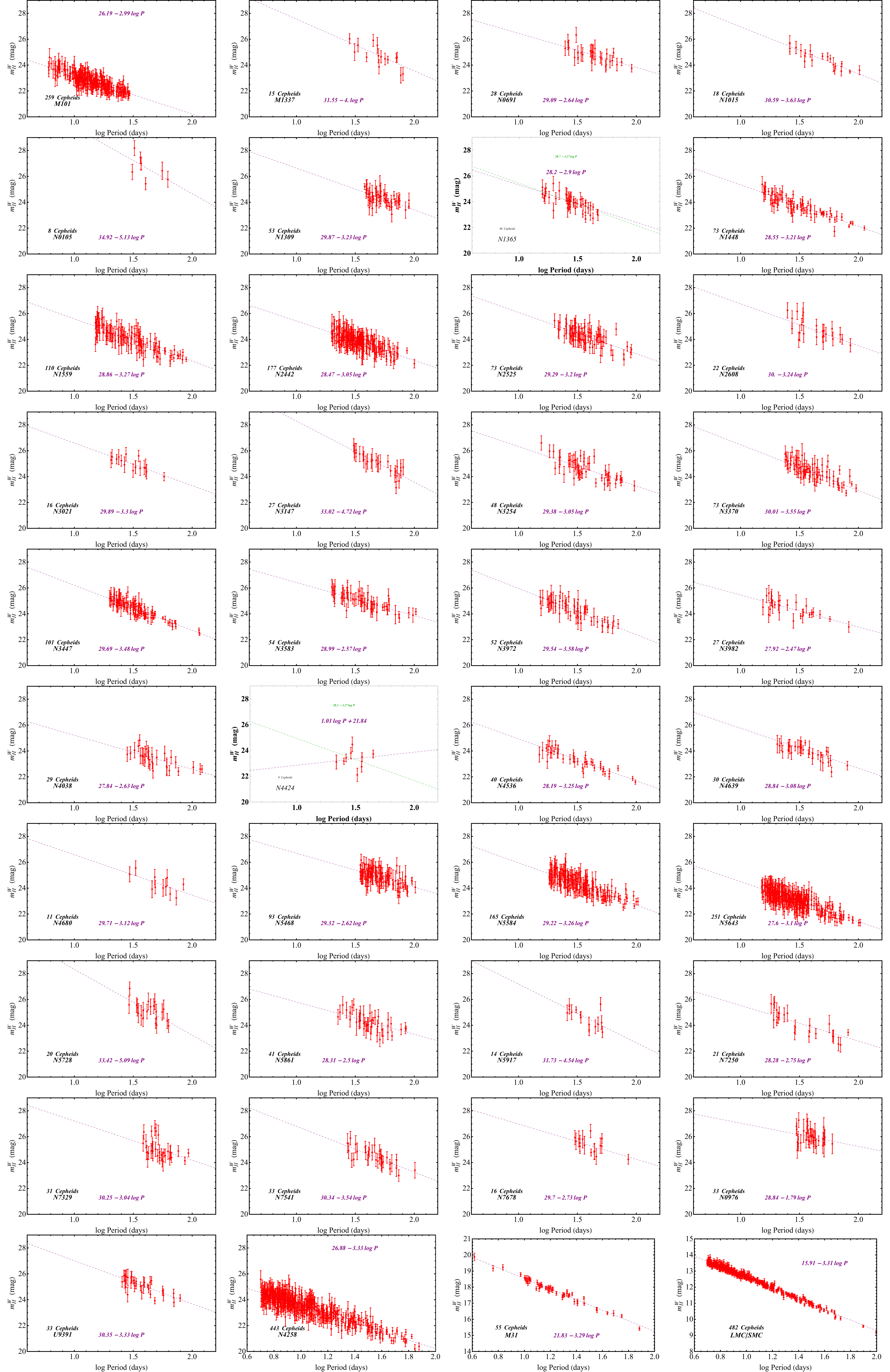}
\par\end{centering}
\caption{Fitting individual slopes $b_W $ of the $m_H^W-\log P$, P-L relations for 40 Cepheid hosts: 37 SnIa+Cepheid hosts, 2 anchors (N4258+LMC/SMC) and the pure Cepheid host M31. In the case of the SnIa+Cepheid host N4424 our fit (purple line) is not in agreement with the fit of R21 (green line). In all other cases the agreement is very good.} 
\label{figballnocov} 
\end{figure*}

\begin{figure*}
\begin{centering}
\includegraphics[width=0.8\textwidth,height=0.8\textheight]{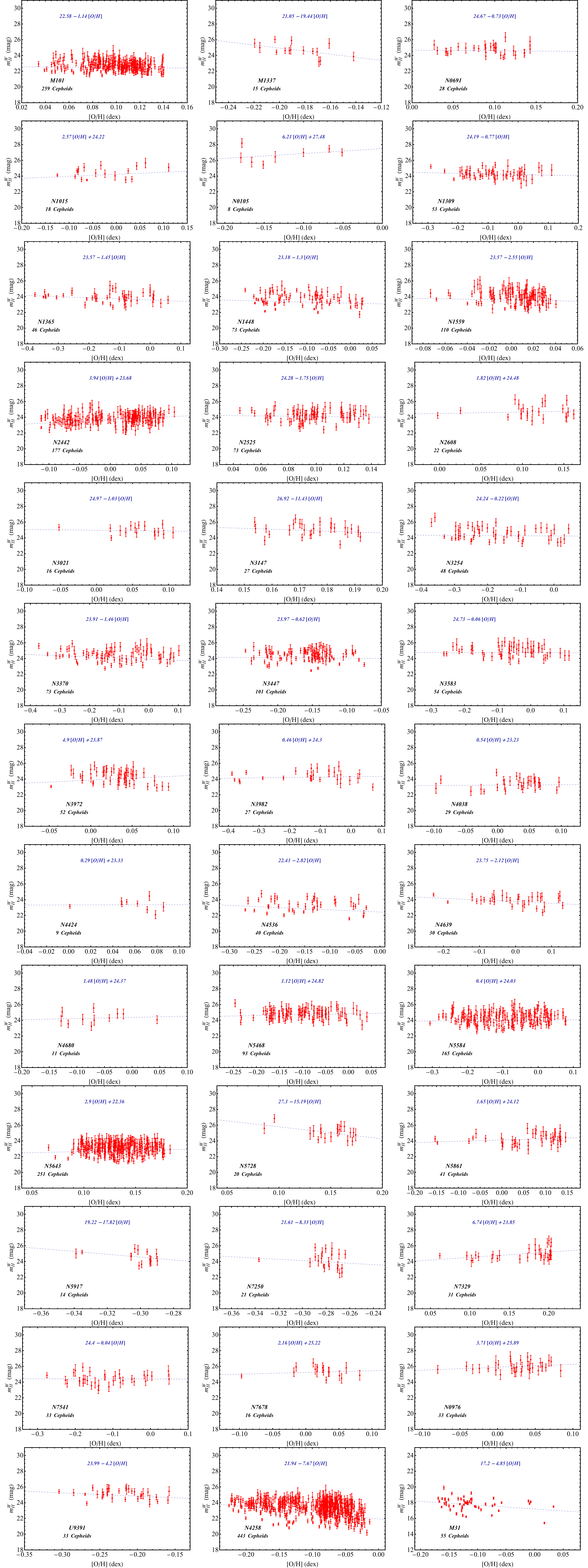}
\par\end{centering}
\caption{Fitting individual slopes $Z_W $ of the $m_H^W$ metallicity relations
to Cepheid data for the hosts where individual Cepheid metallicity data were  provided by R21 (only an average metallicity was provided for LMC/SMC and thus these Cepheid hosts are not included in the plot). Thus the 39 hosts shown correspond to the 37 SnIa+Cepheid hosts, the anchor N4258 and the pure Cepheid host M31. } 
\label{figzallnocov} 
\end{figure*}

\begin{figure*}
\begin{centering}
\includegraphics[width=0.8\textwidth]{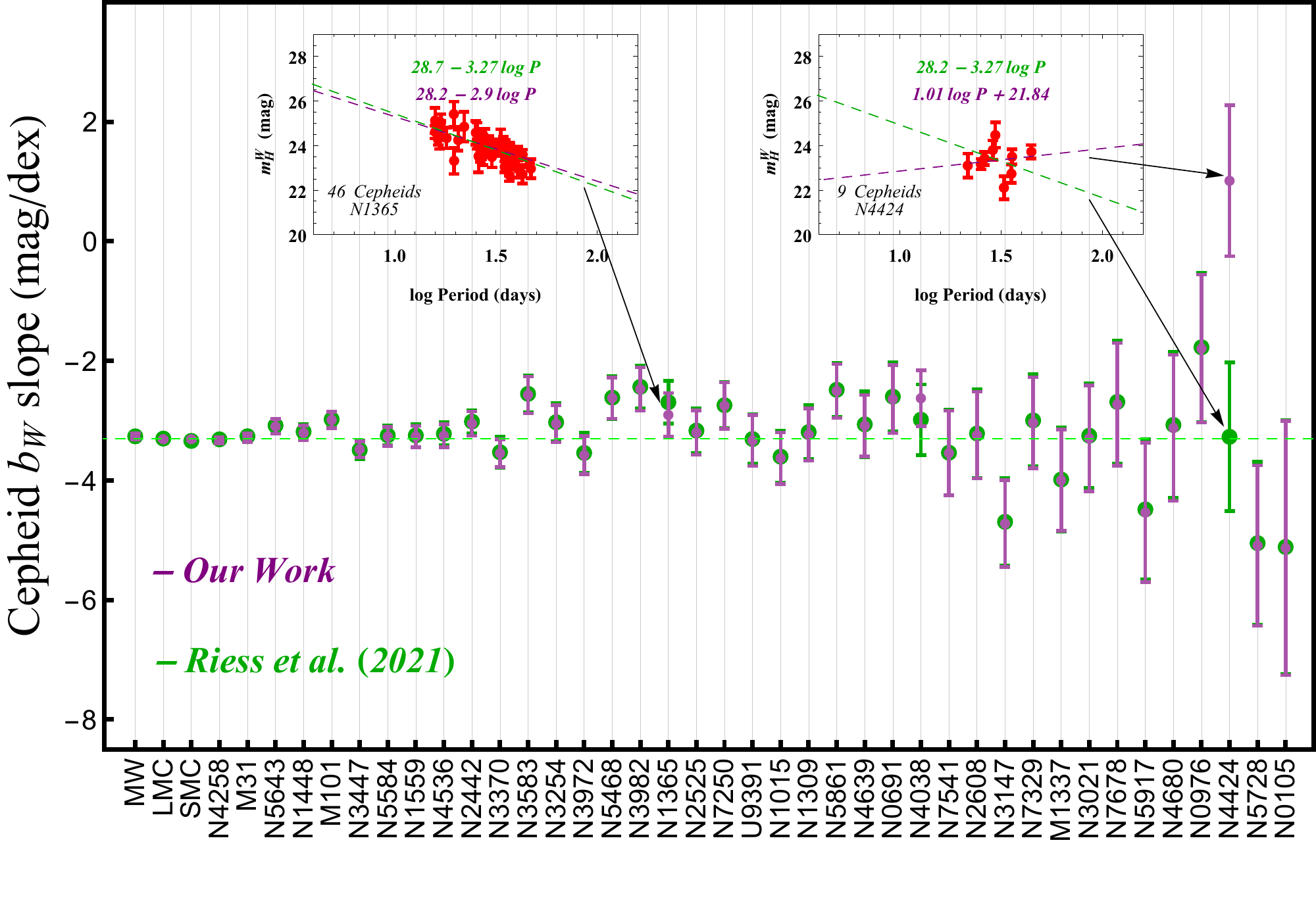}
\par\end{centering}
\caption{Independently-fitted slopes $b_W $ of the $m_H^W$
P-L relations (without background covariance of any pair of Cepheids as was done in R21 i.e. taking into account only diagonal terms of the covariance matrix). The points in green are the corresponding points shown in Fig. 10 of R21 (obtained using plot digitizer). Our fits are in excellent agreement with R21 with the exception of 3 points corresponding to hosts N4424, N4038 and N1365. For N4424 there is no point shown in Fig. 10 of R21 but in the corresponding $m_H^W-\log P$ plot of R21 (their Fig. 9 where the magnitudes decrease upwards on the vertical axis) the indicated best fit line is the green line shown in the upper right inset plot corresponding to the green point. The correct best fit however corresponds to the purple line of the upper right inset plot with slope given by the purple point pointed by the arrow. Similarly the best fit line corresponding to N1365 shown in the upper left inset plot is identical to our purple point and not to the green point. For the Milky Way (MW) we have used directly the values provided in R21 obtained from Gaia EDR3+HST. } 
\label{figb102bnocov} 
\end{figure*}

We use the released Cepheid P-L data ($m_H^W-\log P$) to find the best fit slope $b_W$ in each one of the 40 Cepheid hosts where the 3130 Cepheid magnitudes are distributed. We thus reproduce Figs. 9 and 10 of R21. Since the released Cepheid magnitude data due not include outliers we focus on the reproduction of the black points of Fig. 10 of R21. Our motivation for this attempt includes the following:
\begin{itemize}
    \item It is evident by inspection of Fig. 10 of R21 (see also Fig. \ref{figb102bnocov} below) that most of the better measured $b_W$ slopes have a smaller absolute slope than the corresponding slopes obtained from anchors+MW hosts. This makes it interesting to further test the homogeneity of these measurements and their consistency with the assumption of a global value of $b_W$ for both anchors and SnIa hosts.
    \item Most of the measurements that include outliers (red points of Fig. 10 of R21) tend to amplify the above mentioned effect ie they have smaller absolute slopes $b_W$ than the slopes obtained from the anchors.
    \item The slope corresponding to the host N4424 is missing from Fig. 10 of R21 while the host M1337 appears to show extreme behavior of the outliers.
\end{itemize}

Based on the above motivation we have recalculated all the $b_W$ slopes of the Cepheid hosts and also extended the analysis to metallicity slopes for the Cepheids of each hosts. The individual metallicity slopes $Z_{W,i}$ for each host have been obtained for the first time in our analysis and thus no direct comparison can be made with R21 for these slopes.

The Cepheid magnitude period and magnitude metallicity data used to calculate the corresponding best fit slopes $b_W$ and $Z_W$ are shown in Figs. \ref{figballnocov} and \ref{figzallnocov} for each host along with the best fit straight lines from which the individual best fit $b_{W,i}$ and $Z_{W,i}$ were obtained. The numerical values of the derived best fit slopes along with the corresponding uncertainties are shown in Table \ref{tab:slopes}.

The derived best fit slopes $b_{W,i}$ for each host along with their standard errors (obtained using Eqs. (\ref{bfpari}) and (\ref{errmati}) ) are shown in Fig. \ref{figb102bnocov}  (purple points) along with the corresponding points of Fig. 10 of R21 (green points).

The best fit values of the slopes $b_{W,i}$ we have obtained using the raw Cepheid data are consistent with Fig. 10 of R21 as shown in Fig. \ref{figb102bnocov}.  The agreement with the results of R21 is excellent except of three points. Two slopes corresponding to the hosts N4038 and N1365 are slightly shifted in our analysis compared to R21 due a small disagreement in the best fit slope and a typo of R21 in transferring the correct slope to Fig. 10. 

In addition, the slope corresponding to the host N4424 is missing in Fig. 10 of R21. In the $m_H^W-\log P$ plot of R21 (their Fig. 9 where the magnitudes decrease upwards on the vertical axis) corresponding to N4424, the indicated best fit line is the green line shown in the upper right inset plot of Fig. \ref{figb102bnocov} corresponding to the green point. The correct best fit however corresponds to the purple line of the upper right inset plot with slope given by the purple point pointed by the arrow. 

Thus our Fig. \ref{figb102bnocov} is in excellent agreement with Fig. 10 of R21 with the exception of three points where our plot corrects the corresponding plot of R21. In any case we stress that these three points do not play a significant role in our conclusion about the inhomogeneities of $b_W$ and $Z_W$ discussed in section \ref{seccephomog}.

\begin{table}
\caption{The best fit individual slopes $b_W$ and $Z_W$ with their standard errors $\sigma$ and scatter uncertainties $\sigma_{scat}$. The scatter uncertainty secures that the model of a universal best fit slope becomes an acceptable model with $\chi_{min}^2/dof \simeq 1$, as described in subsection \ref{seccephomog}. The number of hosts in the Table is $N=42$ including the Milky Way (MW). For LMC and SMC the metallicities were not provided individually for each Cepheid in R21 and thus no $Z_W$ slope could be estimated in our analysis.}
\label{tab:slopes} 
\vspace{2.5mm}
\setlength{\tabcolsep}{0.6em}
\begin{adjustwidth}{0cm}{1.5cm}
{\footnotesize\begin{tabular}{ccccccccc} 
\hhline{=========}
  & & & & & & & & \\
Galaxy & $\qquad D^{a}\quad $ & $b_W$&$\sigma$ &$\sigma_{scat}$ &$Z_W$&$\sigma$ &$\sigma_{scat}$\\ 
  &[Mpc]   & [mag/dex] & [mag/dex]   & [mag/dex] & [mag/dex]&[mag/dex] & [mag/dex] \\ 
 & & & & & & & & \\
\hhline{=========}
 & & & & & & & & \\
 M101& 6.71& -2.99& 0.14& 0.18& -1.14& 0.86& 3.2\\
 M1337& 38.53& -4.& 0.85& 0.18& -19.44& 6.27& 3.2\\
 N0691& 35.4& -2.64& 0.57& 0.18& -0.73& 2.54& 3.2\\
 N1015& 35.6& -3.63& 0.43& 0.18& 2.57& 1.3& 3.2\\
 N0105& 73.8& -5.13& 2.13& 0.18& 6.21& 3.8& 3.2\\
 N1309& 29.4& -3.23& 0.44& 0.18& -0.77& 0.59& 3.2\\
 N1365& 22.8& -2.9& 0.37& 0.18& -1.45& 0.39& 3.2\\
 N1448& 16.3& -3.21& 0.12& 0.18& -1.3& 0.39& 3.2\\
 N1559& 19.3& -3.27& 0.18& 0.18& -2.55& 1.62& 3.2\\
 N2442& 20.1& -3.05& 0.2& 0.18& 3.94& 0.58& 3.2\\
 N2525& 27.2& -3.2& 0.37& 0.18& -1.75& 2.74& 3.2\\
 N2608& 35.4& -3.24& 0.73& 0.18& 1.82& 2.33& 3.2\\
 N3021& 26.6& -3.3& 0.89& 0.18& -1.03& 2.85& 3.2\\
 N3147& 42.5& -4.72& 0.73& 0.18& -11.43& 9.13& 3.2\\
 N3254& 24.4& -3.05& 0.31& 0.18& -0.22& 0.57& 3.2\\
 N3370& 24& -3.55& 0.23& 0.18& -1.46& 0.4& 3.2\\
 N3447& 20.8& -3.48& 0.13& 0.18& -0.62& 0.66& 3.2\\
 N3583& 34.4& -2.57& 0.31& 0.18& -0.06& 0.62& 3.2\\
 N3972& 15.1& -3.58& 0.32& 0.18& 4.9& 1.58& 3.2\\
 N3982& 19& -2.47& 0.36& 0.18& 0.46& 0.47& 3.2\\
 N4038& 29.3& -2.63& 0.47& 0.18& 0.54& 2.07& 3.2\\
 N4424& 16.4& 1.01& 1.26& 0.18& 0.29& 4.33& 3.2\\
 N4536& 31.9& -3.25& 0.19& 0.18& -2.82& 0.56& 3.2\\
 N4639& 19.8& -3.08& 0.51& 0.18& -2.12& 0.64& 3.2\\
 N4680& 42.1& -3.12& 1.22& 0.18& 1.48& 3.27& 3.2\\
 N5468& 46.3& -2.62& 0.35& 0.18& 1.12& 0.74& 3.2\\
 N5584& 28& -3.26& 0.16& 0.18& 0.4& 0.36& 3.2\\
 N5643& 20.7& -3.1& 0.12& 0.18& 2.9& 1.17& 3.2\\
 N5728& 44.9& -5.09& 1.34& 0.18& -15.19& 5.71& 3.2\\
 N5861& 30.3& -2.5& 0.45& 0.18& 1.65& 0.72& 3.2\\
 N5917& 31& -4.54& 1.17& 0.18& -17.82& 6.56& 3.2\\
 N7250& 12.8& -2.75& 0.39& 0.18& -8.33& 4.44& 3.2\\
 N7329& 46.8& -3.04& 0.76& 0.18& 6.74& 1.86& 3.2\\
 N7541& 34.4& -3.54& 0.71& 0.18& -0.04& 1.08& 3.2\\
 N7678& 46.6& -2.73& 1.03& 0.18& 2.16& 2.27& 3.2\\
 N0976& 60.5& -1.79& 1.24& 0.18& 3.71& 2.73& 3.2\\
 U9391& 29.4& -3.33& 0.43& 0.18& -4.2& 1.51& 3.2\\
 N4258& 7.4& -3.32& 0.06& 0.18& -7.67& 0.31& 3.2\\
 M31& 0.86& -3.29& 0.07& 0.18& -4.85& 0.35& 3.2\\
LMC$^b$& 0.05& -3.31&0.017& 0.18& -& -& 3.2\\
SMC$^b$& 0.06&  -3.31&0.017& 0.18& -& -& 3.2\\
MW$^c$& 0& -3.26& 0.05& 0.18& -0.2& 0.12& 3.2\\

&&&&&&&\\
\hhline{=========}
&&&&&&&\\
\end{tabular} }
\end{adjustwidth}
{\footnotesize NOTE - (a) Distances from \href{https://ned.ipac.caltech.edu/}{NASA/IPAC Extragalactic Database}.  (b) No individual Cepheid metallicities were provided for LMC and SMC in R21 and thus we could not estimate the slope $Z_W$ for these hosts. (c) For the Milky Way (MW) we have used directly the values provided in R21 obtained from Gaia EDR3+HST.}
\end{table}

\renewcommand{\theequation}{\thesection.\arabic{equation}}
\renewcommand{\thetable}{\thesection.\arabic{table}}
\section[\appendixname~\thesection]{Model selection criteria}
\label{AppendixD}
\setcounter{equation}{0} 
\setcounter{table}{0} 

Various methods for model selection have been developed  and model comparison techniques used \cite{Liddle:2004nh,Liddle:2007fy,Arevalo:2016epc,Kerscher:2019pzk}.  The reduced $\chi^2$ is a very popular method for model comparison. This is defined by 
\be
\chi_{red}^2=\frac{\chi_{min}^2}{dof}
\ee
where $\chi_{min}^2$ is the minimum $\chi^2$ and $dof=N-M$ is typically the number of degrees of freedom (with $N$ is the number of datapoints used in the fit and $M$ is the number of free  parameters) for each model.

The model selection methods like  Akaike Information Criterion (AIC) \cite{akaike1974new} and the Bayesian Information Criterion (BIC) \cite{Schwarz:1978tpv} that penalize models with additional parameters are used. For a model with $M$ parameters and a dataset with $N$ total observations these are defined through the relations  \cite{Liddle:2004nh,Liddle:2007fy,Arevalo:2016epc}
\be 
AIC=-2ln\mathcal{L}_{max}+2M=\chi_{min}^2+2M
\label{aic}
\ee
\be 
BIC=-2ln\mathcal{L}_{max}+Mln{N}=\chi_{min}^2+Mln{N}
\label{bic}
\ee 
where $\mathcal{L}_{max}\equiv e^{-\chi_{min}^2/2}$  (e.g. \cite{John:2002gg,Nesseris:2012cq}) is the maximum likelihood of the model under consideration.

The "preferred model" is the one which minimizes AIC and BIC. The absolute values of the AIC and BIC are not informative. Only the relative values between  different competing models are relevant. Hence when comparing one model versus the baseline-SH$0$ES we can use the model differences  $\Delta$AIC and $\Delta$BIC.

The differences $\Delta$AIC and $\Delta$BIC  with respect to the baseline-SH$0$ES model defined as
\be 
\Delta AIC=AIC_i-AIC_s=\Delta \chi_{min}^2+2 \Delta M
\label{daic}
\ee
\be 
\Delta BIC=BIC_i-BIC_s=\Delta \chi_{min}^2+\Delta M (ln{N})
\label{dbic}
\ee 
where the subindex i refers to value of AIC (BIC) for the model i and $AIC_s$ ($BIC_s$) is the value of AIC (BIC) for the baseline-SH$0$ES model.  Note that a positive value of $\Delta$AIC or $\Delta$BIC means a preference for baseline-SH$0$ES model. 

According to the calibrated Jeffreys’ scales \cite{Jeffreys:1961} showed in the Table \ref{jefsc} (see also Refs. \cite{Liddle:2004nh,Nesseris:2012cq,BonillaRivera:2016use,Perez-Romero:2017njc,Camarena:2018nbr}) a range $0<|\Delta AIC|<2$ means that the two comparable  models have about the same support from the data, a range  $4<|\Delta AIC|<7$ means this support is considerably less for the model with the larger $AIC$ while for $|\Delta AIC|>10$ the model with the larger $AIC$ have no support i.e. the model is practically irrelevant. Similarly, for two competing models a range $0<|\Delta BIC|<2$ is regarded as weak evidence, a range $2<|\Delta BIC|<6$ is regarded as positive evidence, while for $|\Delta BIC|>6$ the evidence is strong against the model with the larger value.

We attribute the difference between $\Delta$AIC and $\Delta$BIC for the models considered to the fact that the BIC penalizes additional parameters more strongly than the AIC as inferred by the Eqs. (\ref{aic}) and (\ref{bic}) for the used dataset with $ln N>2$  (see Refs. \cite{Liddle:2004nh,Arevalo:2016epc,Rezaei:2019xwo}.

\begin{table}
\caption{The interpretation of differences $\Delta AIC$ and $\Delta BIC$ according to the calibrated Jeffreys’ scale \cite{Jeffreys:1961} (see also Refs. \cite{Liddle:2004nh,Nesseris:2012cq,BonillaRivera:2016use,Perez-Romero:2017njc,Camarena:2018nbr}). However, it should be noted that the Jeffreys’ scale has to be interpreted with care \cite{Nesseris:2012cq} because it has been shown to lead to diverse qualitative conclusions.}
\label{jefsc} 
\vspace{2mm}
\setlength{\tabcolsep}{0.1em}
\begin{adjustwidth}{0 cm}{1.cm}
\begin{tabular}{cccc|cccc}  
\hhline{========}
	&&    &&   &&   &   \\
\multicolumn{3}{c}{$\Delta AIC$ }  &&  \multicolumn{4}{c}{$\Delta BIC$ } \\
   &&     &&  &&    &   \\
    \hline 
    	&&    &&   &&   &  \\
\multicolumn{3}{c}{\footnotesize Level of empirical support for
the model  with the smaller $AIC$}  &&  \multicolumn{4}{c}{\footnotesize Evidence against the model with the larger $BIC$} \\
   &&     &&  &&    &  \\
  \hline 
  &&     &&  &&    &  \\ 
  $\;$0-2  	&$\qquad$  4-7 &$ >10$   &&$\quad$ 0-2  &$\quad$2-6& $\quad$6-10  &$ >10$  \\ 
 
$\,\,$ \footnotesize Substantial$\;$ & $\qquad$ \footnotesize Strong & \footnotesize Very strong&& $\quad$ \footnotesize Weak  &$\quad$ \footnotesize Positive& $\quad$ \footnotesize Strong &$\quad$ \footnotesize Very strong  \\ 
 &&     &&  &&    &  \\
       \hhline{========}   
\end{tabular} 
\end{adjustwidth}
\end{table}

\renewcommand{\theequation}{\thesection.\arabic{equation}}
\renewcommand{\thetable}{\thesection.\arabic{table}}
\section[\appendixname~\thesection]{}
\label{AppendixE}
\setcounter{equation}{0} 
\setcounter{table}{0} 

Here we present in a comprehensive and concise manner the 3130 Cepheid data that appear in the vector $\bf{Y}$ and in the modeling matrix $\bf{L}$. The sequence is the same as the one appearing in the top 3130 entries of the vector $\bf{Y}$ released in the form of fits files by R21. Even though no new information is presented in the following Table \ref{tab:hoscep} compared to the released fits files of R21, the concise presentation of these data may make them more useful in various applications and studies.\\

\vspace{1.8mm}
\setlength{\tabcolsep}{1.7em}
\begin{adjustwidth}{-2. cm}{2.5cm}
{\footnotesize \begin{longtable}[c]{  c c c c c c } 
\caption{Data for 3130 Cepheids in the host SnIa host galaxies and in the anchor or supporting galaxies N4258, M31, LMC, SMC from fits files in R21. An electronic version of the complete  table is available at the \href{https://github.com/FOTEINISKARA/A-reanalysis-of-the-SH0ES-data-for-H_0}{A reanalysis of the SH0ES data for $H_0$}  GitHub repository under the MIT license.} \\
\label{tab:hoscep}\\
\hhline{======}
 & & & & & \\
Galaxy & Ranking & ${\bar m}_H^W$&$\sigma$ &[P] &[O/H]\\ 
  &in Vector Y& [mag] & [mag]   & [dex] & [dex] \\ 
 & & & & & \\
\hhline{======}
 & & & & & \\
 M1337&  1& 27.52& 0.39& 0.45& -0.2\\
 M1337&  2& 28.05& 0.47& 0.65& -0.19\\
 M1337&  3& 26.56& 0.35& 0.6& -0.2\\
 M1337&  4& 26.97& 0.27& 0.78& -0.2\\
 M1337&  5& 27.01& 0.47& 0.72& -0.18\\
 M1337&  6& 26.29& 0.5& 0.9& -0.17\\
 M1337&  7& 26.12& 0.6& 0.89& -0.17\\
 M1337&  8& 26.94& 0.45& 0.75& -0.17\\
 M1337&  9& 27.78& 0.61& 0.69& -0.16\\
 M1337&  10& 26.17& 0.55& 0.7& -0.14\\
 M1337&  11& 26.63& 0.65& 0.49& -0.22\\
 M1337&  12& 27.4& 0.39& 0.85& -0.19\\
 M1337&  13& 27.26& 0.53& 0.52& -0.22\\
 M1337&  14& 27.34& 0.34& 0.85& -0.17\\
 M1337&  15& 27& 0.54& 0.7& -0.19\\
 N0691&  1& 26.59& 0.57& 0.4& 0.09\\
 N0691&  2& 27.1& 0.46& 0.61& 0.06\\
 N0691&  3& 26.79& 0.47& 0.71& 0.04\\
 N0691&  4& 26.83& 0.5& 0.78& 0.14\\
 N0691&  5& 26.23& 0.44& 0.51& 0.05\\
 N0691&  6& 27.16& 0.48& 0.63& 0.1\\
 N0691&  7& 26.89& 0.46& 0.7& 0.03\\
 N0691&  8& 26.75& 0.59& 0.43& 0.03\\
 N0691&  9& 27.11& 0.28& 0.88& 0.04\\
 N0691&  10& 26.99& 0.68& 0.62& 0.1\\
 N0691&  11& 27.6& 0.35& 0.81& 0.07\\
 N0691&  12& 27.33& 0.38& 0.75& 0.08\\
 N0691&  13& 26.6& 0.53& 0.54& 0.14\\
 N0691&  14& 26.34& 0.45& 0.74& 0.09\\
 N0691&  15& 26.91& 0.3& 0.96& 0.11\\
 N0691&  16& 26.79& 0.49& 0.84& 0.13\\
 N0691&  17& 27.08& 0.48& 0.63& 0.1\\
 N0691&  18& 27.07& 0.59& 0.63& 0.1\\
 N0691&  19& 26.69& 0.63& 0.53& 0.07\\
 N0691&  20& 27.94& 0.58& 0.49& 0.11\\
 N0691&  21& 26.15& 0.62& 0.48& 0.12\\
 N0691&  22& 26.46& 0.43& 0.74& 0.11\\
 N0691&  23& 27.16& 0.6& 0.42& 0.14\\
 N0691&  24& 26.96& 0.38& 0.8& 0.14\\
 N0691&  25& 26.59& 0.44& 0.65& 0.04\\
 N0691&  26& 26.61& 0.52& 0.5& 0.09\\
 N0691&  27& 26.21& 0.6& 0.62& 0.11\\
 N0691&  28& 27.42& 0.65& 0.65& 0.1\\
 ...& ...& ...& ...& ...& ...\\
\hhline{======}
\end{longtable} }

\end{adjustwidth}

\begin{adjustwidth}{-\extralength}{0cm}

\printendnotes[custom]

\reftitle{References}




\externalbibliography{yes}
\bibliography{Bibliography.bib}

\end{adjustwidth}


\end{document}